\documentclass[11pt, superscriptaddress,nofootinbib]{revtex4-2}
\usepackage{graphics,epsfig}
 \usepackage{float}
\usepackage{subfloat}
\usepackage{graphicx}
 \usepackage{subfig}
\usepackage{amsmath} 
\usepackage{hyperref}
\usepackage{color}
\begin{document}

\title{Charged $AdS$ Black Holes in $4D$ Einstein--Gauss--Bonnet  Massive Gravity}
\author{Prosenjit Paul}
\email{prosenjitpaul629@gmail.com}
\affiliation{Indian Institute Of Engineering Science and Technology (IIEST), Shibpur, WB 711103, India}
\author{Sudhaker Upadhyay\footnote{Corresponding author}\footnote{Visiting Associate, IUCAA Pune, Maharashtra-411007, India.}}\email{sudhakerupadhyay@gmail.com}
\affiliation{Department of Physics, K. L. S. College, Magadh University, Nawada, Bihar 805110, India} 
\affiliation{School of Physics, Damghan University, Damghan, 3671641167, Iran}

\author{Dharm Veer Singh\footnote{Visiting Associate, IUCAA Pune, Maharashtra-411007, India.}}
\email{veerdsingh@gmail.com}

\affiliation{Department of Physics, Institute of Applied Science and Humanities, GLA University, Mathura, 281406 India}

\begin{abstract}
 We investigate Einstein--Gauss--Bonnet--Maxwell massive gravity in $4D$ AdS background and find an exact black hole solution. The horizon structure of the black holes studied. Treating the cosmological constant as pressure and Gauss-Bonnet  coupling parameters, and massive gravity parameters as variables, we drive the first law of black hole thermodynamics. To study the global stability of the black holes we compute the Gibbs free energy. The local stability of the black hole is also studied through specific heat. We analyze the effects of graviton mass and Gauss-Bonnet  coupling parameters on the phase transition of the black holes. Finally, the effects of graviton mass and massive gravity parameters on the Joule-Thomson expansion of the black hole 
 are studied.  
 \end{abstract}
 \maketitle

\section{Introduction}\label{sec:1}
General Relativity (GR) is a theory of gravitation that help us understand gravitational waves, gravitational lensing, an effect of gravity on time known as gravitational time dilation and black holes. Although GR is not a complete theory of quantum gravity, the simplest theory of gravity describes various astrophysical phenomena. To complete the GR, people try to modify it in various ways, for instance, by adding the higher order term to the Einstein-Hilbert action of GR but still, a complete theory is missing. Some 
 other examples of higher order gravity theory are  scalar-tensor theories \cite{Barrabes:1997kk,Cai:1996pj,Capozziello:2005bu,Sotiriou:2006hs,Moffat:2005si,Faraoni:2007yn},  Lovelock gravity \cite{Lovelock:1971yv,Lovelock:1972vz,Deruelle:1989fj}, regular black holes \cite{010,11,12,13} and  brane world cosmology \cite{Cline:2000xn,Nihei:2004xv,Demetrian:2005sr}.  

Lovelock theories \cite{Lovelock:1971yv,Lovelock:1972vz} is a special theory of higher-order gravity in $4D$ spacetime that preserves diffeomorphism invariance, metricity and second-order equations of motion. From Lovelock's theories of gravity, Gauss-Bonnet  gravity can be obtained in higher dimensions \cite{Lanczos:1938sf}. The Gauss-Bonnet  term does not contribute to the dynamics of the theory in four dimensions but rather  contributes to the dynamics when the dimensions of spacetime are greater than four. In recent days,  Glavan and Lin \cite{glavan2020einstein} found the solution to the Einstein-Gauss--Bonnet  field equation in four dimensions by rescaling Gauss-Bonnet  coupling parameter $\alpha$ by $\alpha/D-4$. However, the charged AdS solution of Einstein-Gauss--Bonnet  theory was found in Ref. \cite{fernandes2020charged}. For a complete discussion on $4D$ Gauss-Bonnet  gravity, see Refs. \cite{Fernandes:2022zrq}. Some other static spherically symmetric black hole solutions and their thermodynamics, phase transition in $4D$ or higher dimensions Einstein-Gauss--Bonnet  gravity studied in Refs. \cite{Hegde:2020xlv,Wei:2020poh,Wang:2020pmb,Singh:2020nwo,Singh:2020xju,EslamPanah:2020hoj,Singh:2021xbk,Godani:2022jwz}. Einstein-Gauss--Bonnet  black hole solution in nonlinear electrodynamics is studied  in Refs. \cite{Hendi:2014lke,Yang:2020jno,Ghosh:2020ijh,Kruglov:2021btd,Kruglov:2021pdp,Kruglov:2021qzd,Kruglov:2021rqf,Kruglov:2021stm,Singh:2022dth}.

Another way of modifying the GR is by adding a mass to the graviton. According to the 
GR, the graviton is a massless spin-2 particle. But one may ask if that is a self-
a consistent theory of massive gravity possible or not. In fact, people have tried to 
answer this question by modifying the Einstein-Hilbert action that describes massive 
graviton. The recent observation of gravitational waves by LIGO put a maximum limit on 
the graviton mass, $m\le 1.2 \times 10^{-22}$eV \cite{LIGOScientific:2016lio}. A 
theory of massive gravity was first constructed by Fierz and Pauli in 1939 
\cite{Fierz:1939ix, Fierz:1939zz}. In the curved background, this theory encounters ghost instabilities \cite{Boulware:1972yco}. A new  nonlinear massive gravity theory proposed by de Rham, Gabadadze and Tolley (dRGT) \cite{deRham:2010ik,deRham:2010kj}
avoids ghost problems. The charged black holes in Gauss--Bonnet  massive gravity 
studied \cite{Hendi:2015pda}. Some other spherically symmetric black holes in massive 
gravity and their thermodynamics have also been studied 
\cite{Hendi:2015bna,Hendi:2016hbe,Upadhyay:2018vfu,Hendi:2018hdo,Singh:2020rnm,Upadhyay:2022axg}.

In our theory, we consider the anti-de-Sitter (AdS) background, i.e. we add a negative cosmological constant. The negative cosmological constant is a crucial ingredient in the AdS/CFT correspondence, a duality between a theory of quantum gravity in  (AdS) space and a conformal field theory (CFT) in one lower dimension. The AdS/CFT correspondence allows for the study of strongly coupled field theories using classical gravity, which is a useful tool for investigating non-perturbative phenomena that cannot be understood through standard perturbative methods. One significant consequence of the negative cosmological constant is that it leads to the presence of a holographic screen at the AdS boundary, which encodes the bulk geometry's information. This holographic principle means that the number of degrees of freedom in the AdS space is proportional to the area of the holographic screen, rather than the volume, as in ordinary theories. The AdS/CFT correspondence, therefore, implies that the degrees of freedom in the AdS space are equivalent to those of the boundary CFT. The negative cosmological constant also plays a critical role in the AdS black hole physics. Black holes in AdS space can have a negative specific heat, which is impossible in flat space. This phenomenon is related to the AdS space's boundary conditions, which force the black hole to lose energy and mass through the AdS boundary, leading to a reduction in temperature. This behavior is known as Hawking-Page phase transition \cite{Hawking:1982dh}, where the black hole is in thermal equilibrium with a thermal AdS space. The AdS/CFT correspondence allows for the study of the thermodynamic behaviour of black holes using the corresponding CFT, providing insights into the nature of black hole thermodynamics.

Recently, researchers have considered the cosmological constant as a variable parameter and linked it to the thermodynamic pressure, which is conjugate to the thermodynamic volume \cite{Wang:2006eb,Kastor:2009wy,Kastor:2010gq,Dolan:2010ha,Dolan:2011xt}. This approach has resulted in an extended phase space, where the black hole mass is regarded as the enthalpy, instead of the internal energy \cite{Kastor:2009wy}. Many studies have explored the thermodynamics and phase transitions of black holes in this extended phase space, revealing new phenomena such as P-V criticality in various black holes spacetime \cite{Kubiznak:2012wp,Gunasekaran:2012dq}.

In AdS spacetime, the black hole mass is naturally treated as the enthalpy, leading to the consideration of Joule-Thomson expansion for the black hole. This expansion investigates isenthalpic curves, which are constant mass curves. Previous investigations of the Joule-Thomson expansion for charged AdS \cite{Okcu:2016tgt} and Kerr AdS \cite{Okcu:2017qgo} black holes have been conducted within the framework of Einstein gravity. However, extended theories of Einstein gravity introduce new physical degrees of freedom, raising questions about their role and physical impact on the Joule-Thomson expansion. In this paper, we investigate how the presence of massive gravity modifies the Joule-Thomson expansion of the charged AdS black hole in Gauss--Bonnet gravity, inspired by recent progress in understanding massive gravity. The approach taken here is relevant not only for charged AdS black holes in Eiantein-Gauss-Bonnet massive gravity but also for those in other alternative theories of gravity where additional gravitational modes emerge. For instance, the Joule-Thomson expansion may be examined for charged AdS black holes in teleparallel $f(T)$ gravity \cite{Capozziello:2019uvk,Nashed:2017fnd}, with $T$ representing torsion, or in $f(R)$ gravity \cite{Nashed:2019tuk} with a nonlinear electrodynamics field.

Since the charged AdS Einstein-Gauss--Bonnet theory in  massive gravity is not studied yet. Therefore, in this paper, we investigate charged Einstein-Gauss--Bonnet  massive gravity and find an exact solution in $4D$ AdS space. we also discuss the horizon structure of charged AdS black hole in $4D$ Einstein--Gauss--Bonnet massive gravity. Moreover, we discuss the thermal properties of this black hole.  To be more precise, 
we compute the entropy and temperature that satisfy the first law of black hole thermodynamics. In AdS space, the mass of the black hole is treated as enthalpy. Furthermore, we analyzed the stability and Van der walls like phase transition of the black holes.  Here, we study the Joule-Thomson expansion of charged AdS black hole in 4D Einstein--Gauss--Bonnet  massive gravity, and the constant mass curves are known as  isenthalpic curves. 
Finally, we investigate the effects of graviton mass and the massive gravity parameters on the Joule-Thomson expansion of charged AdS black holes in $4D$ Einstein--Gauss--Bonnet  massive gravity.

The paper is organized as follows. In section \ref{sec:2}, 
we discuss the action describing the Einstein--Gauss--
Bonnet--Maxwell massive gravity in $4D$ AdS space and
their field equations. Here, we find the exact black hole 
solution. The effects of graviton mass on the horizon 
structure of the black hole are also depicted. In section 
\ref{sec:3}, we study the first law of black hole 
thermodynamics  and the effects of graviton mass on Hawking 
temperature. To investigate the global stability of the 
black holes, we compute the Gibbs free energy. Next, in 
section \ref{sec:4}, the effects of graviton mass on the 
local stability of the black hole are studied. The Van der 
walls-like phase transition of the black hole is analyzed 
in section \ref{sec:5}. We numerically investigate the 
effects of the graviton mass, the charge of the black hole 
and Gauss--Bonnet  coupling parameter on the critical 
parameters (namely, critical volume, critical pressure and 
critical volume) of the black hole. The  effects of 
critical parameters on the phase transition of the black 
hole are also studied. we investigate the Joule-Thomson 
expansion of the black hole  in section \ref{sec:6}. Here, 
we analyze the effects of graviton mass and massive gravity 
parameters on the constant mass curve and inverse curve. 
Finally, we compute Joule-Thomson thermodynamic coefficient 
as a function of the black hole horizon radius.

\section{Einstein-Gauss--Bonnet  Massive Gravity in {4D}}\label{sec:2}
\begin{figure}[hbt]
\centering
\subfloat[ $M=5$ and $\alpha=0.5$]{\includegraphics[width=6.5cm,height=5.0cm]{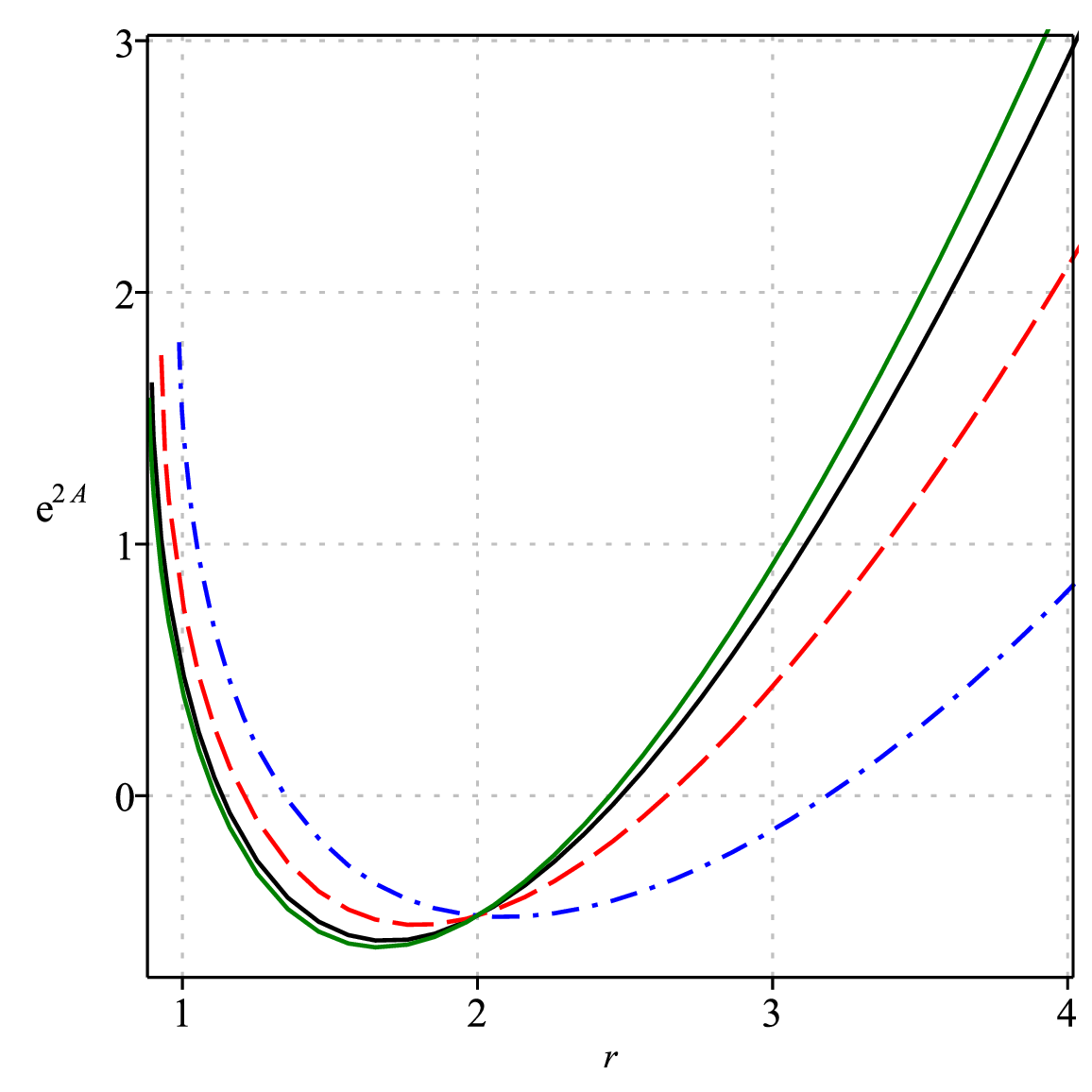}}\hfill
\subfloat[$M=5$ and $\alpha=0.8$]{\includegraphics[width=6.5cm,height=5.0cm]{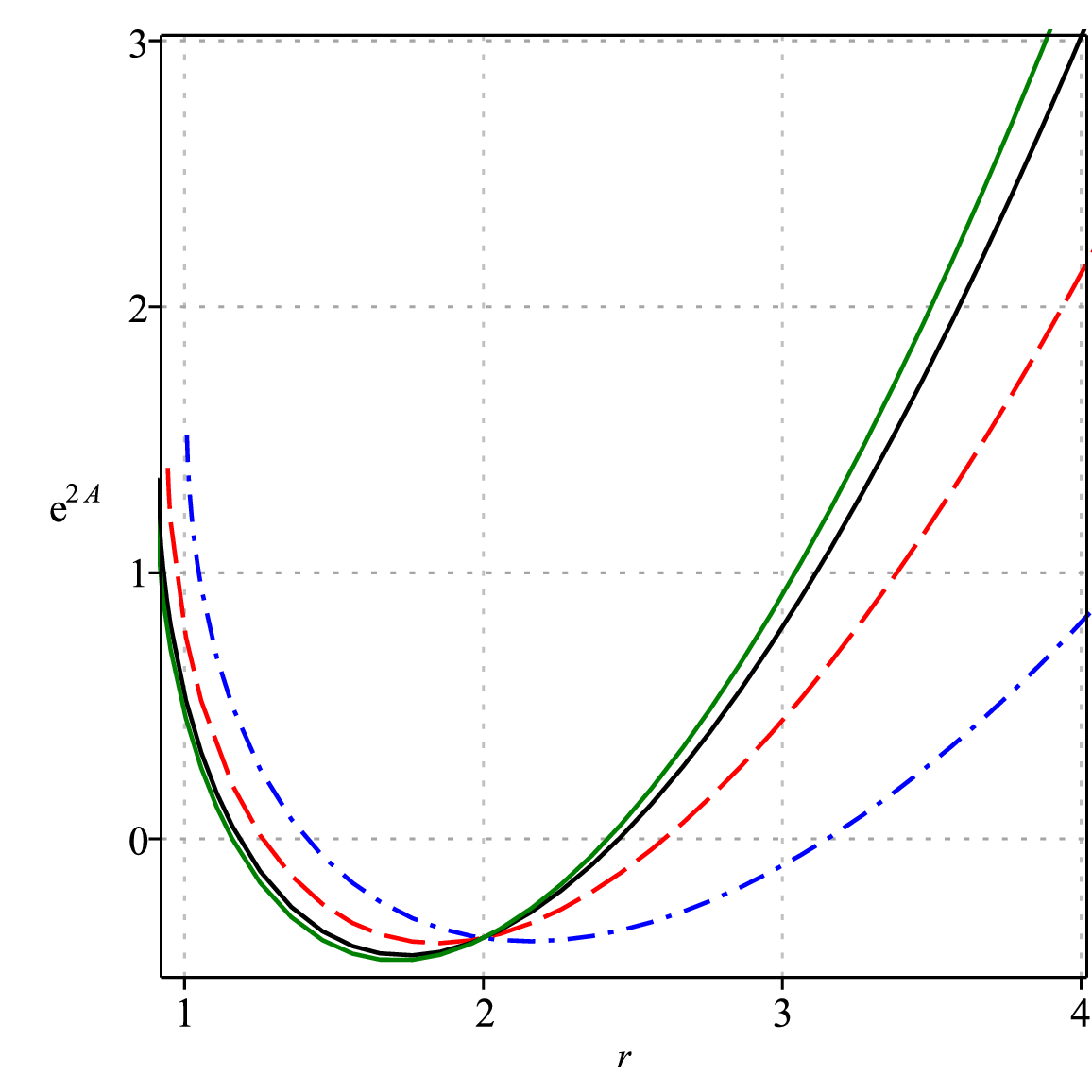}}\hfill
\subfloat[$M=3$ and $\alpha=0.5$]{\includegraphics[width=6.5cm,height=5.0cm]{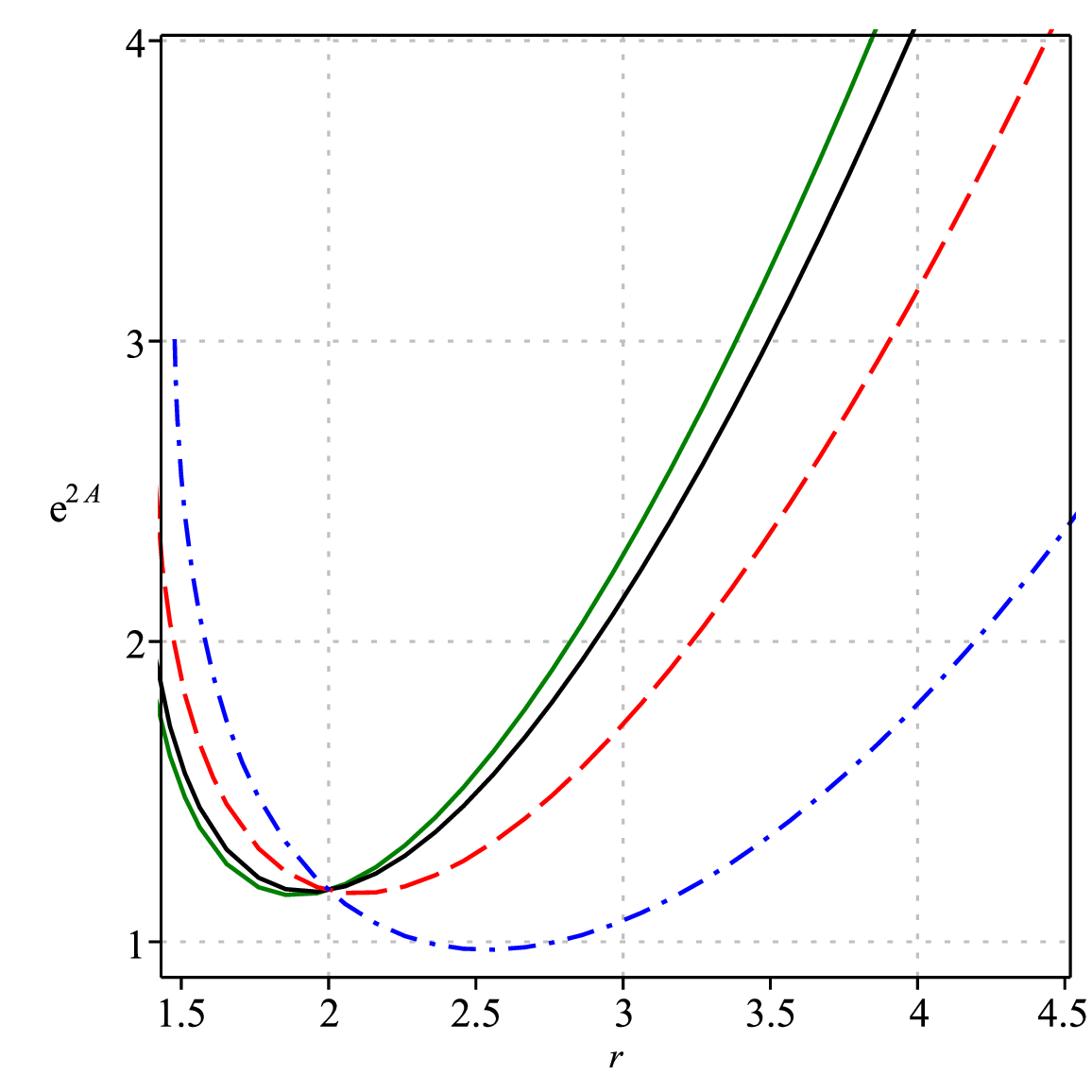}}\hfill
\subfloat[$M=1$ and $\alpha=0.5$]{\includegraphics[width=6.5cm,height=5.0cm]{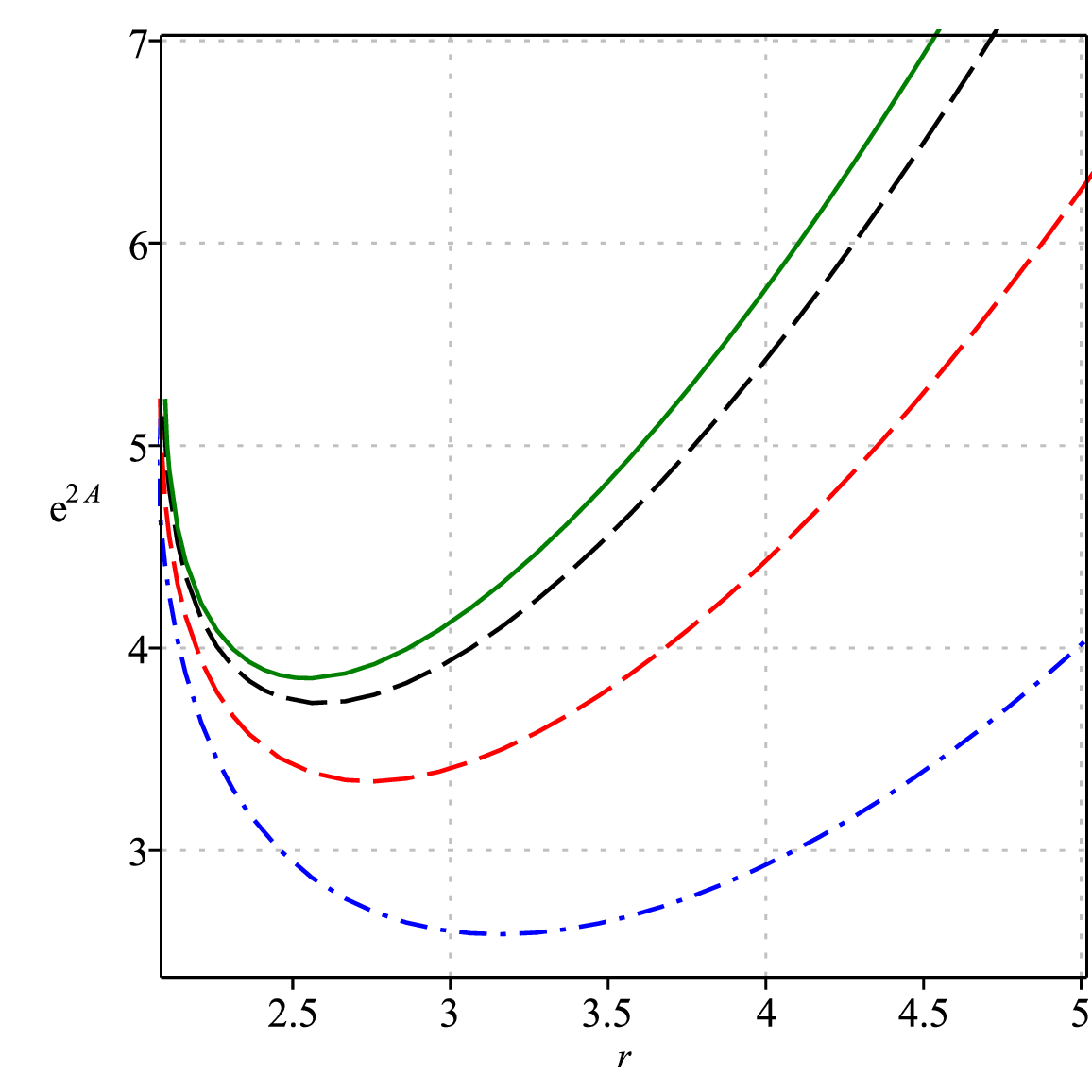}}\hfill
\caption{$m=0.0$ denoted by solid green line, $m=0.5$ denoted by Solid black line, $m=1.0$ denoted by dash red line and  $m=1.5$ denoted by dash dot blue line with $l=2$, $Q=3$, $c=1$, $c_{1}=-1$ and $c_2=1$.}\label{fig:1}
\end{figure}

The action for Einstein--Maxwell--Gauss--Bonnet  massive 
gravity with a negative cosmological constant in $D$ dimensions  is 
given by 
\begin{equation}
S= \frac{1}{16 \pi} \int d^{D}x \sqrt{-g} \Biggr[R -2 \Lambda + \alpha 
\mathcal{G} - F_{\mu \nu}F^{\mu \nu} + m^2 \sum_{i} c_{i} \mathcal{U}_{i}
(g,h)    \Biggr],
\end{equation}
where $g$ is determinant of the metric $g_{\mu \nu}$, $R$ is Ricci 
scalar, $\alpha$ is Gauss--Bonnet  coupling parameter, $\mathcal{G}= 
R_{\mu \nu \rho \sigma} R^{\mu \nu \rho \sigma} -4 R_{\mu \nu } R^{\mu 
\nu } + R^2$ is the Gauss--Bonnet  term, $R_{\mu \nu \rho \sigma}$ is  
Riemann tensor, $R_{\mu \nu }$ is Ricci tensor and $F_{\mu 
\nu}=\partial_{\mu}{A_{\nu}}-\partial_{\nu}{A_{\mu}}$ is Maxwell tensor. 
Apart from that $m$ is a parameter related to graviton mass, $ h_{\alpha \nu}$  is a fixed symmetric tensor and usually is called the reference metric, $c_{i}(i=1,2,3,4)$ are constant\footnote{In order to have a self-consistent massive gravity theory, the coupling parameters $c_i$ might be required to be negative if the squared mass of the graviton is positive. However, in the AdS spacetime, the coupling parameters $c_i$ can still take the positive values. This is because the fluctuations of the fields with the negative squared masses in the AdS spacetime could still be stable if their squared masses obey the corresponding
Breitenlohner–Freedman bounds.} \cite{nam2020effect}  and 
$\mathcal{U}_{i}(g,h)$ is  symmetric polynomials of eigenvalues of 
matrix $\mathcal{K}_{\nu}^{\mu}= \sqrt{g^{\mu \alpha} h_{\alpha \nu}}$, 
given by
\begin{equation} \label{eq:2.2}
\begin{split}
\mathcal{U}_{1} & = \bigr[  \mathcal{K} \bigr],\\
\mathcal{U}_{2} & = \bigr[  \mathcal{K} \bigr]^{2} -  \bigr[  
\mathcal{K}^2 \bigr],\\
\mathcal{U}_{3} & = \bigr[  \mathcal{K} \bigr]^{3} - 3\bigr[  \mathcal{K}
\bigr] \bigr[  \mathcal{K}^2 \bigr] + 2  \bigr[  \mathcal{K}^{3} 
\bigr], \\
\mathcal{U}_{4} & = \bigr[\mathcal{K} \bigr]^{4} - 6\bigr[\mathcal{K}^2 
\bigr] \bigr[  \mathcal{K}\bigr]^2 + 8 \bigr[\mathcal{K}^{3}\bigr] \bigr[\mathcal{K}\bigr] + 3 \bigr[\mathcal{K}^2\bigr]^2 -6 \bigr[\mathcal{K}^4\bigr],
\end{split}
\end{equation}
where parentheses $[...]$ represents trace of the matrix $\mathcal{K}_{\nu}^{\mu}$. In $D=4$ dimensions the Gauss--Bonnet  term does not contribute to the dynamics, so we rescale the Gauss--Bonnet  coupling parameter $\alpha \to \alpha/(D-4)$ \cite{glavan2020einstein}, therefore, the action takes the following form:
\begin{equation}\label{eq:2.3}
S= \frac{1}{16 \pi} \int d^{D}x \sqrt{-g} \left[R -2 \Lambda + \frac{\alpha}{D-4} \mathcal{G} - F_{\mu \nu}F^{\mu \nu} + m^2 \sum_{i} c_{i} \mathcal{U}_{i}(g,h)    \right].
\end{equation}
Now, we consider a static and spherically symmetric solution of the form 
\begin{equation}\label{eq:2.4}
    ds^{2}= - e^{2A(r)} dt^2 +  e^{2B(r)} dr^2 +r^2 d{\Omega}_{D-2}^2,
\end{equation}
and following the Ref. \cite{Cai:2014znn} we take the  reference metric as
\begin{equation}\label{eq:2.5}
    h_{\mu \nu}= diag\bigl( 0, 0, c^2, c^2\sin^{2}\theta   \bigl),
\end{equation}
where $c$ is a dimensionless positive constant. The reference metric $h_{\mu \nu}$ is a rank two symmetric tensor. Physically, $h_{\mu \nu}$ corresponds to the background metric around which fluctuations take the Fierz–Pauli form. By using equations \eqref{eq:2.2} and \eqref{eq:2.5}, we obtain
\begin{equation} \label{eq:2.6}
\begin{split}
\mathcal{U}_{1} & = \frac{(D-2)c}{r}, \\
\mathcal{U}_{2} & = \frac{(D-2)(D-3)c^{2}}{r^{2}},\\
\mathcal{U}_{3} & = \frac{(D-2)(D-3)(D-4)c^{3}}{r^{3}}, \\
\mathcal{U}_{4} & = \frac{(D-2)(D-3)(D-4)(D-4)c^{4}}{r^{4}}.
\end{split}
\end{equation}
Substituting the metric and the electrostatic potential in action \eqref{eq:2.3}, 
the first integral exists     \cite{fernandes2020charged}
\begin{equation}\label{eq:2.7}
   \phi^{\prime}(r)= - \frac{Q}{r^{D-2}} e^{A+B}, 
\end{equation}
and taking the limit $D \to 4$ and using the relation $\Lambda=-3/l^2$ we obtain
\begin{equation}\label{eq:2.8}
S= \frac{\Sigma_{2}}{16 \pi} \int dt dr 2 e^{A+B}  \biggr[ r^{3} \psi \Bigl( 1+ \alpha \psi\Bigl) + \frac{r^{3}}{l^2} + \frac{Q^2 }{r}  + m^2 \Bigl\{ \frac{c_{1} c r^{2}}{2} + {c_{2}c^2 r}\Bigl\}   \biggr]^{\prime},
\end{equation}
where prime denotes differentiation with respect to r, $\Sigma_{2}=\frac{2{\pi}^\frac{3}{2}}{\Gamma\Bigl( {1+\frac{1}{2}}\Bigl)}$ and $\psi= r^{-2} \Bigl( 1- e^{-2B} \Bigl)$ with 
\begin{equation}\label{eq:2.9}
    e^{A+B}=1.
\end{equation}
If we choose $m=0$ or $c=0$ then equation \eqref{eq:2.8} reduced to action in massless gravity. Now, using the action \eqref{eq:2.8}, we obtain solution as 
\begin{equation}\label{eq:2.10}
  \psi  \Bigl( 1+ \alpha \psi\Bigl) + \frac{1}{l^2} + \frac{Q^2 }{r^4}  + \frac{m^2}{r^3} \Bigl\{ \frac{c_{1} c r^{2}}{2} + {c_{2}c^2 r}\Bigl\} - \frac{8 \pi M}{\Sigma_{2}r^3}=0,
\end{equation}
where $ M$ is the integration constant  related to the mass of the black hole. Therefore, the exact solution is 
\begin{equation}\label{eq:2.11}
e^{2A}= e^{-2B}= 1+ \frac{r^2}{2\alpha} \Biggr[1 \pm \sqrt{1+4\alpha \biggl\{ \frac{2M}{r^3} - \frac{Q^2}{r^4} - \frac{1}{l^2} - \frac{m^2}{2r^2}\Bigl( c c_{1} r + 2 c^2 c_{2}\Bigl)  \biggl\} }  \Biggr].
\end{equation}
The negative branch corresponds to the $4D$ charged AdS EGB massive black hole, whereas the +ve branch does not lead to a physically meaningful solution because the positive sign in the mass term indicates graviton instabilities, so we only take the negative branch of equation \eqref{eq:2.11}. For the chargeless limit, our solution reduces to a black hole solution as obtained in Ref. \cite{Upadhyay:2022axg} 
\begin{equation}\label{eq:2.12}
e^{2A}= e^{-2B}= 1+ \frac{r^2}{2\alpha} \Biggr[1 - \sqrt{1+4\alpha \biggl\{ \frac{2M}{r^3} - \frac{1}{l^2} - \frac{m^2}{2r^2}\Bigl( c c_{1} r + 2 c^2 c_{2}\Bigl)  \biggl\} }  \Biggr].
\end{equation}
In the limit $\alpha \to 0$, equation \eqref{eq:2.11} reduces to the charged AdS black hole in massive gravity \cite{Nam:2020gud}
\begin{equation}\label{eq:2.13}
e^{2A}= e^{-2B}=1- \frac{2M}{r} +\frac{Q^2}{r^2} + \frac{r^2}{l^2}+ \frac{m^2}{2} \bigl(  c c_{1} r + 2 c^2 c_{2}   \bigl).
\end{equation}
Also, in the massless limit, the above equation reduces to Reissner–Nordstr\"om AdS solution 
\begin{equation}\label{eq:2.14}
    e^{2A}= e^{-2B}=1- \frac{2M}{r} +\frac{Q^2}{r^2} + \frac{r^2}{l^2}.
\end{equation}
Now, we apply a massless limit to equation \eqref{eq:2.11} and obtain charged AdS
Black Hole in $4D$ Einstein-Gauss--Bonnet gravity \cite{fernandes2020charged} as
\begin{equation}\label{eq:2.15}
    e^{2A}= e^{-2B}= 1+ \frac{r^2}{2\alpha} \Biggr[1 \pm \sqrt{1+4\alpha \biggl\{ \frac{2M}{r^3} - \frac{Q^2}{r^4} - \frac{1}{l^2}   \biggl\} }  \Biggr].
\end{equation}
To find the position of the event horizon of charged AdS Black Hole in $4D$ Einstein-Gauss--Bonnet gravity, we set equation \eqref{eq:2.15} equal to zero and obtain \cite{fernandes2020charged}
\begin{equation}\label{eq:2.16}
    1-\frac{2M}{r} + \frac{Q^2+\alpha}{r}+ \frac{r^2}{l^2}=0.
\end{equation}
In the absence of cosmological constant, we obtain 
\begin{equation}\label{eq:2.17}
    r_{\pm}= M \pm \sqrt{M^2-Q^2-\alpha}.
\end{equation}
For Einstein-Gauss--Bonnet  massive gravity with a nonvanishing cosmological constant the expression for $r_{+}$ is complicated, so we do not represent it here. From equation \eqref{eq:2.17}, we can say that the black hole solution in $4D$ EGB massless gravity exists if and only if $M>M_{*}$ with $M_{*}^2= Q^2 + \alpha$. We will loosely follow the condition $M>M_*$ for charged AdS Einstein-Gauss--Bonnet  massive gravity black holes. In Fig. \ref{fig:1}, we plot the metric function (negative branch) of charged AdS Einstein--Gauss--Bonnet massive gravity black holes for different values of $\alpha$ and M. From Fig. \ref{fig:1} (a) and Fig. \ref{fig:1} (b), it is clear that the black hole has two horizons, as the value of graviton mass increases the position of the outer horizon increases. The position of the event horizon increases as massive gravity parameters increase. In table \ref{table:1} two roots of the metric function \eqref{eq:2.11} are estimated, the position of the horizon slowly decreases as Gauss--Bonnet coupling parameter increases. In Fig. \ref{fig:1} (c) and Fig. \ref{fig:1} (d), we plot the metric function, and it is clear that there are no horizon and no black hole solutions.

\begin{table}[hbt]
\centering
\begin{tabular}{ |p{1.5cm}|p{1.5cm}|p{1.5cm}| } 
\hline
\multicolumn{3}{|c|}{\textbf{Fig. \ref{fig:1}(a)}} \\
\hline
\textbf{$m$} & \textbf{$r_{-}$} & \textbf{$r_{+}$} \\ [0.5ex]  
\hline
0.0 & 1.1118 & 2.4489  \\ \hline
0.5 & 1.1337 & 2.4867  \\ \hline
1.0 & 1.2061 & 2.6419  \\ \hline
1.5 & 1.3470 & 3.1853  \\ \hline
\multicolumn{3}{|c|}{\textbf{Fig. \ref{fig:1}(b)}} \\
\hline
\textbf{$m$} & \textbf{$r_{-}$} & \textbf{$r_{+}$} \\ [0.5ex]  
\hline
0.0 & 1.1597 & 2.4165  \\ \hline
0.5 & 1.1833 & 2.4532  \\ \hline
1.0 & 1.2609 & 2.6055 \\ \hline
1.5 & 1.4091 & 3.1500 \\ \hline

\end{tabular}
\caption{}
\label{table:1}
\end{table}

\begin{figure}[hbt]
    \centering
    \includegraphics[width=6.5cm,height=5.5cm]{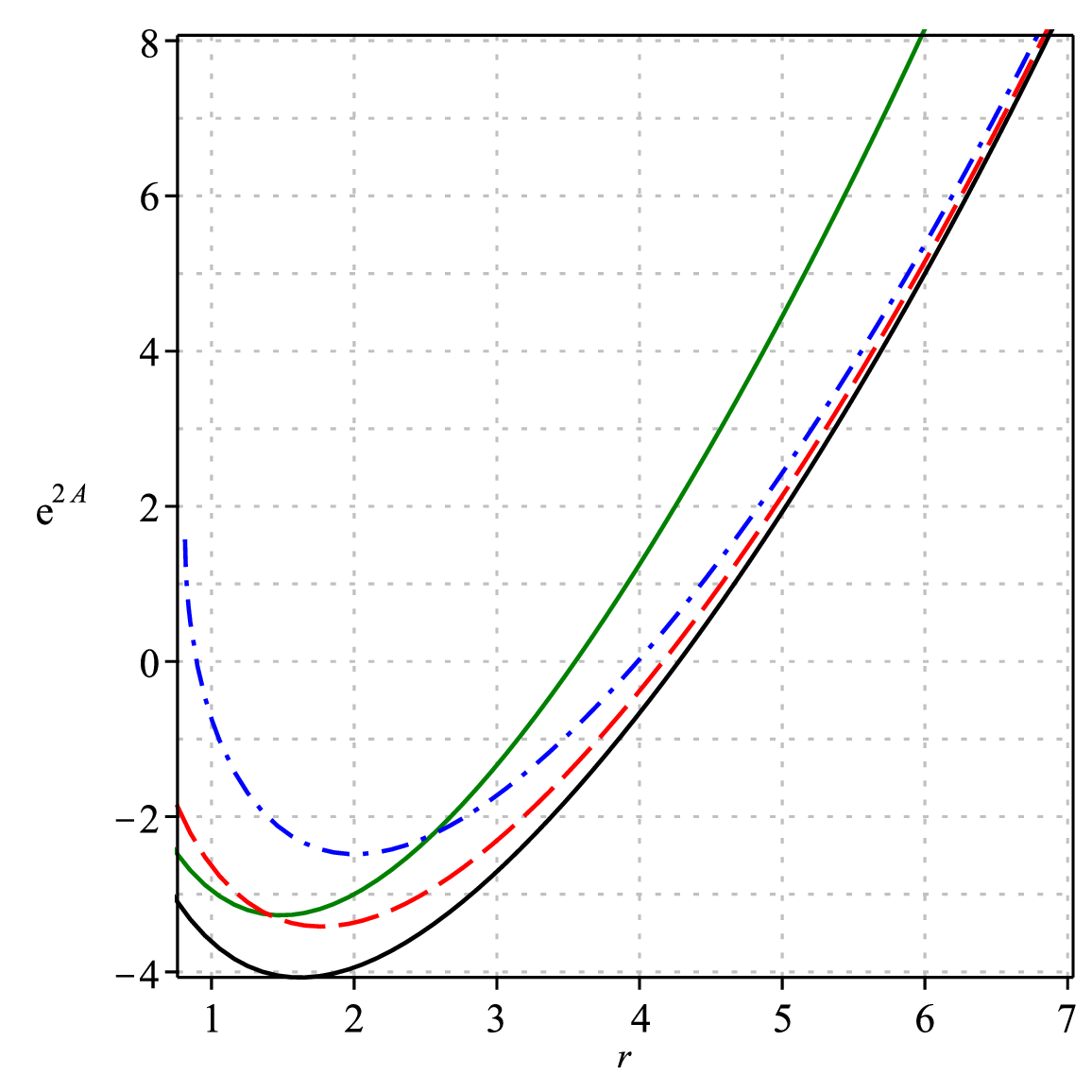}
    \caption{$m=0.0$ and $Q=2$ denoted by solid green line, $m=1$ and $Q=2$ denoted by Solid black line, $m=1$ and $Q=3$ denoted by dash red line, $m=1$ and $Q=4$ denoted by dash dot blue line with $M=10$, $\alpha=0.5$ with $l=2$, $c=1$, $c_{1}=-1$ and $c_2=1$.}
    \label{fig:2}
\end{figure}

In Fig. \ref{fig:2}, we plot the metric function (negative branch) of charged AdS Einstein--Gauss--Bonnet massive gravity in $4D$ for different values of the charge. From the figure, it is clear that the position of the outer horizon is the smallest for the massless case. As we increase the charge and keep the graviton mass fixed, the position of the outer horizon decrease but the position of the outer horizon is still greater than the massless one which is represented by the solid green line.
\section{Black Hole Thermodynamics}\label{sec:3}  
In this section, we study the thermodynamics of charged AdS black holes in $4D$ Einstein-Gauss--Bonnet  massive gravity. The physical mass of the black holes can be obtained from the metric function \eqref{eq:2.11} by setting $\left. e^{-2B}\right|_{r=r_+}=0$ as 
\begin{equation}\label{eq:3.1}
M=\frac{1}{2 r_{+} } \biggr[\frac{ r_{+}^{4}}{l^2} + r_{+}^{2}+Q^{2}+\alpha + m^2 r_{+}^2 \Bigl(\frac{  c c_{1} r_{+}}{2}+ c^{2} c_{2}  \Bigl)  \biggr].
\end{equation}
The Hawking temperature of the black holes can be obtained from the relation
\begin{equation}\label{eq:3.2}
 T_{H}= \frac{f^{\prime}(r_{+})}{4 \pi}. 
\end{equation}
For the metric   \eqref{eq:2.11},  Hawking temperature reads
\begin{equation}\label{eq:3.3}
T_{H}= \frac{3 r_{+}^{4} +l^{2} \bigl( r_{+}^{2}-Q^{2}-\alpha  + m^{2}r_{+}^2( c c_{1} r_{+}^{3}+c^{2} c_{2} ) \bigl) }{4 \pi r_{+} l^{2} (r_{+}^{2}+2 \alpha )}.
\end{equation}
In the massless limit, the Hawking temperature of charged AdS black holes in Einstein-Gauss--Bonnet  $4D$ massive gravity  reduces to the Hawking temperature \cite{fernandes2020charged} of charged AdS black holes in Einstein--Gauss--Bonnet $4D$ gravity as
\begin{equation}\label{eq:3.4}
    T_{H}=\frac{3 r_{+}^{4} +l^{2} \bigl( r_{+}^{2}-Q^{2}-\alpha   \bigl) }{4 \pi r_{+} l^{2} (r_{+}^{2}+2 \alpha )}.
\end{equation}
If we take the limit $\alpha \to 0$, the Hawking temperature  \eqref{eq:3.4}   reduces to the Hawking temperature of Reissner–Nordstr\"om AdS black holes as
\begin{equation}
    T_{H}= \frac{3 r_{+}^{4} +l^{2} \bigl( r_{+}^{2}-Q^{2} \bigl) }{4 \pi r_{+}^3 l^{2} }.
\end{equation}
If we further take the chargeless limit, then the above equation reduces to the Hawking temperature of Schwarzschild AdS black holes. 
  \begin{figure}[hbt]
\centering
\subfloat[$\alpha=0.5$, $Q=1$ and $c_{1}=0$]{\includegraphics[width=6.5cm,height=5.5cm]{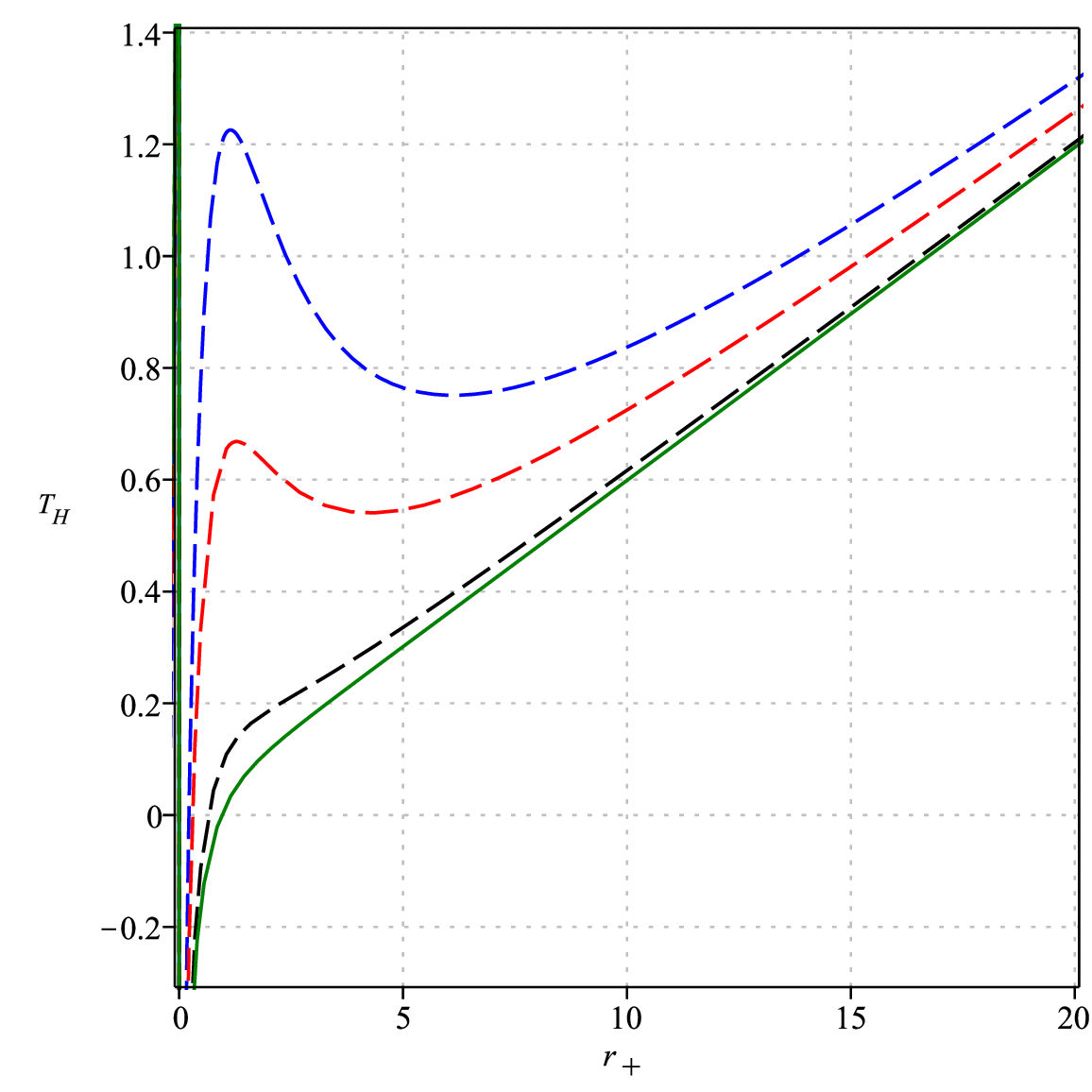}}\hfill
\subfloat[$\alpha=0.8$, $Q=1$ and $c_{1}=0$]{\includegraphics[width=6.5cm,height=5.5cm]{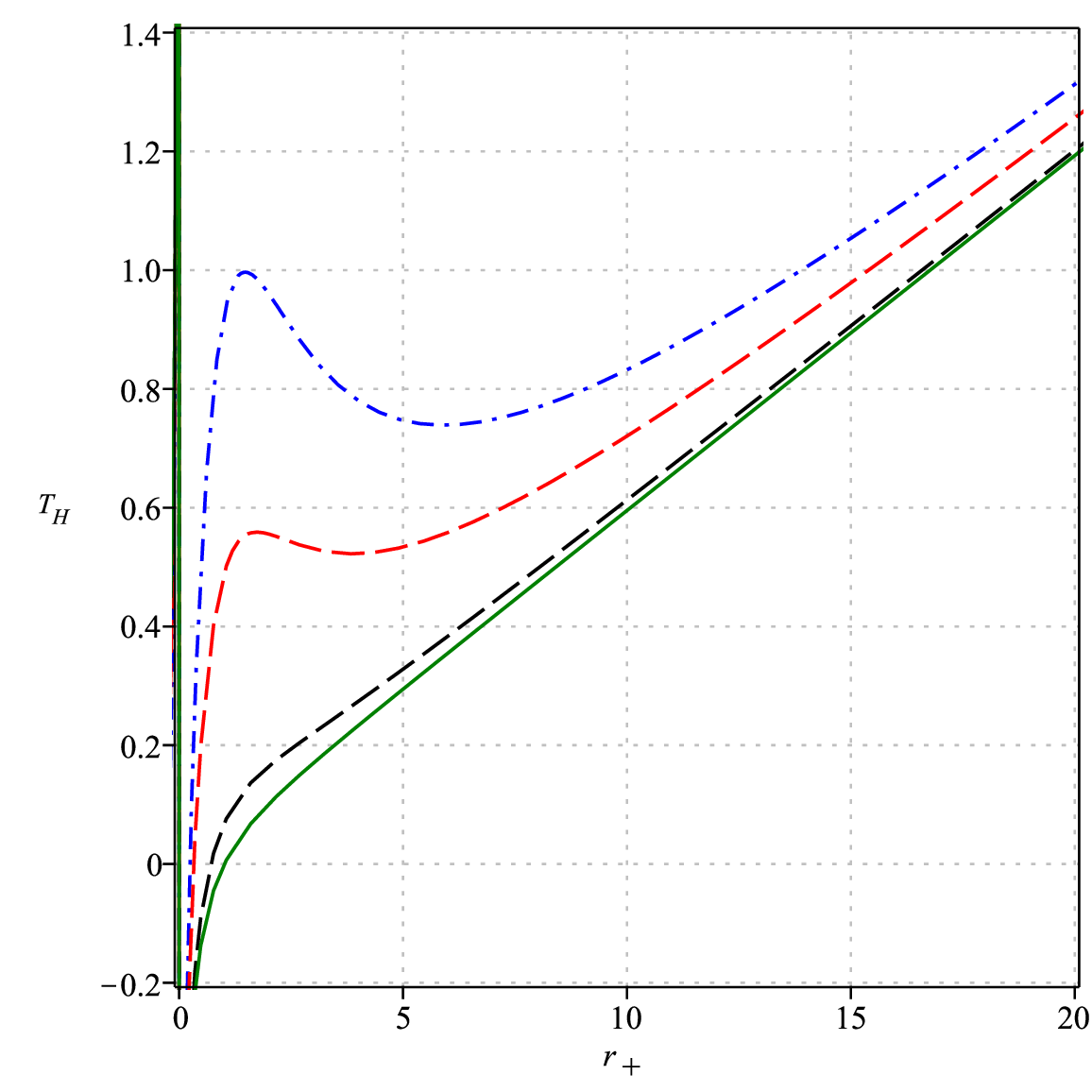}}\hfill
\caption{$m=0$ denoted by solid green line, $m=1.5$ denoted by dash black line, $m=4.0$ denoted by dash red line,  $m=5.5$ denoted by dash dot blue line and $m=10.0$ denoted by dash dot gold line with  $l=2$, $c=1$ and $c_2=1$.}\label{fig:3}
\end{figure}
\begin{figure}[hbt]
\centering
\subfloat[$\alpha=0.5$, $Q=5.0$ and $c_{1}=-1$]{\includegraphics[width=6.5cm,height=5.5cm]{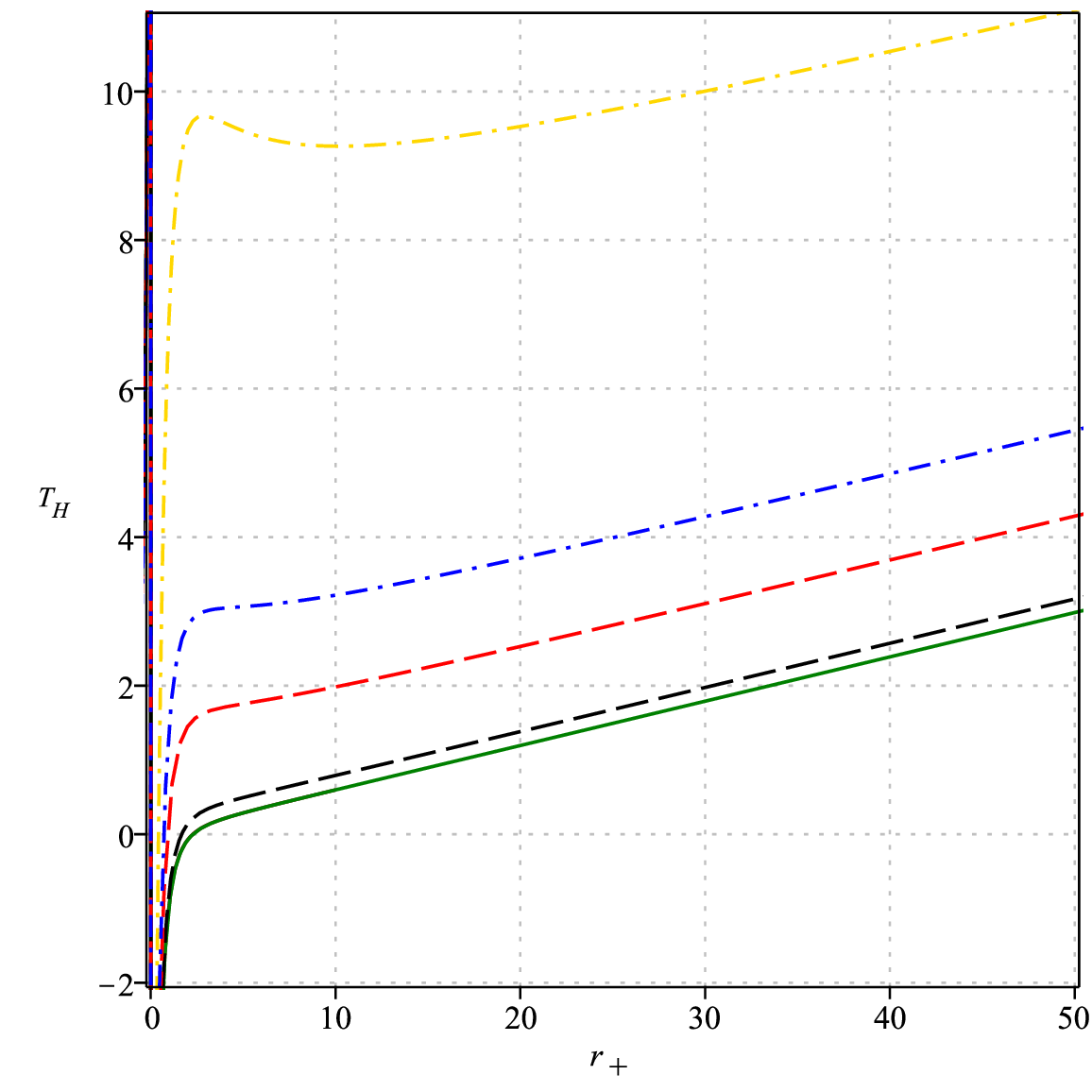}}\hfill
\subfloat[$\alpha=0.5$, $Q=10.0$ and $c_{1}=-1$]{\includegraphics[width=6.5cm,height=5.5cm]{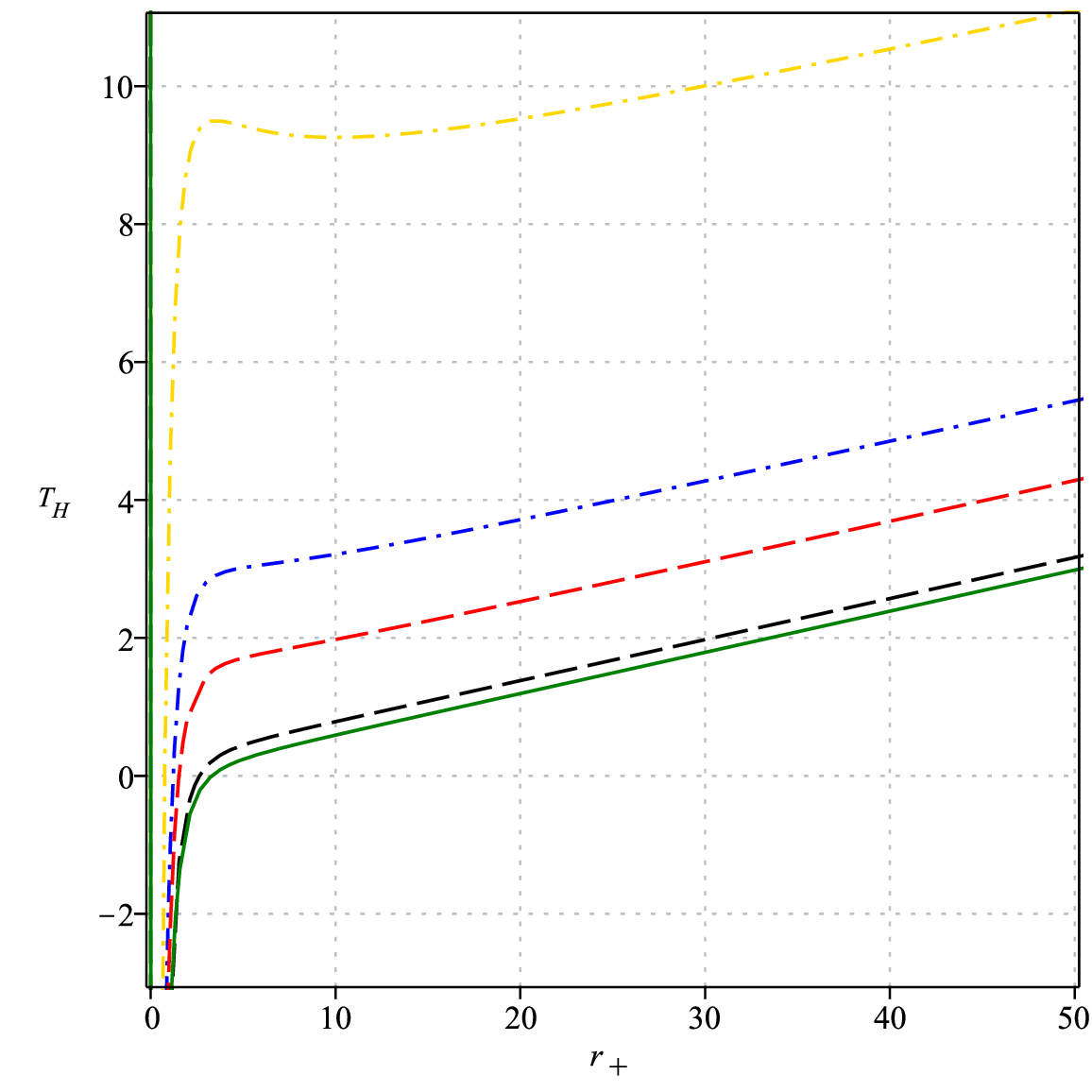}}\hfill
\caption{$m=0.0$ denoted by solid green line, $m=1.5$ denoted by dash black line, $m=4.0$ denoted by dash red line,  $m=5.5$ denoted by dash dot blue line and $m=10.0$ denoted by dash dot gold line with  $l=2$, $c=1$ and $c_2=1$.}\label{fig:4}
\end{figure}
In Fig. \ref{fig:3} and Fig. \ref{fig:4}, we plot the Hawking temperature of the black holes with respect to $r_{+}$ for different values of $\alpha$ and charge. In Fig. \ref{fig:3} (a) and Fig. \ref{fig:3} (b), the Hawking temperature is plotted for different values of $\alpha$. From the Fig., it is clear that for a critical value of horizon radius (say, $r_{+}^{min}$) the Hawking temperature is zero, and if we increase the horizon radius from $r_{+}^{min}$ then Hawking temperature increases. Further increase of horizon radius from $r_{+}^{min}$ Hawking temperature leads to attaining a local maximum for a particular value of $r_{+}$ (say, $r_{+}^{b}$) and local maxima are slowly getting absent as graviton mass decreases.  Hawking temperature attains a minimum for a particular value of horizon radius (say, $r_{+}^{a}$ and $r_{+}^{a}>r_{+}^{b}>r_{+}^{min}$) and the minima  slowly disappear if we decrease the graviton mass. After attaining the minima if we further increase the horizon radii then Hawking temperature again increases. In Fig. \ref{fig:4} (a) and Fig. \ref{fig:4} (b), we plot the Hawking temperature with respect to the black hole horizon for different values of charge. From Fig., it is clear that the behavior is the same as Fig. \ref{fig:3} but the effects of increasing charge are local maxima and the minima now disappear. To observe the local maxima and the minima, we have to increase the graviton mass further, as shown by the gold dashed dot line in Fig. \ref{fig:4}. 

In Fig. \ref{fig:6}, the effects of charge on the Hawking temperature are shown. The inclusion of charge slowly decreases the position of local maxima and minima.

To find the entropy of the black hole, we use the relation $dM=T_{H} dS$.
Using the Hawking temperature and mass of the black hole, we obtain the entropy as
\begin{equation}\label{eq:3.6}
    S= \pi r_{+}^2 + 4\pi \alpha\ln(r_{+}) +S_{0},
\end{equation}
where $S_{0}$ is integration constant. From the above equation, one can say that the inclusion of electric charge has no effect on the entropy of the black hole. To derive the first law of black hole thermodynamics, we  treat the massive gravity parameters $c_{1}$ and $c_{2}$ as thermodynamics variables, and the corresponding potential is $\mathcal{C}_{1}$ and $\mathcal{C}_{2}$. Apart from that, the potential corresponding to Einstein-Gauss--Bonnet  parameter $\alpha$ is $\mathcal{A}$. The thermodynamic pressure is defined as $P=3/8 \pi l^2$. Therefore, the first law of black hole thermodynamics in extended phase space takes the following form:
\begin{equation}\label{eq:3.7}
    dM= T_{H} dS + \Phi dQ +V dP + {\mathcal{A}} d\alpha   + {\mathcal{C}_{1}}  dc_{1} + {\mathcal{C}_{2}}  dc_{2}. 
\end{equation}
Now, from the first law of black hole, one can find the potential and 
volume as 
\begin{eqnarray}\label{eq:3.8}
    \Phi &=& \biggl( \frac{\partial{M}}{\partial{Q}} \biggl)_{S,P,\alpha,c_{1},c_{2}}= \frac{Q}{r_{+}},\\
    V  &=&  \biggl( \frac{\partial{M}}{\partial{P}} \biggl)_{S,Q,\alpha,c_{1},c_{2}}= \frac{4}{3} \pi r_{+}^3,
\\
    \mathcal{A}  &=&  \biggl( \frac{\partial{M}}{\partial{\alpha}} \biggl)_{S,Q,P,c_{1},c_{2}}= \frac{1}{2r_{+}},
\\
    \mathcal{C}_{1}  &=&  \biggl( \frac{\partial{M}}{\partial{c_{1}}} \biggl)_{S,Q,P,\alpha,c_{2}}= \frac{c m^2 r_{+}^2}{4},
\\
    \mathcal{C}_{2}  &=&  \biggl( \frac{\partial{M}}{\partial{c_{2}}} \biggl)_{S,Q,P,\alpha,c_{1}}= \frac{c^2 m^2 r_{+}}{2}.
\end{eqnarray}
\section{Global Stability: Gibbs Free Energy}
\begin{figure*}[hbt]
\begin{tabular}{c c c c}
\includegraphics[width=.5\linewidth]{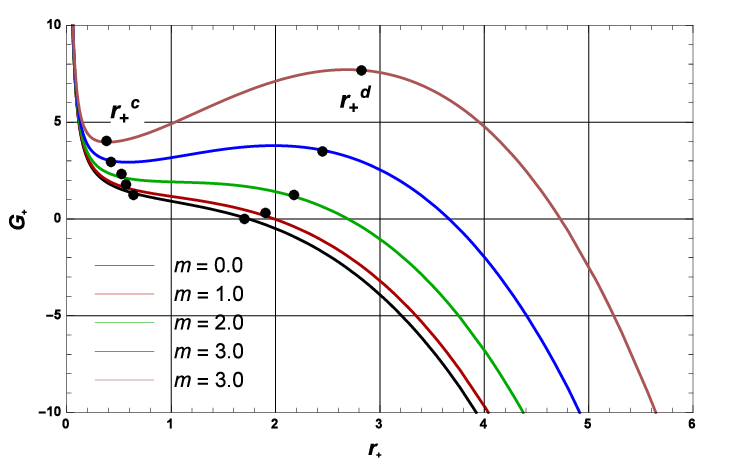}
\includegraphics [width=.5\linewidth]{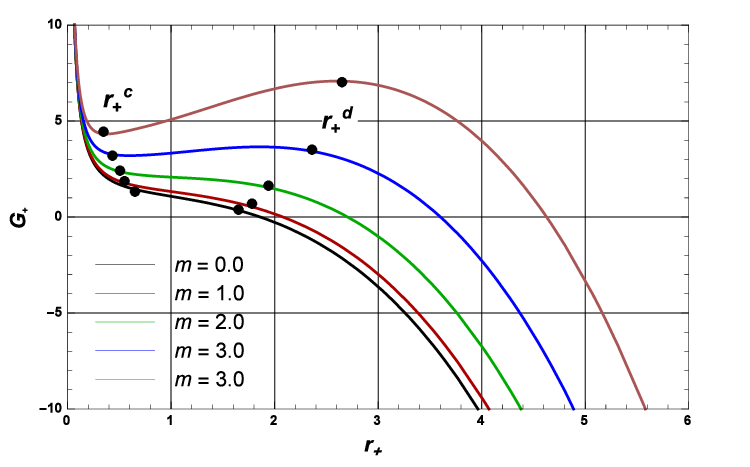}\\
\includegraphics [width=.5\linewidth]{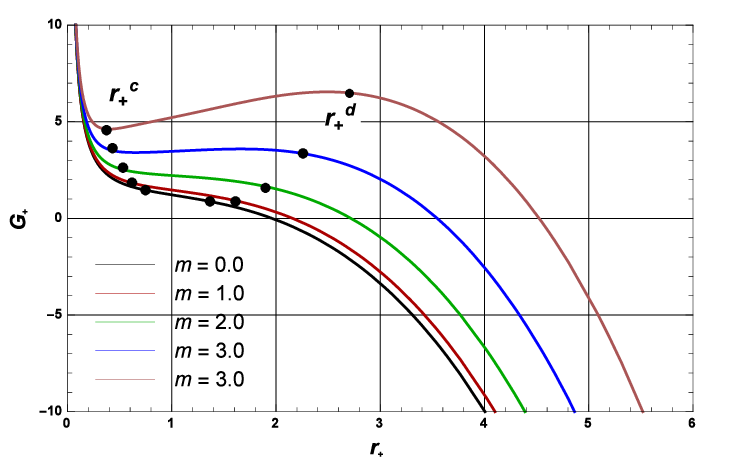}
\includegraphics [width=.5\linewidth]{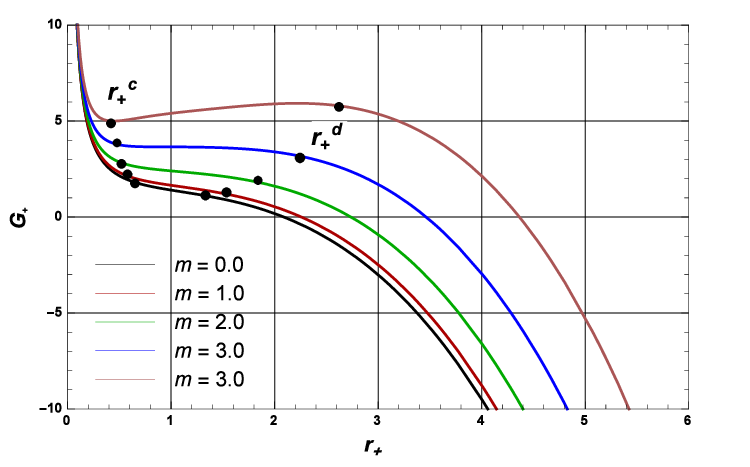}
\end{tabular}
\caption{Plot of Gibbs free energy vs. horizon radius with different values of graviton mass ($m=0,1,2,3,4$) with the fixed value of Gauss-Bonnet coupling $(\alpha=0.1,.3,0.5, 0.8)$ and $l=2$, $c=1$, $c_{1}=-1$ and $c_2=1$. }
\label{fig:6}
\end{figure*}

\begin{figure*}[hbt]
\begin{tabular}{c c c c}
\includegraphics[width=.5\linewidth]{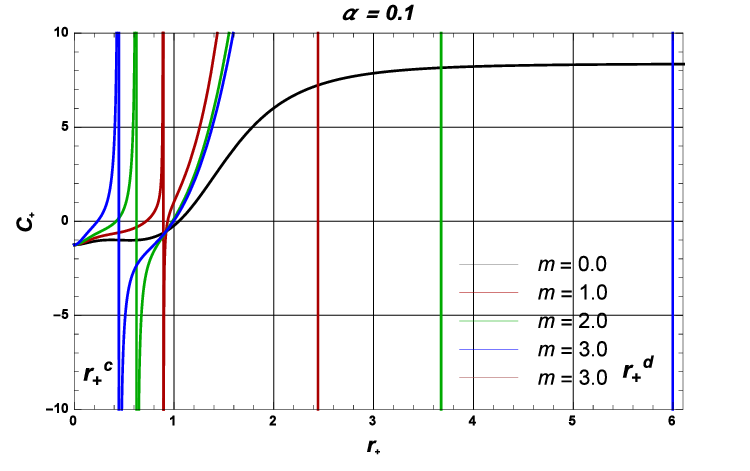}
\includegraphics [width=.5\linewidth]{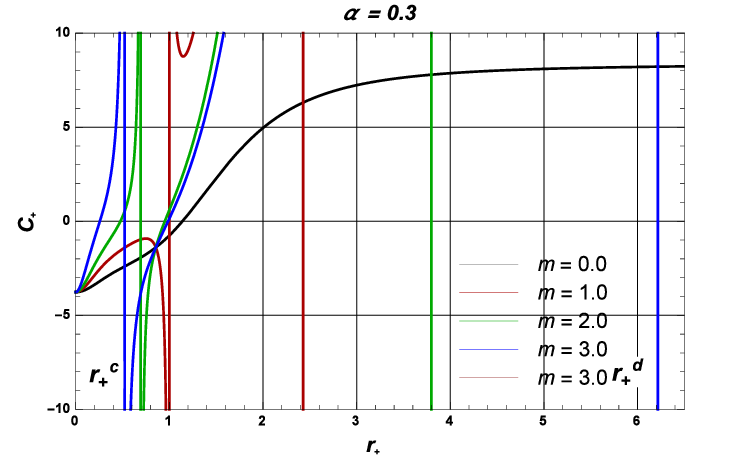}\\
\includegraphics [width=.5\linewidth]{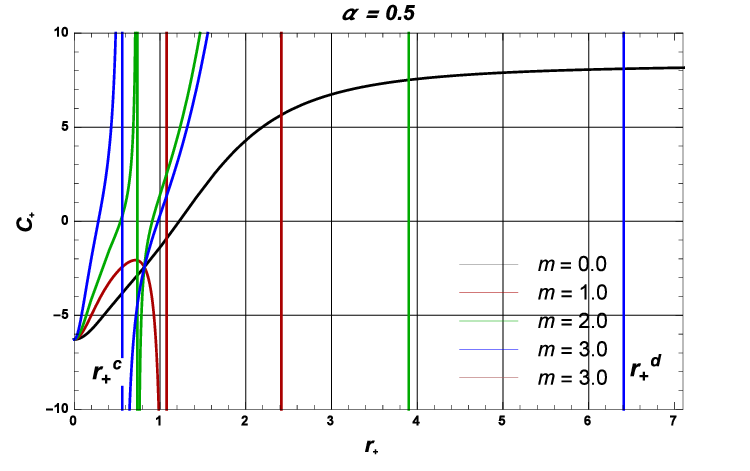}
\includegraphics [width=.5\linewidth]{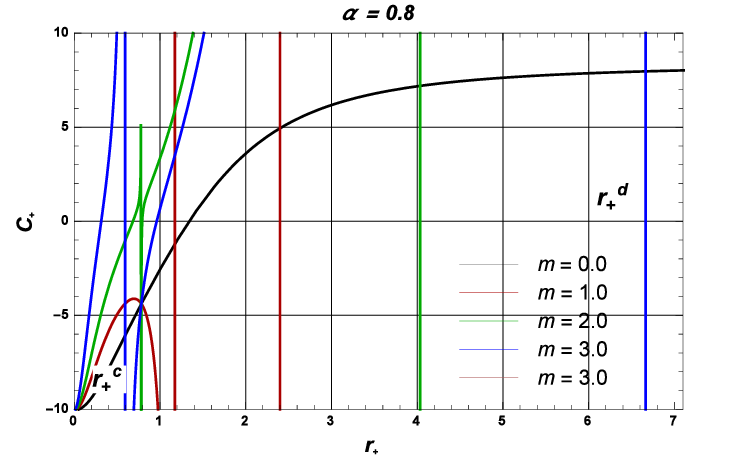}
\end{tabular}
\caption{Plot of heat capacity vs. horizon radius with different values of graviton mass ($m=0,1,2,3,4$) with the fixed value of Gauss-Bonnet coupling $(\alpha=0.1,.3,0.5, 0.8)$ and $l=2$, $c=1$, $c_{1}=-1$ and $c_2=1$. }
\label{fig:7}
\end{figure*}

To study the global stability of the black holes we find Gibbs free energy 
\begin{equation}\label{eq:3.13}
    G= M-T_{H}S-Q \Phi.
\end{equation}
Using equations \eqref{eq:3.1}, \eqref{eq:3.3}, \eqref{eq:3.6} and \eqref{eq:3.7}, we obtain
\begin{eqnarray}
    G&=& \frac{2 r_{+}^{4} +l^{2} \bigl( 2 r_{+}^{2}+2 Q^{2}+2 \alpha +m^{2}  r_{+}^{2}(c c_{1} r_{+}+2 c^{2} c_{2} )  \bigl) }{4 r_{+} l^{2} }\nonumber\\
   &-&\frac{\left( 3 r_{+}^{4} +l^{2} \left( r_{+}^{2}-Q^{2}-\alpha + m^{2}r_{+}^{2}( c c_{1} r_{+}+c^{2} c_{2} )   \right)  \right) \left( 4 \pi  \alpha  \ln  (r_{+} )+\pi  r_{+}^{2} \right)}{4\pi r_{+} l^{2} (r_{+}^{2}+2 \alpha )  }-\frac{Q^{2}}{r_{+}}. 
\end{eqnarray}
In Fig. \ref{fig:6} and Fig. \ref{fig:7}, we plot the Gibbs free energy for different values of $\alpha$ and charge of the black hole. From Fig. \ref{fig:6}, it is  clear that the Gibbs free energy is zero for two critical values of horizon radius (namely, $r_{+}^{c}$, $r_{+}^{d}$ and $r_{+}^{c}>r_{+}^{d}$). The Gibbs free energy is positive between $r_{+}^{c}$ and $r_{+}^{d}$. The positive part of the Gibbs free energy increases as the graviton mass increases and the positive part of the Gibbs free energy decrease as the graviton mass decrease. The positive part of the Gibbs free energy attains its smallest value in the massless limit. If we further increase the horizon radius $r_{+}>r_{+}^{d}$ then Gibbs free energy goes to the negative value. In Fig. \ref{fig:7}, we plot the effects of charge on the Gibbs free energy. Keeping the graviton mass small $m\le 2$, if we increase the charge of the black hole then Gibbs free energy is completely negative (Fig. \ref{fig:7}(a) and \ref{fig:7}(b)). The solid cyan points represent $r_{+}^d$ and solid black points represent $r_{+}^c$. Finally, one can say that the positive part of the Gibbs free energy slowly disappears due to the inclusion of charge for small graviton mass $m \le 2$. As graviton mass increases from $m=2$ the behavior is similar to that in Fig. \ref{fig:6}. 

\section{Local Stability: Heat Capacity}\label{sec:4}
In this section, we study the local thermodynamical stability of the black hole. We compute the specific heat of the black holes. The local stability of Einstein-Gauss--Bonnet  $4D$ AdS black holes is  studied in Ref. \cite{Hegde:2020xlv}.  The specific heat of black holes in nonlinear electrodynamics investigated in  Refs. \cite{Ghosh:2020ijh}--\cite{Kruglov:2021stm}. The local thermodynamical stability of the black holes can be analyzed from the sign of the specific heat. If heat capacity $C_{\Phi}<0$ then the black holes are thermodynamically unstable and for $C_{\Phi}>0$ then black holes are thermodynamically stable. The heat capacity of the black holes is defined as
\begin{equation}\label{eq:4.1}
    C_{\Phi}= T_{H} \Biggl( \frac{dS}{dT_{H}}  \Biggl)_{\Phi}.
\end{equation}
Using equations \eqref{eq:3.3} and \eqref{eq:3.6}, this reads
\begin{equation}\label{eq:4.2}
C_{\Phi}=\frac{2 \pi (r_{{+}}^{2}+2 \alpha )^{2}  \Biggl(3 r_{{+}}^{4} + \biggl(  r_{{+}}^{2}-Q^{2}-\alpha + m^{2} r_{+}^2 \bigl( c c_{1} r_{{+}}+c^{2} c_{2}\bigl) \biggl) l^{2}\Biggl) }{ \Bigl( -(c^{2} c_{2} m^{2}+1) r_{{+}}^{4}+4 c m^{2} c_{1} \alpha  r_{{+}}^{3}+(2 \alpha  c^{2} c_{2} m^{2}+3 Q^{2}+5 \alpha ) r_{{+}}^{2}+2 Q^{2} \alpha +2 \alpha^{2}\Bigl) l^{2}+3 r_{{+}}^{6}+18 \alpha  r_{{+}}^{4}}.
\end{equation} 
\begin{figure}[hbt]
\centering
\subfloat[$\alpha=0.1$]{\includegraphics[width=6.5cm,height=5.5cm]{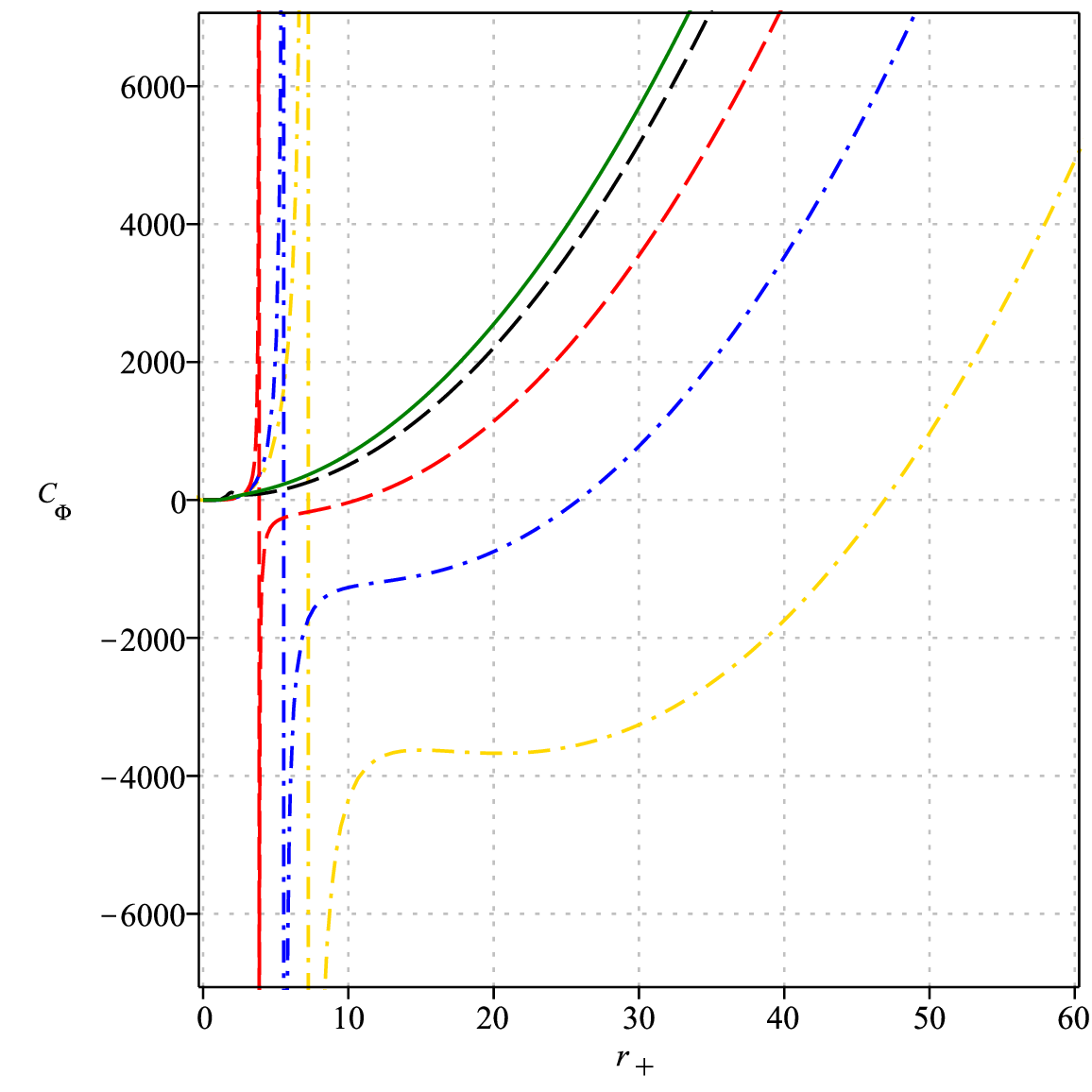}}\hfill
\subfloat[$\alpha=0.5$]{\includegraphics[width=6.5cm,height=5.5cm]{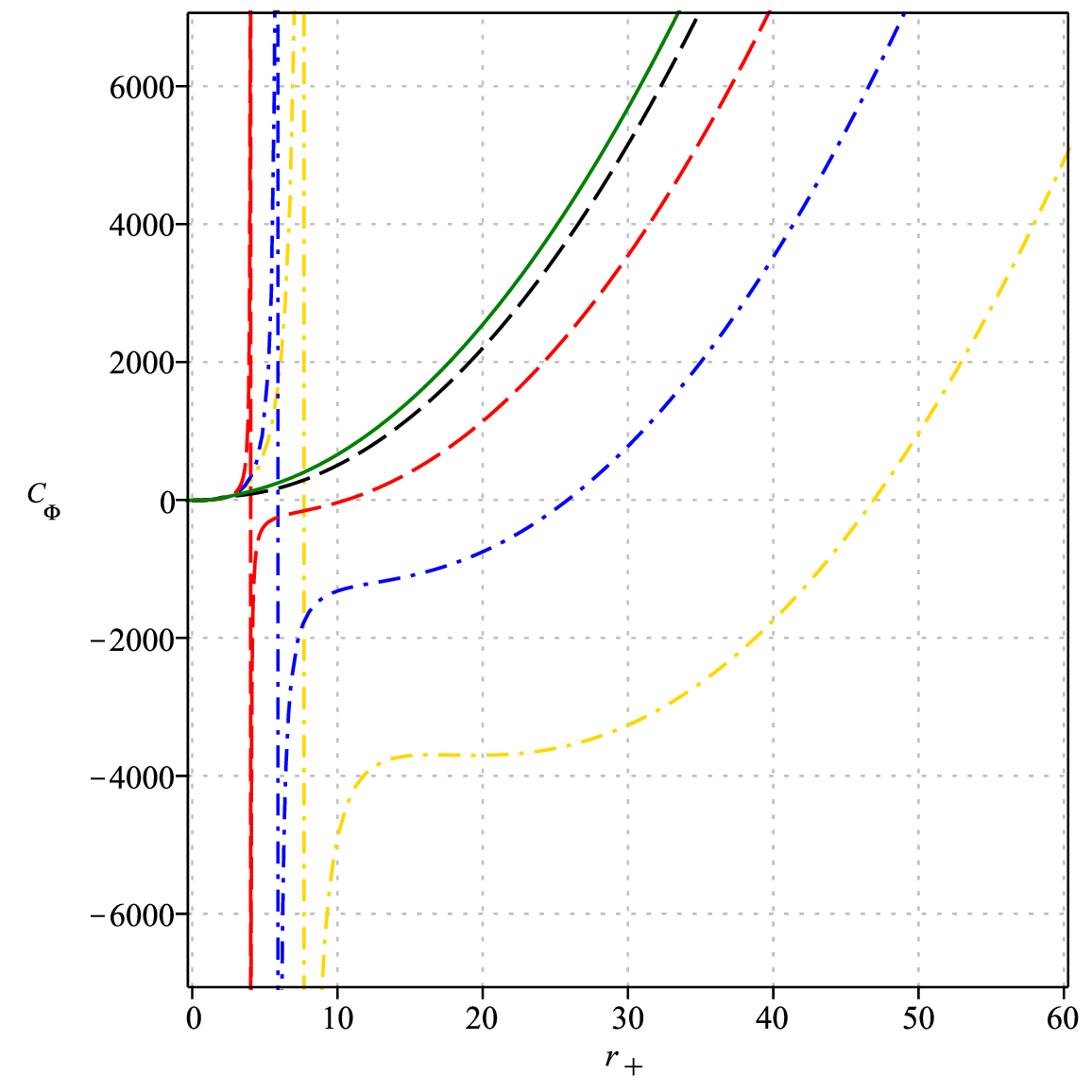}}\hfill
\caption{$m=0.0$ denoted by solid green line, $ m=1.0$ denoted by dash black line, $m=2.0$ denoted by dash red line,  $m=3.0$ denoted by dash dot blue line and $m=4.0$ denoted by dash dot gold line with $Q=1$, $l=3$, $c=1$, $c_{1}=-1$ and $c_2=1$.}\label{fig:8}
\end{figure}
In Fig. \ref{fig:8}, we plot the specific heat for two different values of the Einstein- Gauss-Bonnet  coupling parameter and it is discontinuous at some critical value of horizon radius for a larger value of graviton mass ($m>1$) which indicates that a second-order phase transition occurs for charged AdS black holes in $4D$ Einstein-Gauss--Bonnet  massive gravity. For the smaller value of graviton mass($m=1$), no divergences occur but the smaller-size black holes are thermodynamically unstable as specific heat is negative.  As the size of the black hole increases a phase transition occurs, i.e., the specific heat of the black hole changes   from a negative value  to a positive value. If we take graviton mass as zero then similar kinds of phenomena occur.

In Ref. \cite{Upadhyay:2022axg}, the specific heat of EGB massive gravity black hole is studied with $Q=0$. In the chargeless case, $Q=0$ two diverging points appear for two critical values of horizon radius which separate three regions, i.e. two-second order phase transition occurs for such a black hole. Between two diverging points, specific heat is negative, which indicates that the black hole is thermodynamically unstable in this region. The inclusion of charge removes one diverging point and we are left with only one diverging point, i.e. in the case of charged black hole only one second-order phase transition occurs. For $m=0$ behavior of specific heat same as $Q=0$ and $Q \neq 0$.

\section{Dynamic Stability: Quasinormal modes}
One of the methods to study the dynamic stability of the black holes is studying the nature of quasinormal modes (QNM)  which are characterized by complex numbers. If the imaginary part of QNM is positive, the black hole is unstable; however, if negative, the black hole is stable.

  We compute the QNM and quasinormal frequency (QNF) of the above black hole using the scalar field perturbation. We need to consider the scalar field $\Phi$ in the background of the black hole (\ref{eq:2.11}). The equation of these perturbations takes the form  
\begin{equation}
\frac{1}{\sqrt{-g}}\partial_{\mu}\left(\sqrt{-g} g^{\mu\nu}\partial_{\nu}\right)\Phi=0,
\label{scalar1}
\end{equation}
where $g^{\mu\nu}$ are the metric components of the metric (\ref{eq:2.4}). The mode decomposition of the scalar perturbation in terms of spherical harmonics is given by 
\begin{equation}
\Phi=\frac{1}{r}\sum_{lm}e^{i\omega t}\phi_{lm}Y^m_{l}(\theta,\phi),
\label{scalar2}
\end{equation}
 here $l$, $m$, $Y^m_{l}$, and $\omega$  are respectively the  angular quantum number, magnetic quantum numbers, spherical harmonic, and the oscillating frequency of the scalar field. Substituting the value of $\Phi$ in Eq. (\ref{scalar1}) and using the tortoise coordinate $dr^{*}=dr/e^{2A}$ , we get the Schrodinger-like form,
 
\begin{equation}
\left(\frac{d^2}{dr^{*^2}}+\omega^2-V(r^{*})\right)\phi=0,
\end{equation}
where $V(r^{*})$ is the effective potential and has the form
\begin{equation}
V(r^{*})=e^{2A}\left( { A'(r)} +\frac{l(l+1)}{r^2}\right),
\end{equation}

where $l$ is the harmonic index. To find the QNF one has to impose boundary conditions  near the event horizon. These   boundary conditions can be written as

\begin{eqnarray}
&&\phi(r_{\star})\to e^{ i\omega r_{\star}},\qquad\qquad r_{\star}\to -\infty,\\
&&\phi(r_{\star})\to e^{- i\omega r_{\star}},\qquad\qquad r_{\star}\to \infty,
\end{eqnarray}

\noindent where the $+$ sign corresponds to ingoing waves at the horizon and $-$ sign corresponds to outgoing waves at the infinity. The frequencies corresponding to the QNM are given by $\omega=\omega_R+i\omega_I$,   whose $\omega_R$ and $\omega_I$ are the oscillating  damping components of the frequency.  We use the WKB approximation to find the QNMs and QNFs of the obtained black hole solution (\ref{eq:2.11}). The WKB formula has the form \cite{schutz1985black,Iyer:1986np,Konoplya:2003ii}
\begin{equation}
i\frac{\omega^2-V_0}{\sqrt{-2V''_0}}=n+\frac{1}{2}.
\end{equation}
 where $V_0$ is the height of the barrier and $V''_0$ is the second derivative of the potential with respect to the tortoise coordinate. The numerical value of QNM and QNF for different values of graviton mass is depicted in Tab. \ref{tr14}

\begin{table}[ht]
 \begin{center}
 \begin{tabular}{| l | l   | l   | l   |  l|   }
\hline
            \hline
  \multicolumn{1}{|c|}{ } &\multicolumn{1}{c|}{$\alpha=0.1$}  &\multicolumn{1}{c|}{$\alpha=0.2$}  &\multicolumn{1}{c|}{$\alpha=0.3$} \\
  \hline
    \multicolumn{1}{|c|}{ $m$} &\multicolumn{1}{c|}{$\omega=\omega_R+i \omega_I$}  &\multicolumn{1}{c|}{$\omega=\omega_R+i \omega_I$}  &\multicolumn{1}{c|}{$\omega=\omega_R+i \omega_I$} \\
            \hline
                \,\,\,\,\,1~~ &~~0.0036 - 0.0036 $i$~~ & ~~0.0011 - 0.001 $i$~~ & ~~0.0012 - 0.0012 $i$~~     \\
                
            \,\,\,\,\,2~~ &~~0.0082 - 0.0091 $i$~~ & ~~0.0036 - 0.0039 $i$~~ & ~~0.0036 - 0.0038 $i$~~     \\
            \,\,\,\,\,3~~ &~~0.0065 - 0.0074 $i$~~  & ~~0.0042 - 0.0047 $i$~~ & ~~0.0054 - 0.0058 $i$~~    \\
            \,\,\,\,\,4~~ &~~0.0052 - 0.0060 $i$~~ & ~~0.0056 - 0.0062 $i$~~ & ~~0.0034 - 0.0036 $i$~~   \\
            \,\,\,\,\,5~~ &~~0.0065 - 0.0071 $i$~~ & ~~0.0043 - 0.0046 $i$~~ & ~~0.0052 - 0.0055 $i$~~   \\
            \hline 
\hline
        \end{tabular}
            \caption{The numerical values of QNMs with different values of  gravitation mass $(m)$ and Gauss-Bonnet coupling $(\alpha)$ with a fixed value of    $M=1, Q=1,c=1,c_1=-1,c_2=1,n=1$, and  $l=10$.}
\label{tr14}
    \end{center}
\end{table}

From the table \ref{tr14}, we can see clearly that the imaginary part of the QNMs in the  obtained black hole solution (\ref{eq:2.11}) is negative. So the black hole solution is stable.

\section{Van der Waals Like Phase Transition}\label{sec:5}
In this section, we study the phase transition of charged AdS black holes in $4D$ Einstein-Gauss--Bonnet  massive gravity. The phase transition of the black holes in massive gravity is studied in Refs. \cite{Fernando:2016qhq, Upadhyay:2022axg}. Hawking-Page phase transition for static and rotating $4D$ Gauss--Bonnet  black hole is studied in Refs. \cite{Su:2019gby,Wang:2020pmb}. The phase transition of charged AdS black hole in Einstein--Gauss--Bonnet  massless gravity is studied in Ref. \cite{Hegde:2020xlv}. Phase transition of charged AdS $4D$ Einstein-Gauss--Bonnet  black hole in nonlinear electrodynamics is also studied \cite{Ghosh:2020ijh}. The Van der Waals equation of state for real fluids is given by  
\begin{equation}\label{eq:5.1}
    P= \frac{T}{v-b} - \frac{a}{v^2},
\end{equation}
where $v$ is the specific volume of the fluids, $a$ represents the interaction between the molecules of the fluids and $b$ describes the non-zero of the molecules. Now, from the Hawking temperature \eqref{eq:3.3}, we obtain 
\begin{equation}\label{eq:5.2}
P=\frac{T}{v}+\frac{8 T \alpha}{v^{3}}+\frac{2 Q^{2}}{v^{4} \pi}-\frac{1}{2 v^{2} \pi}+\frac{2 \alpha}{v^{4} \pi} -\frac{c^{2} c_{2} m^{2}}{2 v^{2} \pi}-\frac{c c_{1} m^{2}}{4 v \pi},
\end{equation}
where specific volume $v$ is defined  by 
\cite{Rajagopal:2014ewa}
\begin{equation}\label{eq:5.3}
    v=\frac{6 V}{A} \approx 2r_{+},
\end{equation}
where $A$ is the area of the black hole. To obtain the 
critical points, we use the following conditions: 
\begin{equation}\label{eq:5.4}
    \Biggl( \frac{\partial{P}}{\partial{v}}\Biggl)_{T_{c},v_{c}}= \Biggl( \frac{\partial^2{P}}{\partial{v}^2}\Biggl)_{T_{c},v_{c}}=0.
\end{equation}
Using equations \eqref{eq:5.2} and \eqref{eq:5.4}, we obtain the condition for critical volume as
\begin{equation}\label{eq:5.5}
-2 (c^{2} c_{2} m^{2}  +1 ) v_{c}^{4}+24 \alpha  c c_{1} m^{2} v_{c}^{3} +48 (\alpha  c^{2} c_{2} m^{2} + Q^{2}+ 2 \alpha) v_{c}^{2}+384 Q^{2} \alpha +384 \alpha^{2}= 0.
\end{equation}
Equation \eqref{eq:5.5} can not be solved analytically, we numerically solved the above equation and estimate the critical points as shown in the tables below.
\begin{table}[h!]
\centering
\begin{tabular}{ |p{1.5cm}|p{1.5cm}|p{1.5cm}|p{1.5cm}|p{1.5cm}| } 
 \hline
 m & ${v_c}$ & $P_{c}$ & $T_{c}$ & $\rho_{c}$   \\ [0.5ex]  \hline
 \hline
 0.0 & 5.4328 & 0.0025 & 0.0380 & 0.3702 \\
 0.1 & 5.4030 & 0.0026 & 0.0369 & 0.3806 \\
 0.2 & 5.3161 & 0.0027 & 0.0337 & 0.4259 \\
 0.3 & 5.1788 & 0.0030 & 0.0285 & 0.5451 \\
 0.4 & 5.0010 & 0.0034 & 0.0213 & 0.7982 \\
 0.5 & 4.7940 & 0.0040 & 0.0123 & 1.5590 \\ [1ex] 
 \hline
\end{tabular}
\caption{Values of critical volume ($v_{c}$), critical pressure ($P_{c}$), critical temperature ($T_{c}$) and $\rho_{c}=P_{c}v_{c}/T_{c}$ for different graviton mass with $Q=1$, $\alpha=0.1$, $c=1$ $c_{1}=-2$ and $c_{2}=0.75$.}
\label{table:2}
\end{table}
\begin{table}[H]
\centering
\begin{tabular}{ |p{1.5cm}|p{1.5cm}|p{1.5cm}|p{1.5cm}|p{1.5cm}| } 
 \hline
 Q & ${v_c}$ & $P_{c}$ & $T_{c}$ & $\rho_{c}$   \\ [0.5ex]  \hline
 \hline
 0.0 & 1.9620 & 0.0206 & 0.0733 & 0.5513 \\ 
 0.2 & 2.1519 & 0.0177 & 0.0659 & 0.5779 \\
 0.4 & 2.6326 & 0.0125 & 0.0501 & 0.6568 \\
 0.6 & 3.2763 & 0.0083 & 0.0345 & 0.7882 \\
 0.8 & 4.0092 & 0.0057 & 0.0219 & 1.0434 \\
 1.0 & 4.7940 & 0.0040 & 0.0123 & 1.5577 \\ [1ex] 
 \hline
\end{tabular}
\caption{Values of critical volume($v_{c}$), critical pressure ($P_{c}$), critical temperature ($T_{c}$) and $\rho_{c}=P_{c}v_{c}/T_{c}$ for different charged of the black hole with $m=0.5$, $\alpha=0.1$, $c=1$ $c_{1}=-2$ and $c_{2}=0.75$.}
\label{table:3}
\end{table}

\begin{table}[H]
\centering
\begin{tabular}{ |p{1.5cm}|p{1.5cm}|p{1.5cm}|p{1.5cm}|p{1.5cm}| } 
 \hline
 $\alpha$ & ${v_c}$ & $P_{c}$ & $T_{c}$ & $\rho_{c}$   \\ [0.5ex]  \hline
 \hline
 0.0 & 0.4827 & 0.3516 & 0.4464 & 0.3801 \\
 0.1 & 2.2693 & 0.0132 & 0.0778 & 0.3834 \\
 0.2 & 3.1460 & 0.0068 & 0.0543 & 0.3939 \\
 0.3 & 3.8142 & 0.0046 & 0.0437 & 0.4014 \\
 0.4 & 4.3728 & 0.0035 & 0.0373 & 0.4103 \\
 0.5 & 4.8610 & 0.0028 & 0.0330 & 0.4124 \\  [1ex] 
 \hline
\end{tabular}
\caption{Values of critical volume ($v_{c}$), critical pressure ($P_{c}$), critical temperature ($T_{c}$) and $\rho_{c}=P_{c}v_{c}/T_{c}$ for different Gauss--Bonnet  coupling parameter with $Q=0.1$, $m=0.2$, $c=1$ $c_{1}=-2$ and $c_{2}=0.75$.}
\label{table:4}
\end{table}
In tables \ref{table:2},  \ref{table:3} and  \ref{table:4}, we numerically solve equation \eqref{eq:5.5} for different values of graviton mass, charge and Einstein-Gauss-Bonnet  coupling parameter, we estimate the value of critical volume ($v_c$), critical pressure ($P_c$), critical temperature ($T_c$) and $\rho_c$. From table \ref{table:2}, we can say that as graviton mass increases from zero then the critical volume ($v_c$),  critical temperature ($T_c$) decreases, critical pressure ($P_c$), and $\rho_c$ increases. The effects of black hole charge on the critical parameters are shown in table \ref{table:3}, keeping the graviton mass fixed. As the charge of the black holes increases critical volume ($v_c$) and $\rho_c$ increase, however,  critical pressure and critical temperature decrease. The effects of the Gauss--Bonnet  coupling parameter ($\alpha$) on the critical parameters are shown in table \ref{table:4}. In Fig. \ref{fig:8}, we plot the Hawking temperature for different values of Gauss--Bonnet  coupling parameter and   charge with $P<P_c$, $P=P_c$ and $P>P_{c}$.
 
\begin{figure}[hbt]
\centering
\subfloat[$\alpha=0.5$, $Q=1$ and $c_{1}=0$]{\includegraphics[width=6.5cm,height=5.5cm]{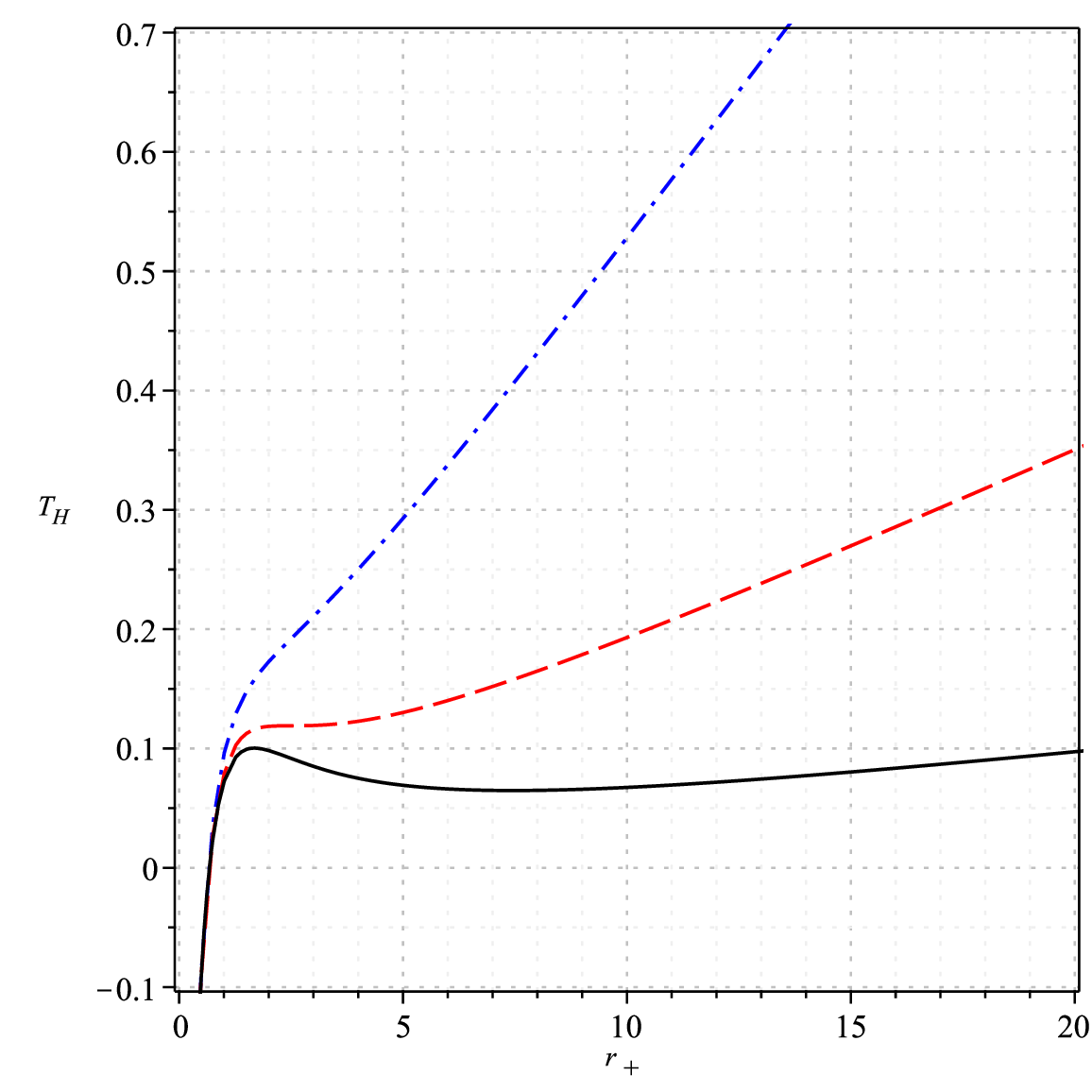}}\hfill
\subfloat[$\alpha=0.8$, $Q=1$ and $c_{1}=0$]{\includegraphics[width=6.5cm,height=5.5cm]{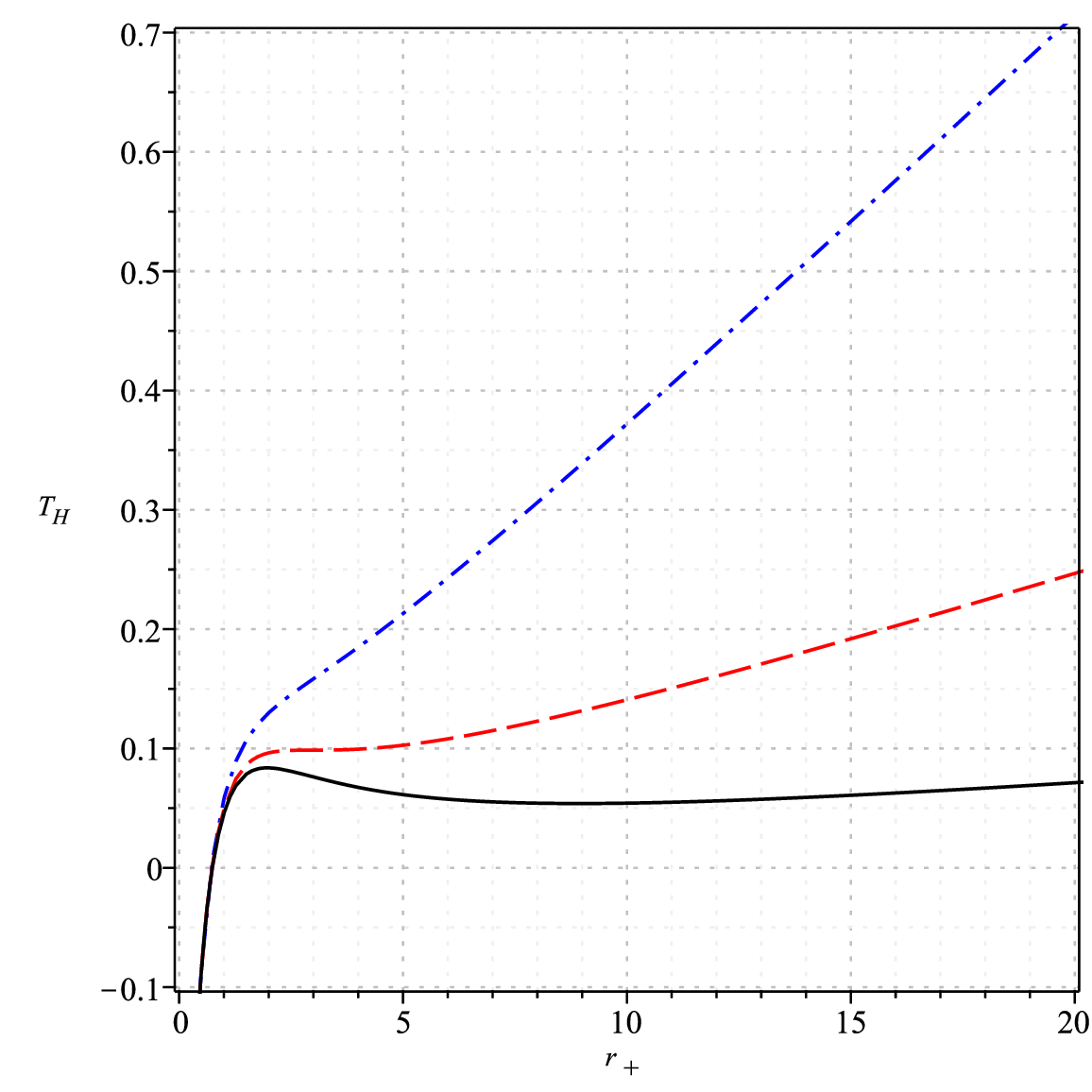}}\hfill
\subfloat[$\alpha=0.5$, $Q=3.0$ and $c_{1}=-0.1$]{\includegraphics[width=6.5cm,height=5.5cm]{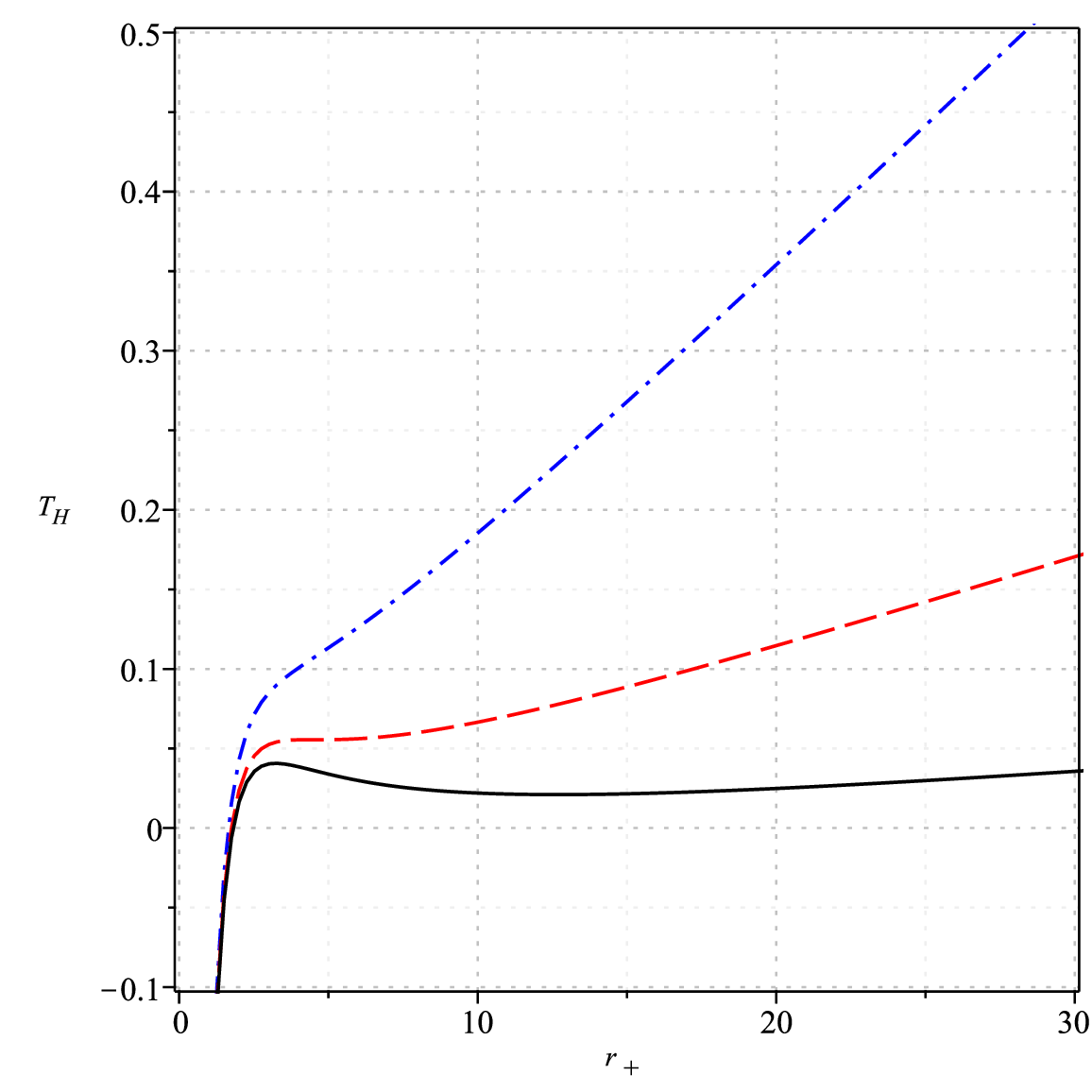}}\hfill
\subfloat[$\alpha=0.5$, $Q=5.0$and $c_{1}=-0.1$]{\includegraphics[width=6.5cm,height=5.5cm]{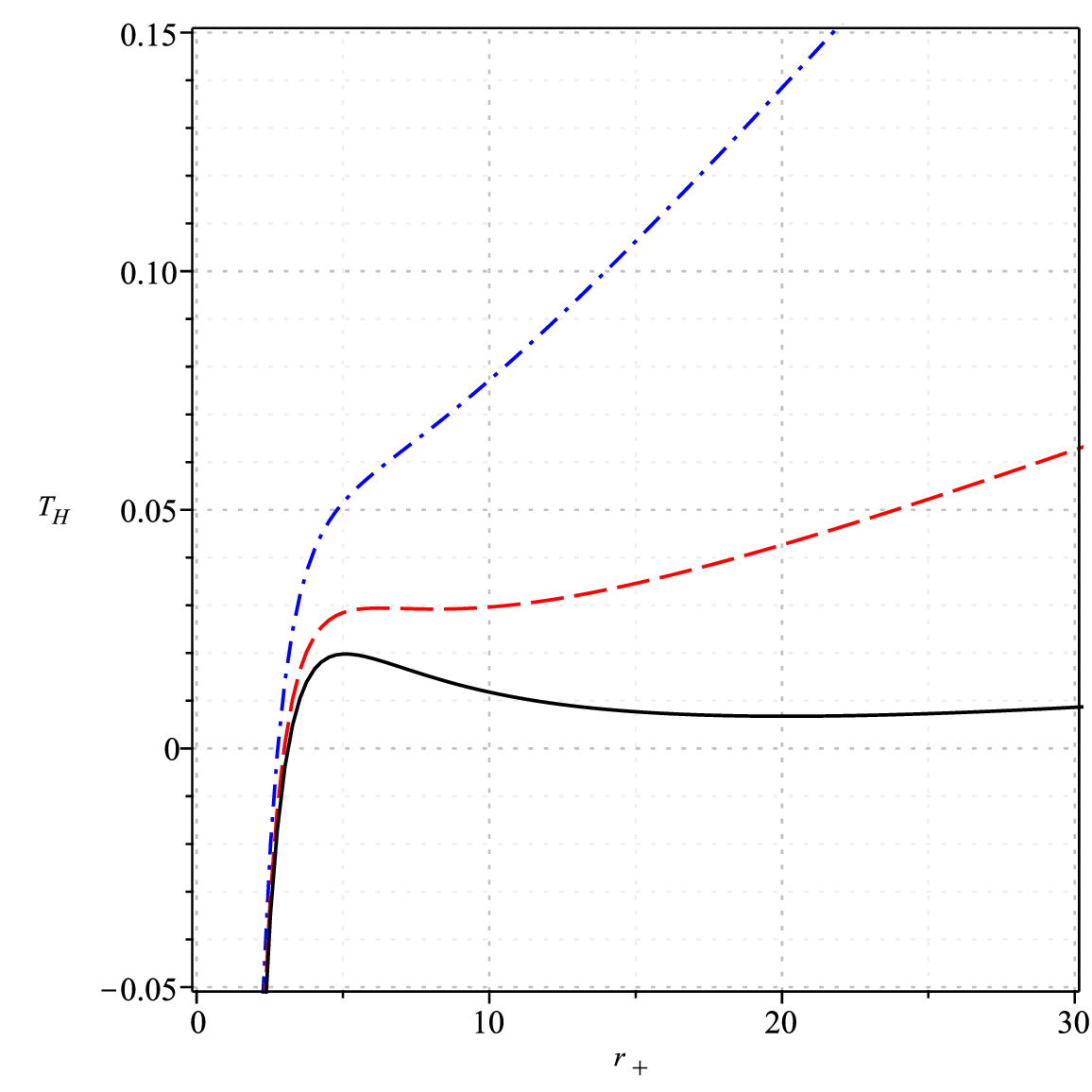}}\hfill
\caption{$P=0.25P_c$ denoted by solid black line, $P=P_c$ denoted by dash red line and $P=3P_c$ denoted by dash dot blue line with $c=1$, $c_2=1$ and $m=1.5$.}\label{fig:9}
\end{figure}

\begin{figure}[hbt]
\centering
\subfloat[$\alpha=0.5$ and $Q=1.0$ ]{\includegraphics[width=6.5cm,height=5.5cm]{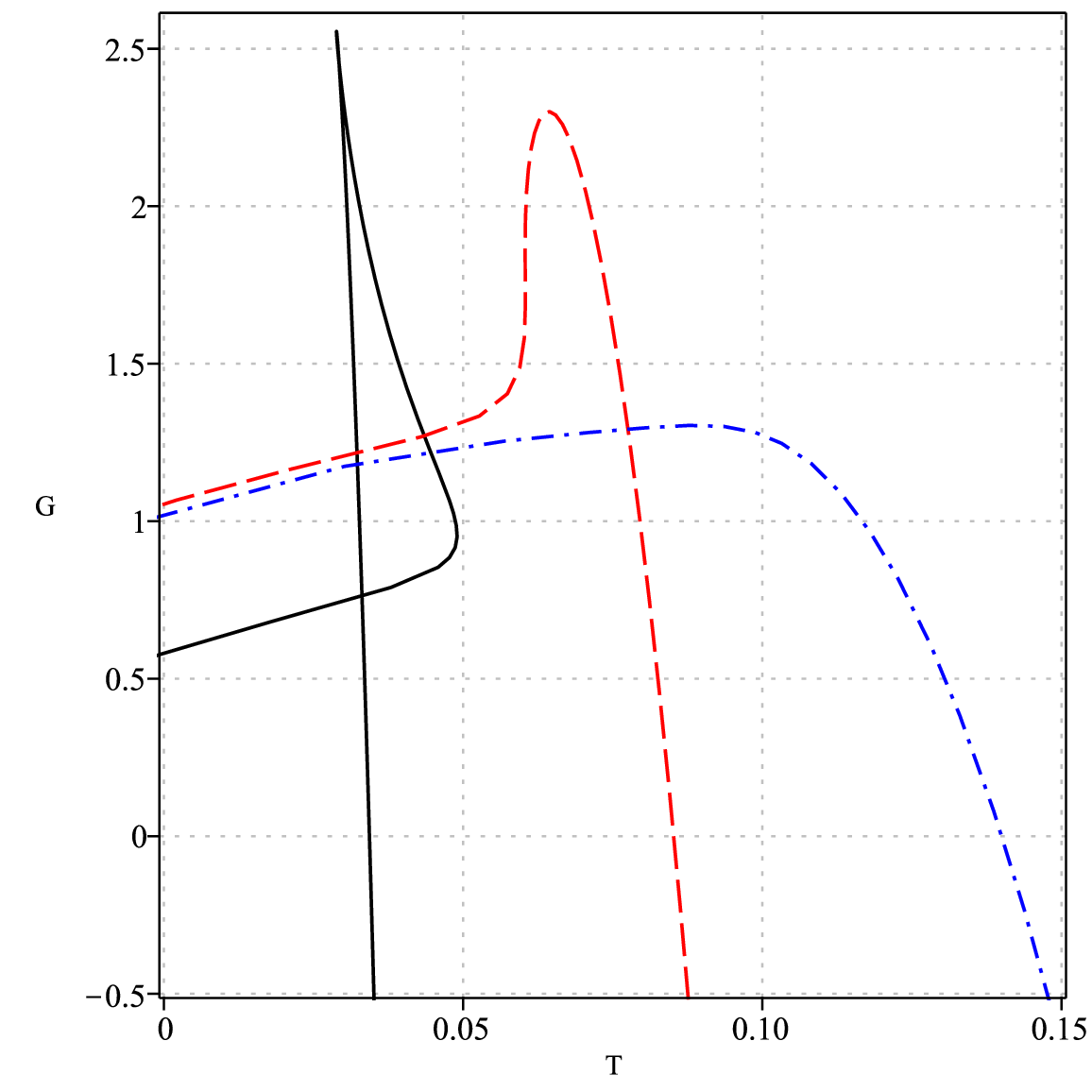}}\hfill
\subfloat[$\alpha=0.8$ and $Q=1.0$]{\includegraphics[width=6.5cm,height=5.5cm]{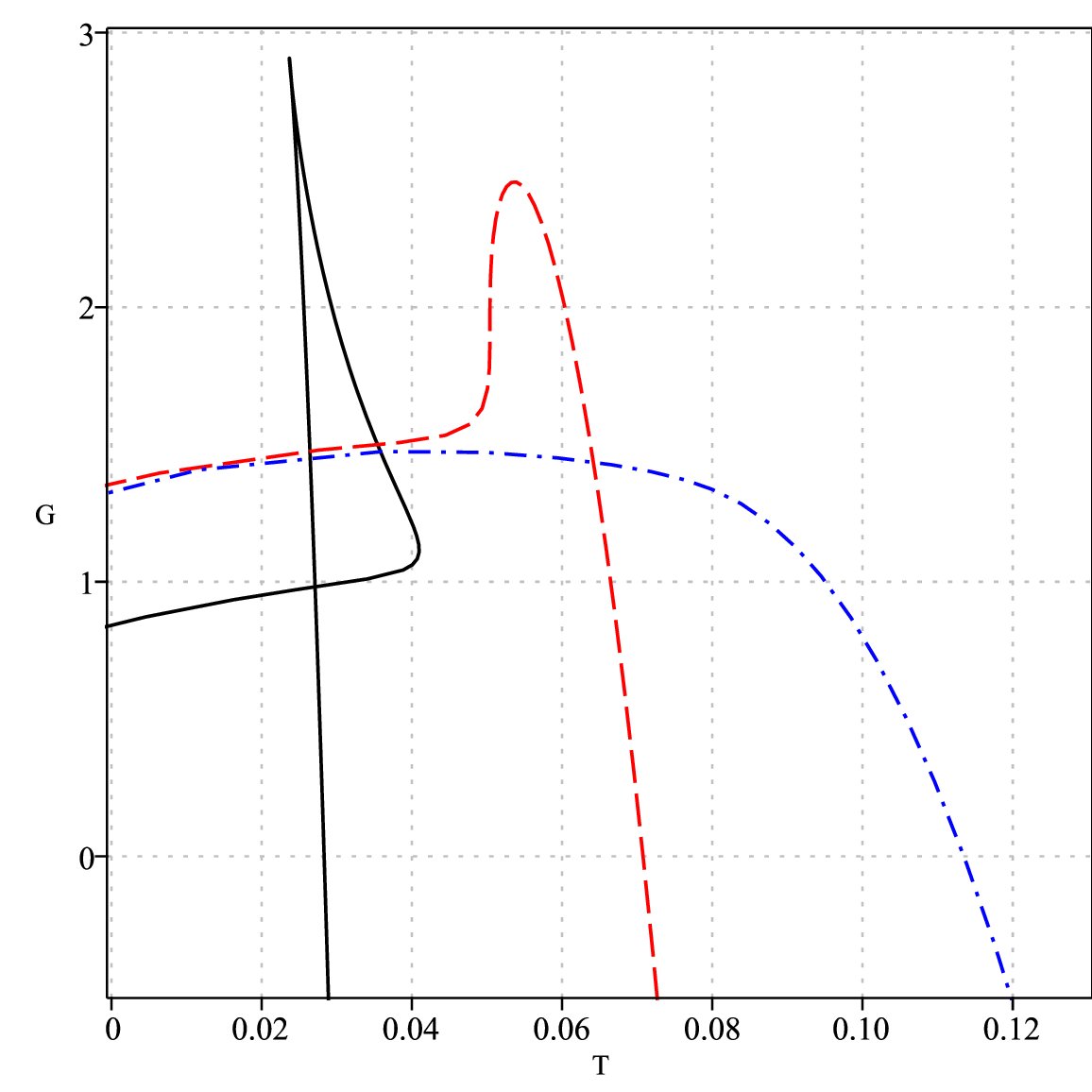}} \hfill\subfloat[$\alpha=0.5$ and $Q=3.0$]{\includegraphics[width=6.5cm,height=5.5cm]{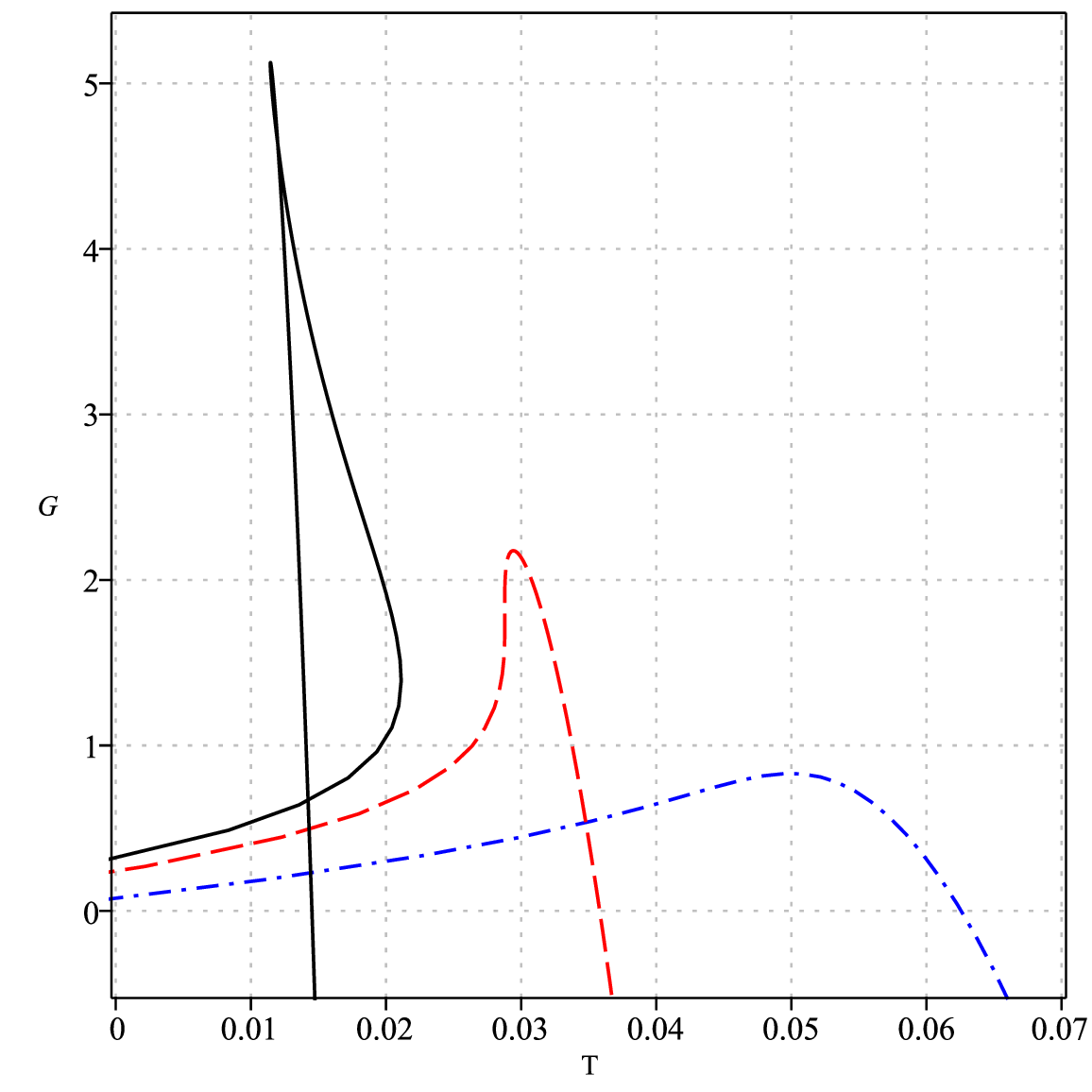}}\hfill
\subfloat[$\alpha=0.5$ and $Q=5.0$]{\includegraphics[width=6.5cm,height=5.5cm]{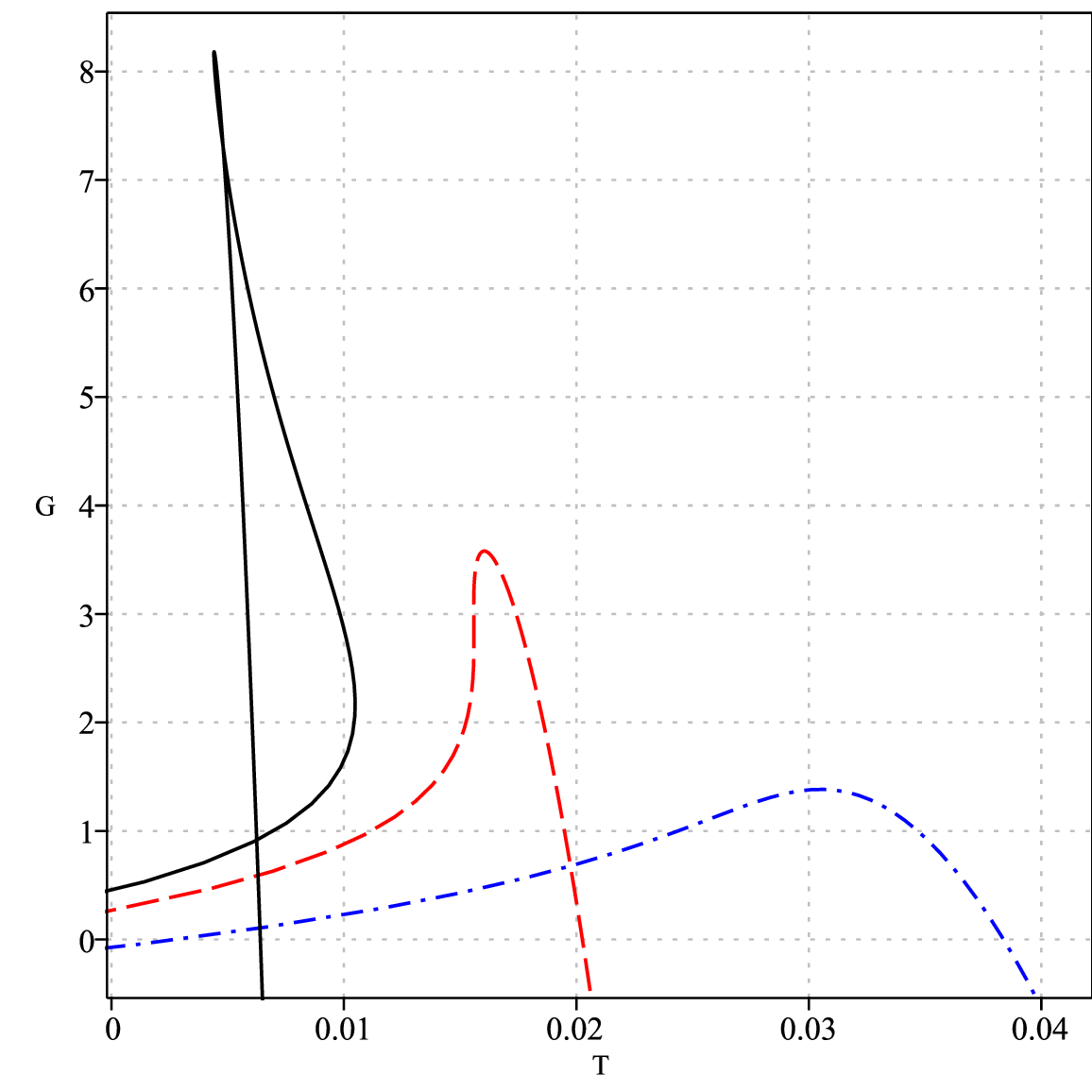}}\hfill
\caption{$P=0.25P_c$ denoted by solid black line, $P=P_c$ denoted by dash red line and $P=3P_c$ denoted by dash dot blue line with $c=1$, $c_{1}=-0.1$, $c_2=1$ and $m=1$.}\label{fig:10}
\end{figure}

 In Fig. \ref{fig:9}(a) and \ref{fig:9}(b), the Hawking temperature is depicted for different values of $\alpha$ keeping pressure fixed. When pressure is less than critical pressure ($P_c$), the curve has two critical points (one  of them is maxima and another is minima). For the pressure equal to the critical pressure ($P_c$), two turning points come to an inflection point and when $P>P_c$ the curve does not attain any turning points. The effects of charge on the Hawking temperature are shown in Fig. \ref{fig:9}(c) and \ref{fig:9}(d). The inclusions of charge basically reduced the position of local maxima and minima when $P<P_c$. Furthermore, if we increase the charge then the position of local maxima and minima decrease by a significant amount (Fig. \ref{fig:9}(d)). The rest of the  behaviour is similar to Fig. \ref{fig:9}(a) and Fig. \ref{fig:9}(b). The behaviour of Hawking temperature of charged less black hole for $P \leq P_c$ and $P>P_c$ is shown in Ref. \cite{Upadhyay:2022axg}. The Hawking temperature of the black hole with $Q=0$ attains local maxima and minima when $P <P_c$. In the chargeless case position of local maxima and minima is higher than the charged black hole.  
In Fig. \ref{fig:10}, we plot Gibbs free energy Vs. temperature for different values of Gauss--Bonnet  coupling parameter and black holes charge with $P<P_c$, $P=P_c$ and $P>P_{c}$. In Fig. \ref{fig:10}(a) and \ref{fig:10}(b),   Gibbs free energy for different values of Gauss--Bonnet  coupling parameters is depicted. When pressure is less than the critical pressure ($P_c$), Gibbs free energy shows swallow tail (triangular shape) behavior, which indicates that the system undergoes a first-order phase transition, i.e., below the critical pressure a transition between small black hole (SBH) and large black hole (LBH) occurs. The Gibbs free energy of LBH is smaller compared to the SBH. At the point of intersection of the curve ($P<P_c$), where first-order phase transition occurs, the entropy of the system is discontinuous  as entropy depends on the horizon radius of the black holes and the radius of the SBH and  LBH is different. For $P=P_c$, the swallow tail  behavior disappears at which a  second-order phase transition occurs. For $P>P_c$ swallow tail behavior completely disappears and no phase transition occurs. A similar kind of behavior is shown in Fig.\ref{fig:10}(c) and Fig.\ref{fig:10}(d) for different values of charge.

\section{Joule-Thomson Expansion}\label{sec:6}
\begin{figure}[hbt]
\centering
\subfloat[$c_{1}=1$ and $c_{2}=1$ ]{\includegraphics[width=6.5cm,height=5.5cm]{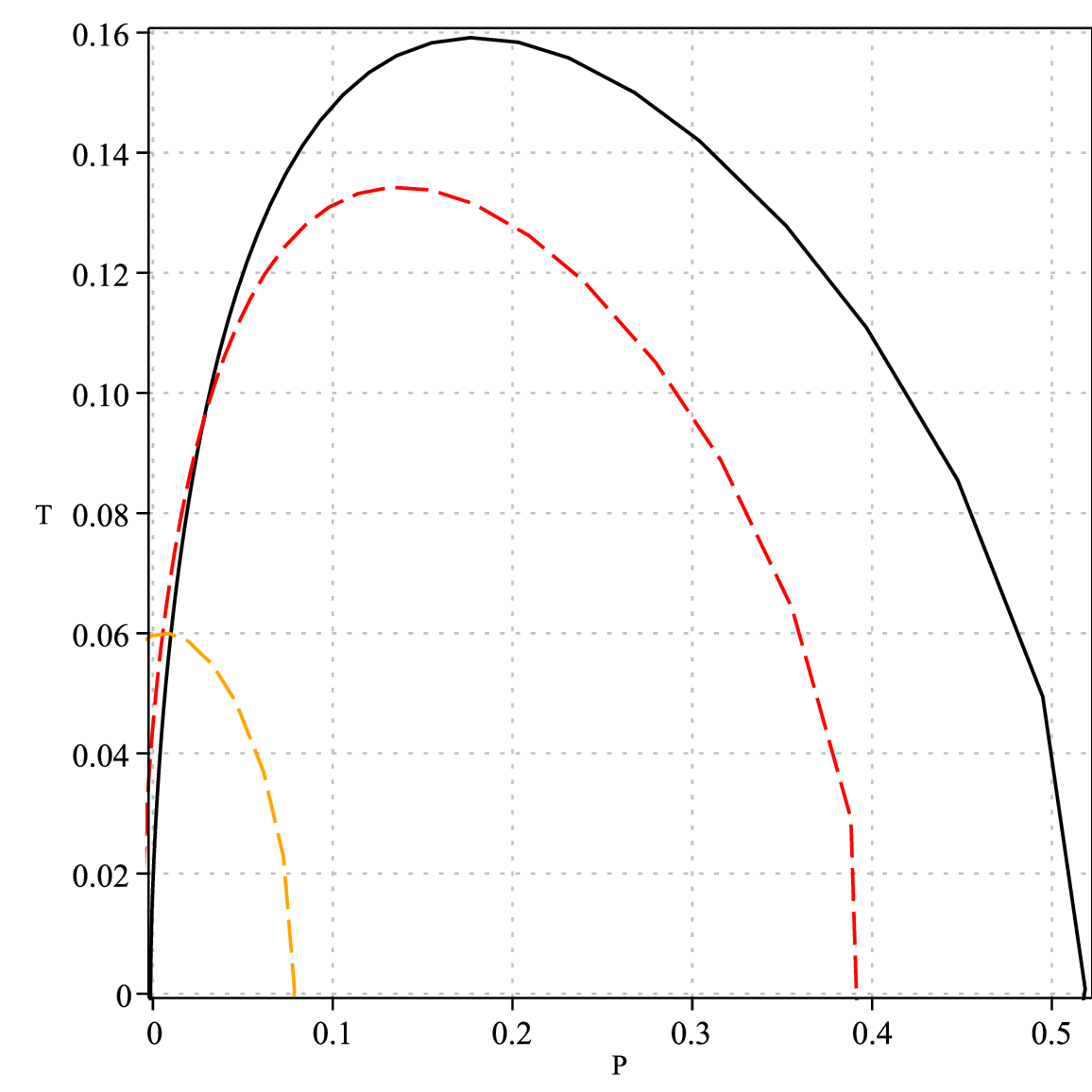}}\hfill
\subfloat[$c_{1}=1$ and $c_{2}=-1$]{\includegraphics[width=6.5cm,height=5.5cm]{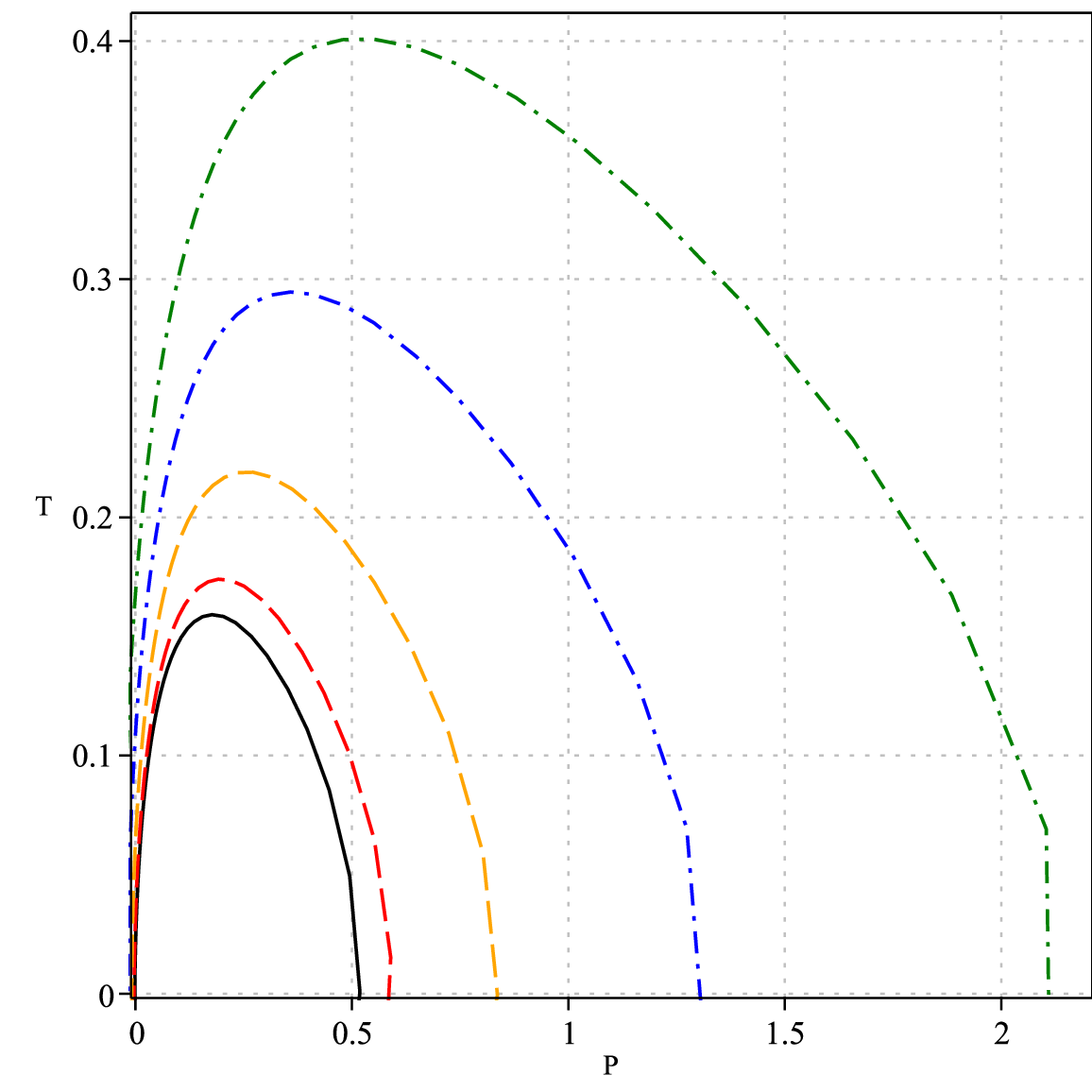}}\hfill
\caption{$m=0$ denoted by solid black line, $m=0.5$ denoted by dash red line, $m=1$ denoted by dash orange line, $m=1.5$ denoted by blue dash dot line and $m=2$ denoted by dash dot green line with $M=2$, $Q=1$, $\alpha=0.5$ and $c=1$.}\label{fig:15}
\end{figure}

\begin{figure}[hbt]
\centering
\subfloat[$c_{1}=-1$ and $c_{2}=1$]{\includegraphics[width=6.5cm,height=5.5cm]{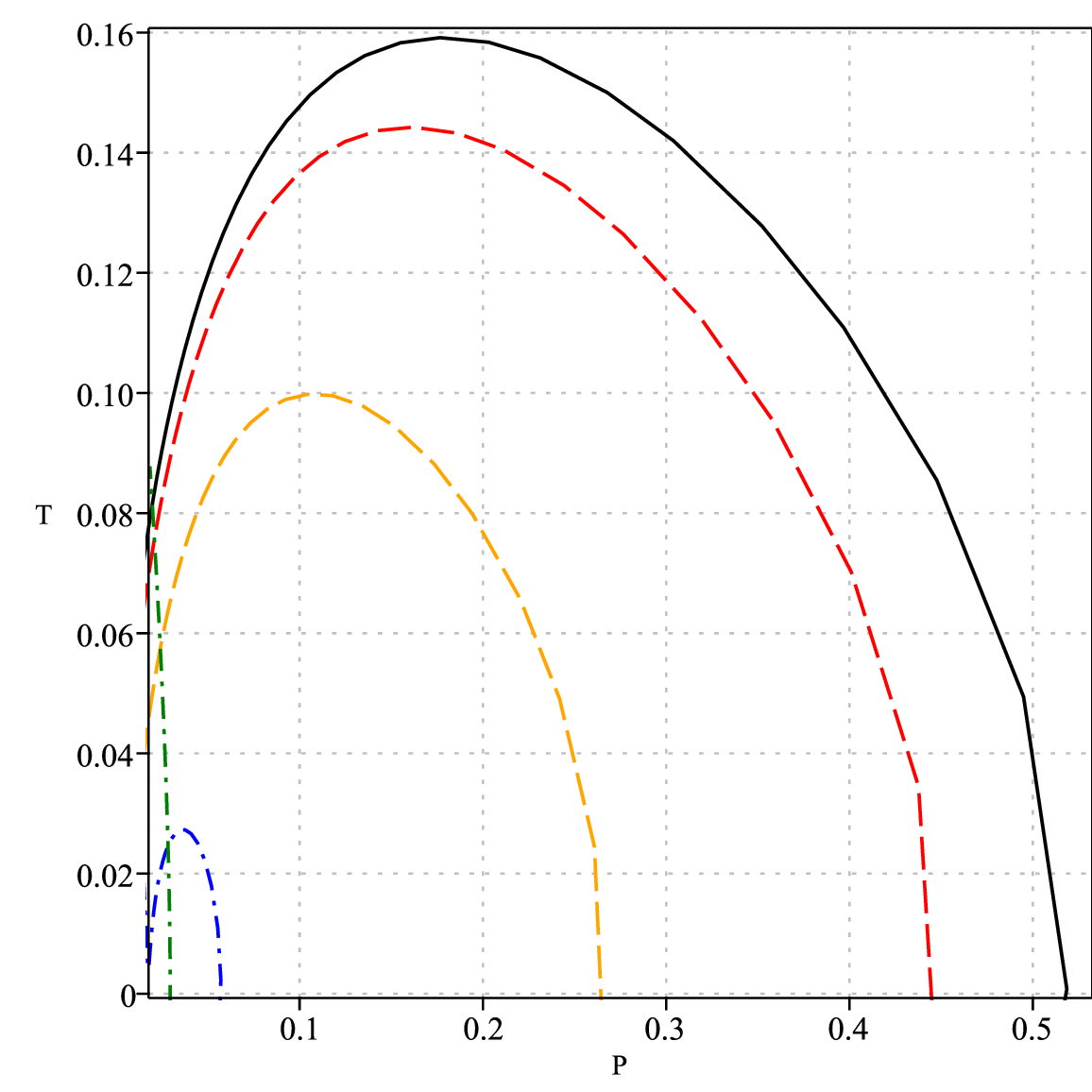}}\hfill
\subfloat[$c_{1}=-1$ and $c_{2}=-1$]{\includegraphics[width=6.5cm,height=5.5cm]{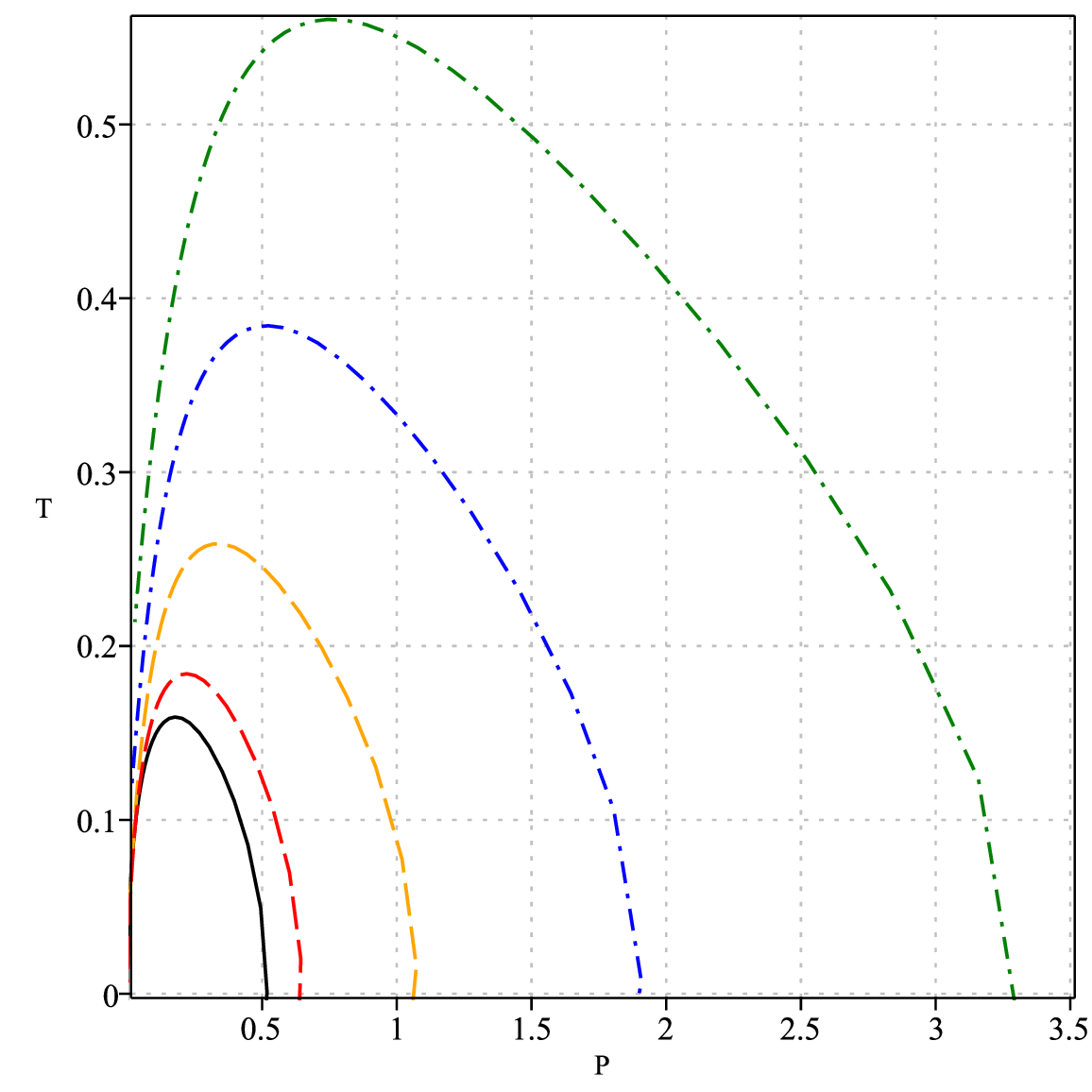}}\hfill
\caption{$m=0$ denoted by solid black line, $m=0.5$ denoted by dash red line, $m=1$ denoted by dash orange line, $m=1.5$ denoted by blue dash dot line and $m=2$ denoted by dash dot green line with $M=2$, $Q=1$, $\alpha=0.5$ and $c=1$.}\label{fig:16}
\end{figure}

In this section, we discuss the effects of massive gravity on the Joule-Thomson expansion of charged AdS black holes in $4D$ Einstein--Gauss--Bonnet massive  gravity. The Joule-Thomson expansion of charged AdS black holes was first studied in Ref. \cite{Okcu:2016tgt}. After that Joule-Thomson expansion of $D$ dimensional black holes was studied in Ref. \cite{Mo:2018rgq}.  Using numerical investigation, Joule-Thomson expansion of Kerr-AdS and Kerr-Newman-AdS is also studied \cite{Okcu:2017qgo,Zhao:2018kpz}. Joule-Thomson expansion of charged AdS black hole in $4D$ Einstein massive gravity discussed in Ref. \cite{Nam:2020gud}. Joule-Thomson expansion of charged AdS $4D$ Einstein massive gravity black hole in Maxwell and Born-Infeld  electrodynamics discussed in Ref. \cite{Hegde:2020xlv,Zhang:2021kha}. The Joule-Thomson thermodynamic coefficient is given by
\begin{equation}\label{eq:6.1}
    \mu_{J}= \biggl( \frac{\partial{T}}{\partial{P}} \biggl)_{M}= \frac{1}{C_{P}} \biggr[  T \Bigl(\frac{\partial{V}}{\partial{P}}\Bigl)_P - V \biggr]=\frac{(\partial{T}/\partial{r_{+}})_M}{(\partial{P}/\partial{r_{+}})_M}.
\end{equation} 
The Joule-Thomson effect is an isenthalpic process, which means that enthalpy remains constant during the process. In the Joule-Thomson process pressure always decrease but the temperature can increase/decrease, thus Joule-Thomson thermodynamic coefficient ($\mu$) can be negative/positive. When $\mu>0$ the Joule–Thomson expansion corresponds to the cooling region of the isenthalpic or constant mass curve and  $\mu<0$ corresponds to the heating region of the isenthalpic or constant mass curve. The Joule - Thomson thermodynamic coefficient vanishes for some particular value of temperature, which is known as inverse temperature ($T_i$) and corresponding pressure is known as inverse pressure ($P_i$). The cooling and heating regions are separated by the set of points ($P_i$, $T_i$) and the curved formed by the set of points ($P_i$, $T_i$) known as the inverse curve. Clearly, the region above the inverse curve is known as the cooling region and the region below the inverse curve is known as the heating region. At inverse temperature sign of Joule–Thomson coefficient changes $\mu_{J}(T_{i})=0$. From the above equation, we obtain inverse temperature
\begin{equation}\label{eq:6.2}
    T_{i}= V \biggl( \frac{\partial{T}}{\partial{V}} \biggl)_{P} = \frac{r_{+}}{3} \biggl( \frac{\partial{T}}{\partial{r_{+}}} \biggl)_{P}.
\end{equation}
From equation \eqref{eq:3.1} and using $P=3/8\pi l^2$, we obtain pressure in terms of black hole mass
\begin{equation}\label{eq:6.3}
    P=\frac{3}{4\pi r_{+}^2}\left[\frac{M}{ r_{+}}-\frac{ Q^{2}}{2 r_{+}^2}-\frac{1}{2}-\frac{ \alpha}{2 r_{+}^2}  -\frac{m^{2} c^{2} c_{2} }{2}-\frac{m^{2} c c_{1}r_{+}}{4 }\right].
\end{equation}
Now, from Hawking temperature and using the relations $P=1/8\pi l^2$,  we obtain
\begin{equation}\label{eq:6.4}
    T=\frac{1}{4 \pi  r (r_{+}^{2}+2 \alpha )}\left[ 8\pi  P  r_{+}^{4}-Q^{2}+r_{+}^{2}-\alpha +c^{2} c_{2} m^{2} r_{+}^{2}+m^{2} c c_{1} r_{+}^{3}\right].
\end{equation}
\begin{figure}[hbt]
\centering
\subfloat[$\alpha=0.5$ ]{\includegraphics[width=6.5cm,height=5.5cm]{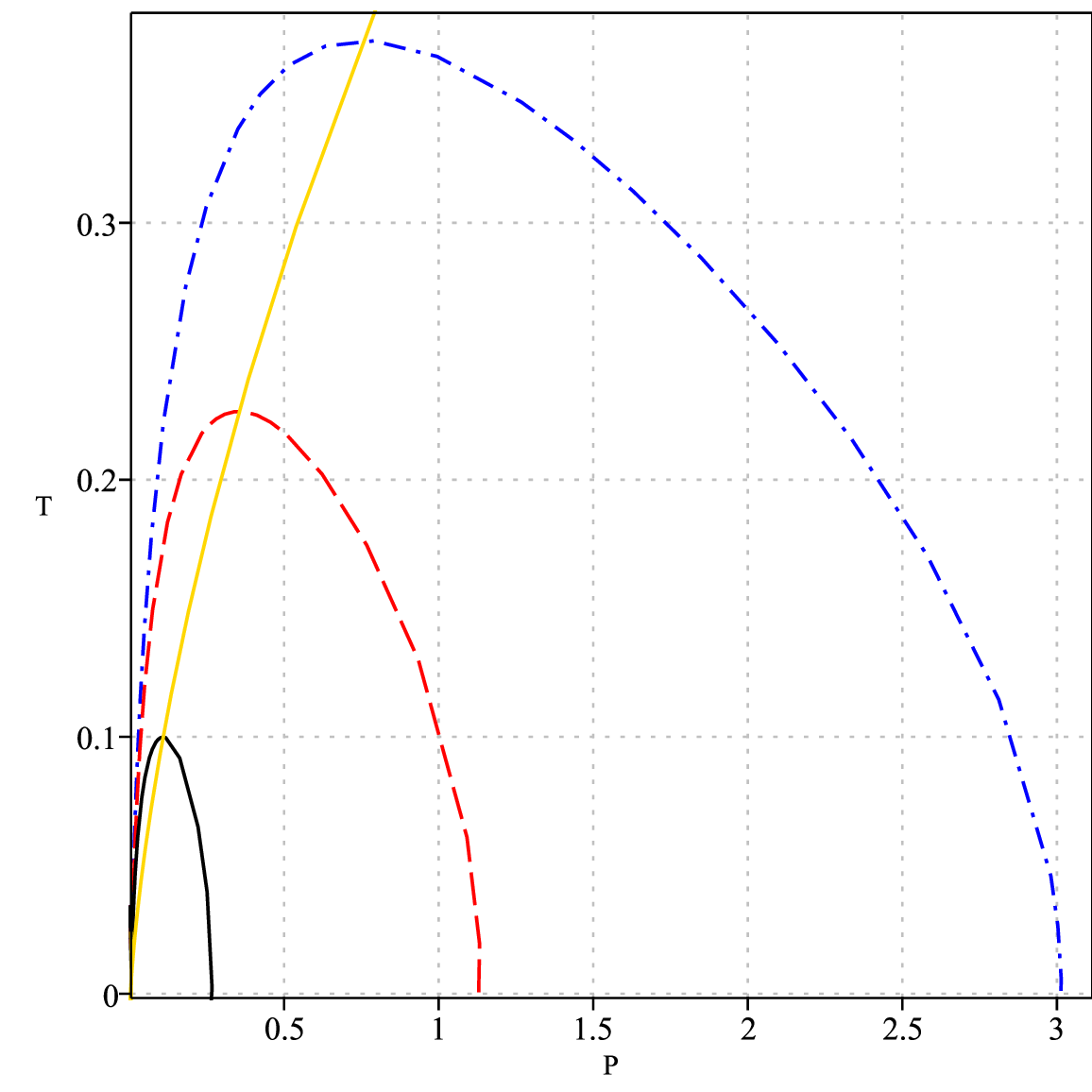}}\hfill
\subfloat[$\alpha=0.8$]{\includegraphics[width=6.5cm,height=5.5cm]{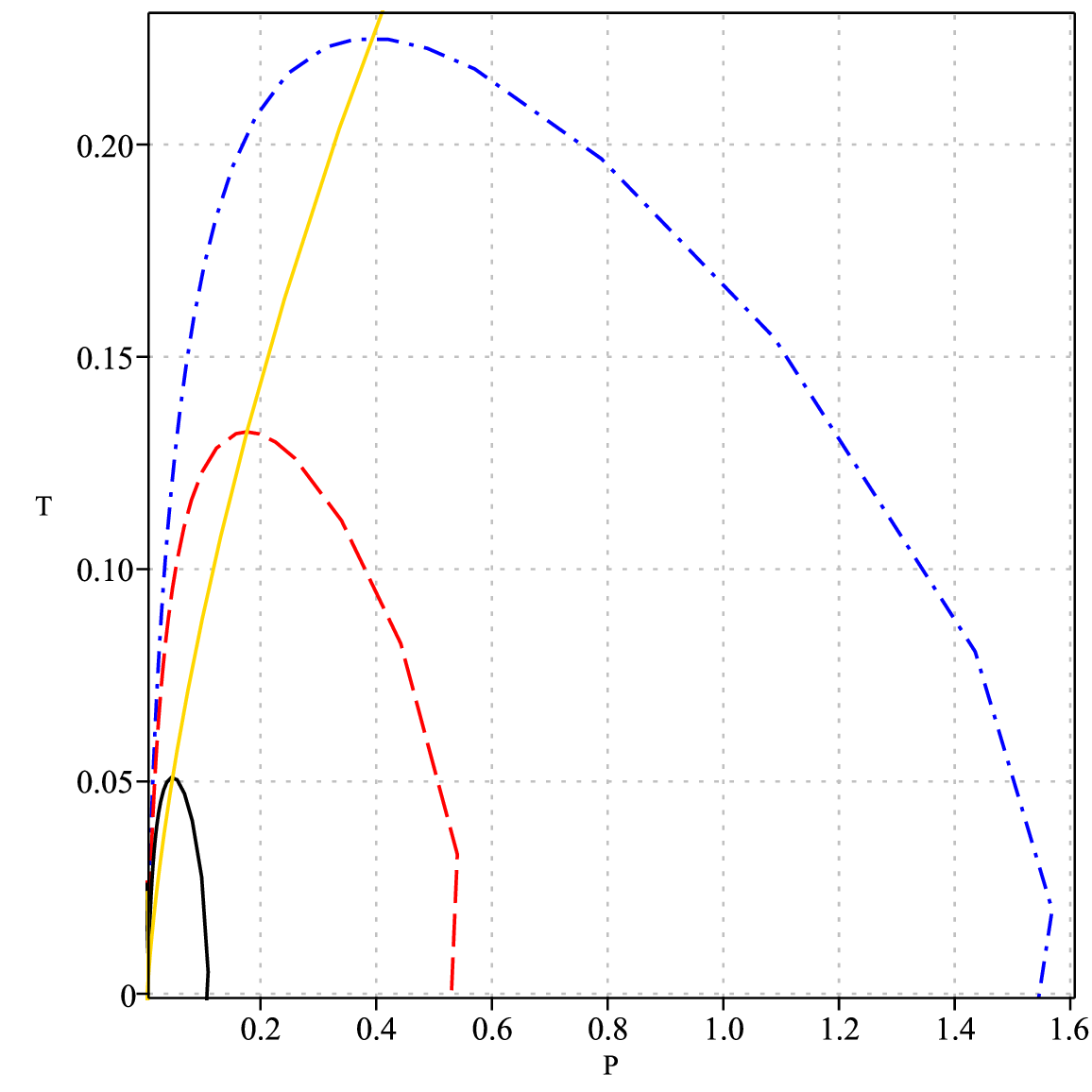}}\hfill
\caption{ $M=2$ denoted by solid black line,  $M=2.5$ denoted by red dash line,  $M=3$ denoted by blue dash-dot line  with $m=1$, $Q=1$, $c=1$, $c_1=-1$ and $c_2=1$. A solid gold line denotes an inverse curve.}\label{fig:11}
\end{figure}
\begin{figure}[hbt]
\centering
\subfloat[$\alpha=0.5$ ]{\includegraphics[width=6.5cm,height=5.5cm]{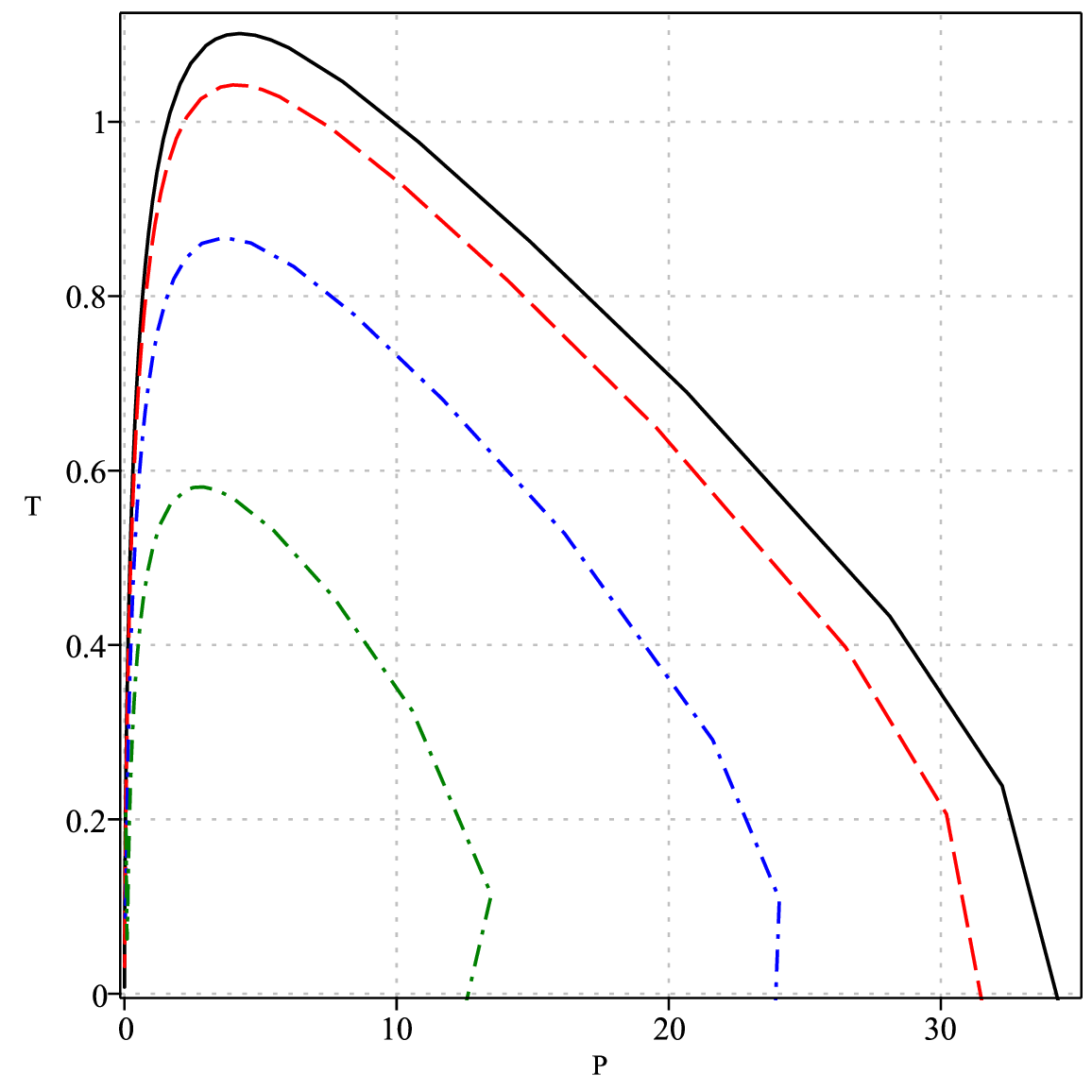}}\hfill
\subfloat[$\alpha=0.8$]{\includegraphics[width=6.5cm,height=5.5cm]{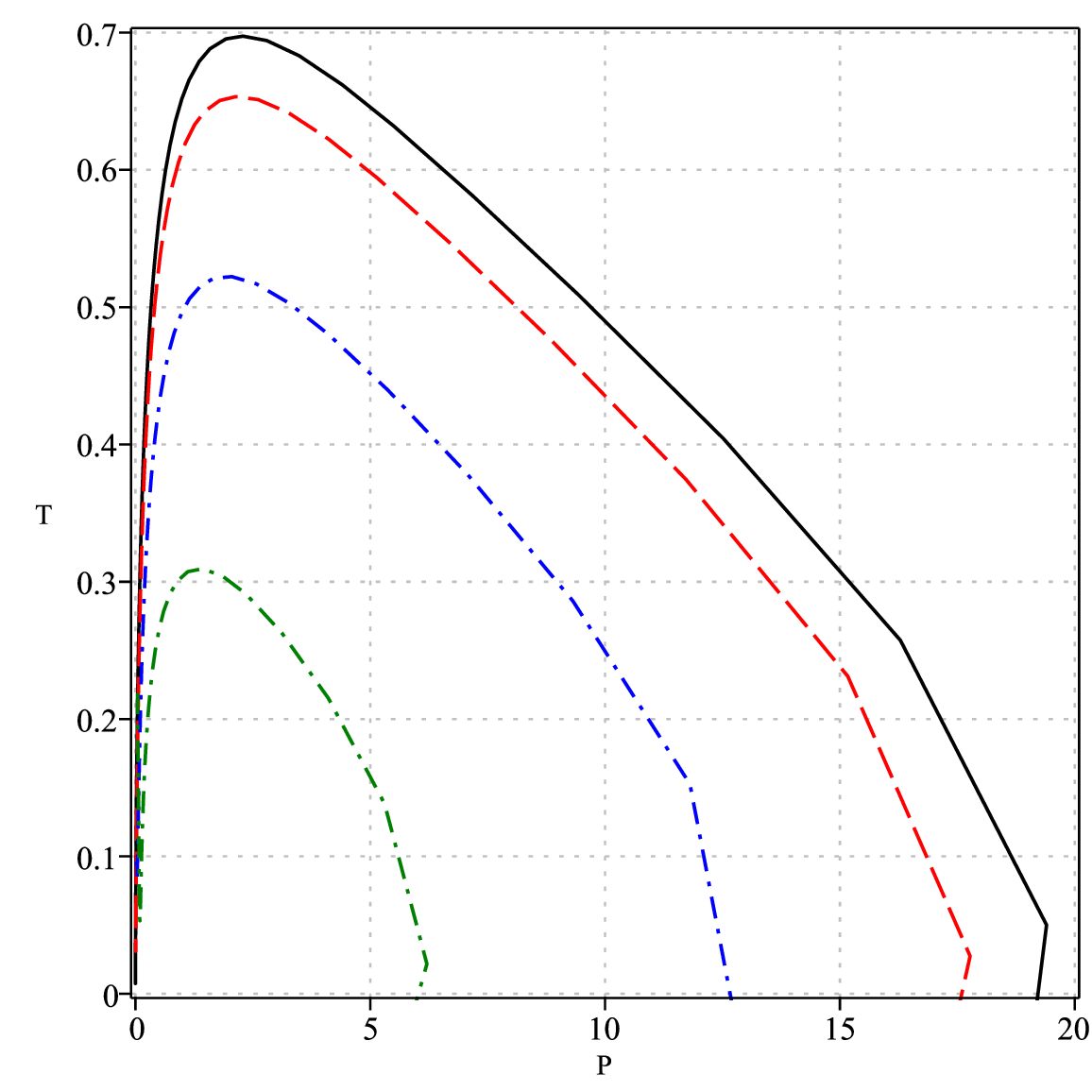}}\hfill
\caption{ $m=0$ denoted by solid black line,  $m=1$ denoted by red dash line,  $m=2$ denoted by blue dash dot line and $m=3$ denoted by green dash dot line with $M=5$, $Q=1$, $c=1$, $c_1=-1$ and $c_2=1$.}\label{fig:12}
\end{figure}
\begin{figure}[hbt]
\centering
\subfloat[$Q=2.0$ ]{\includegraphics[width=6.5cm,height=5.5cm]{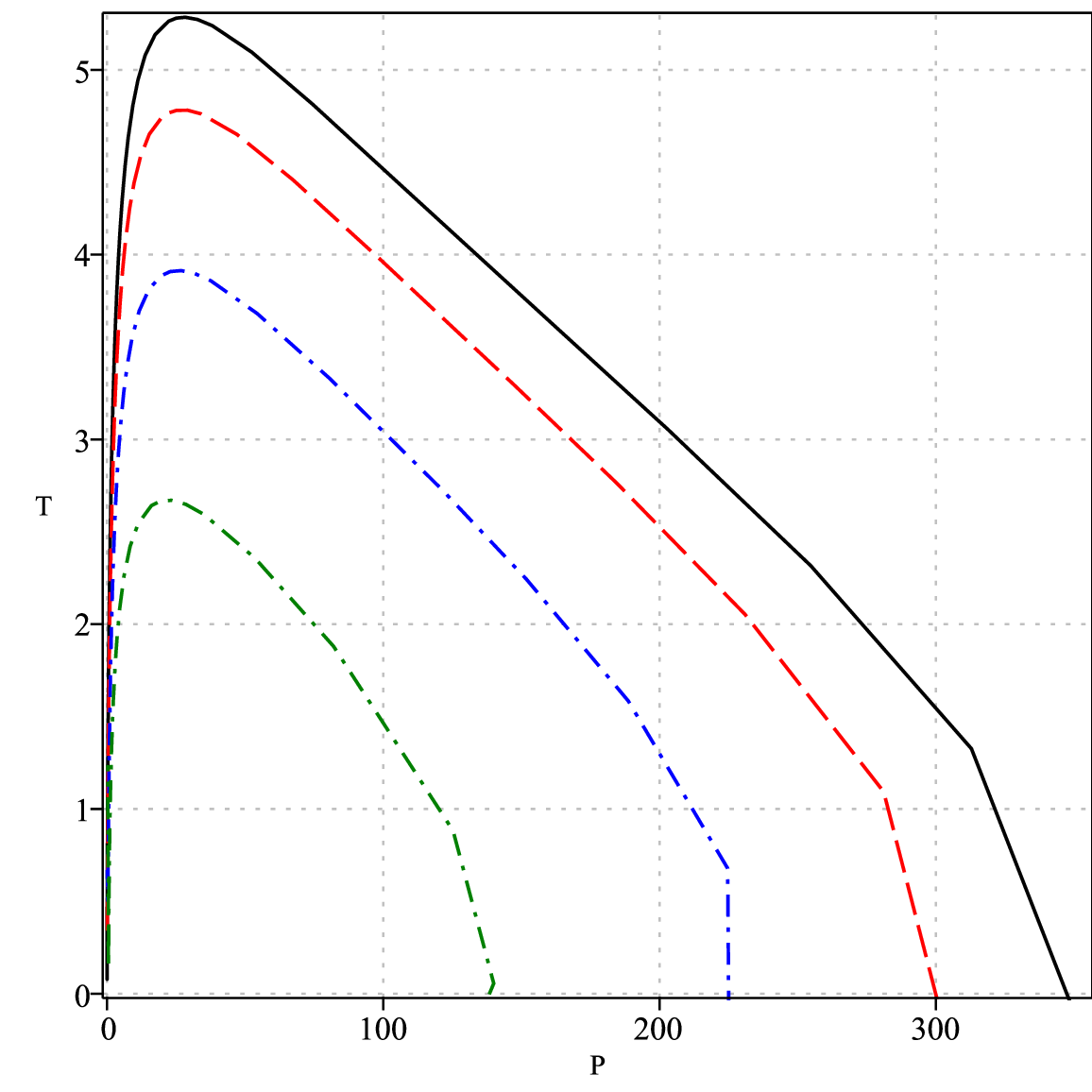}}\hfill
\subfloat[$Q=4.0$]{\includegraphics[width=6.5cm,height=5.5cm]{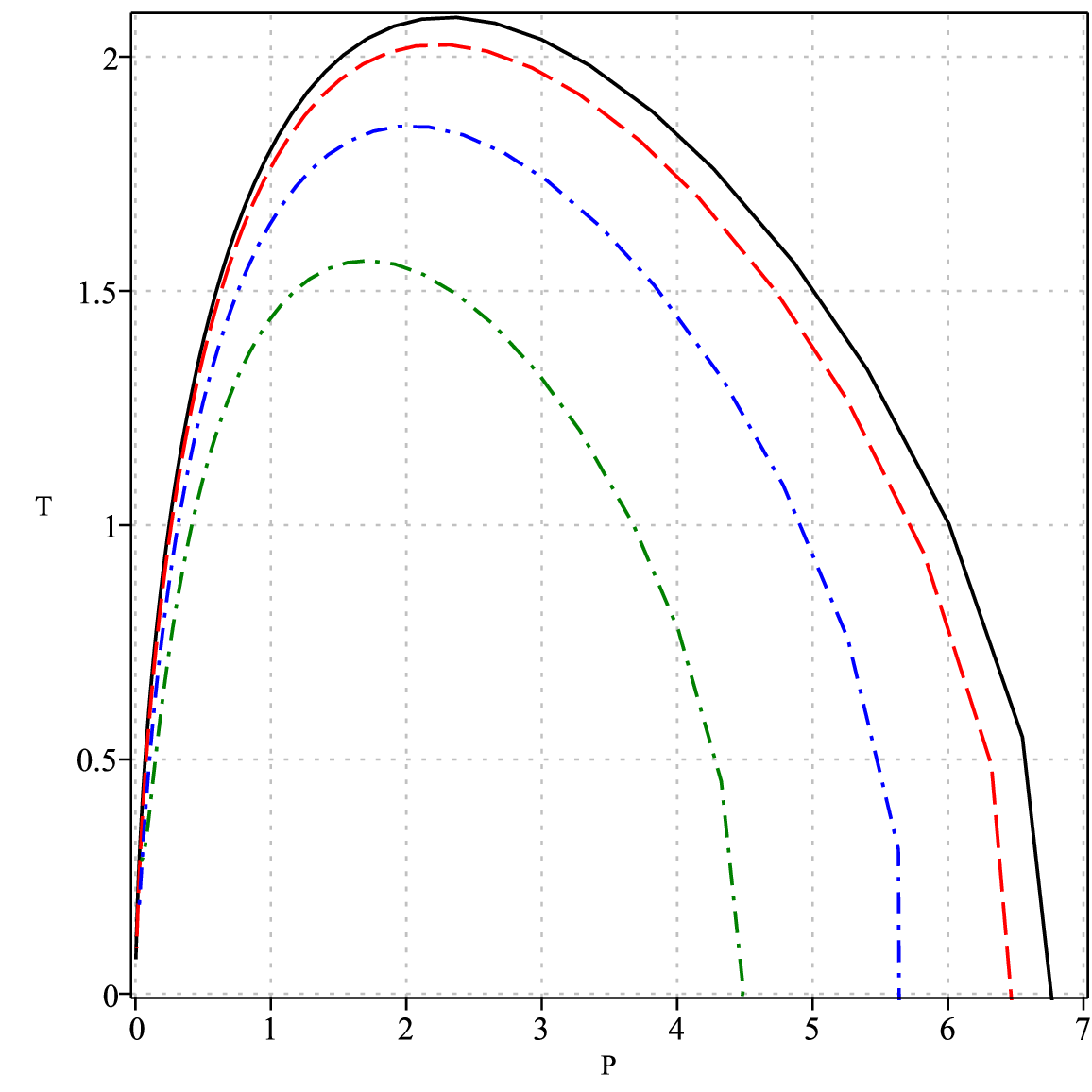}}\hfill
\caption{$M=20$, $\alpha=0.5$ and $c=1$.  Left panel: $m=0$ denoted by the solid black line,  $m=3$ denoted by a red dash line,  $m=5$ denoted by a blue dash-dot line and $m=7$ denoted by an orange dash-dot line.  Right panel: $m=0$ denoted by the solid black line,  $m=1$ denoted by a red dash line,  $m=2$ denoted by a blue dash-dot line and $m=3$ denoted by an orange dash-dot line.}\label{fig:13}
\end{figure}
\begin{figure}[hbt]
\centering
\subfloat[$c_1=0$ ]{\includegraphics[width=6.5cm,height=5.5cm]{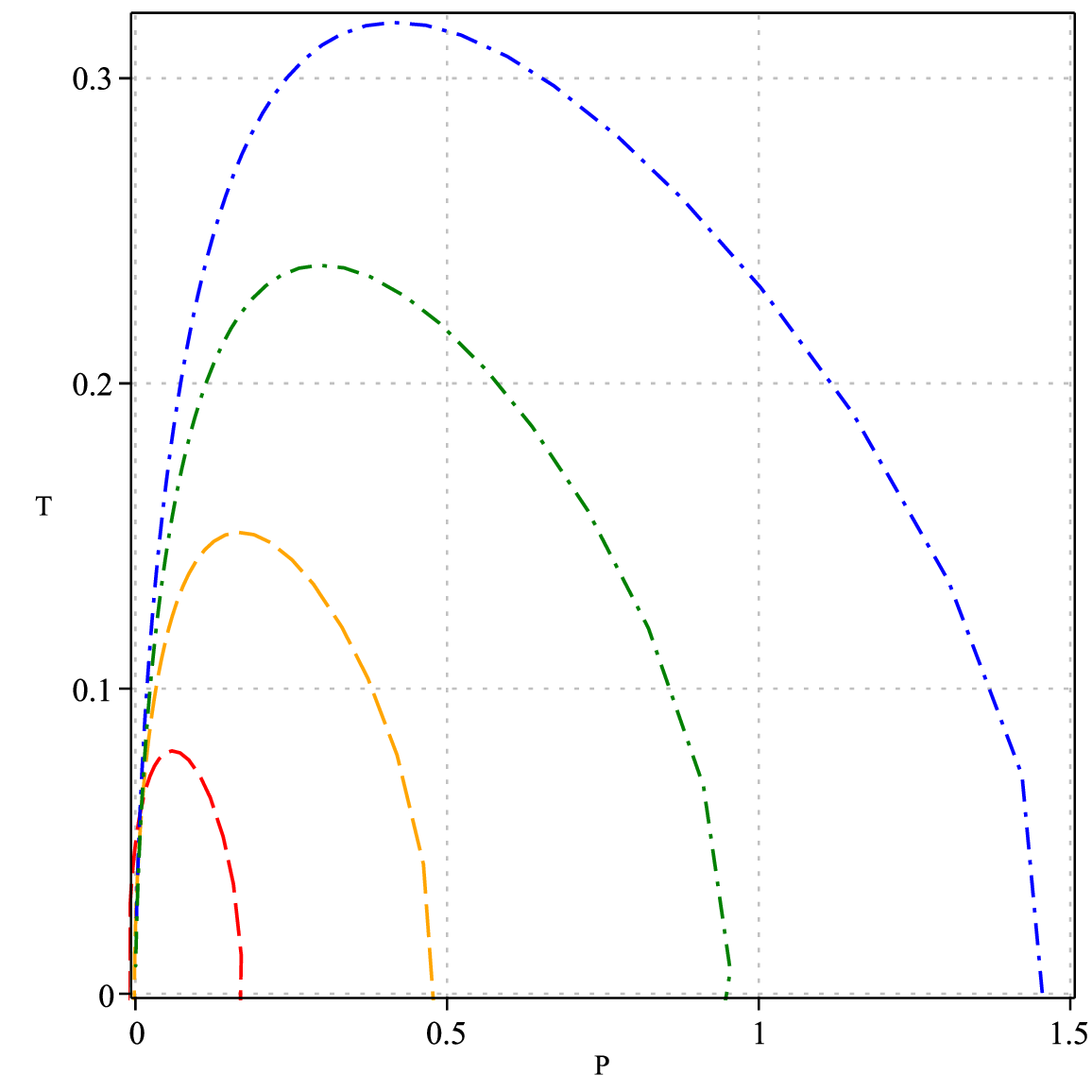}}\hfill
\subfloat[$c_2=0$]{\includegraphics[width=6.5cm,height=5.5cm]{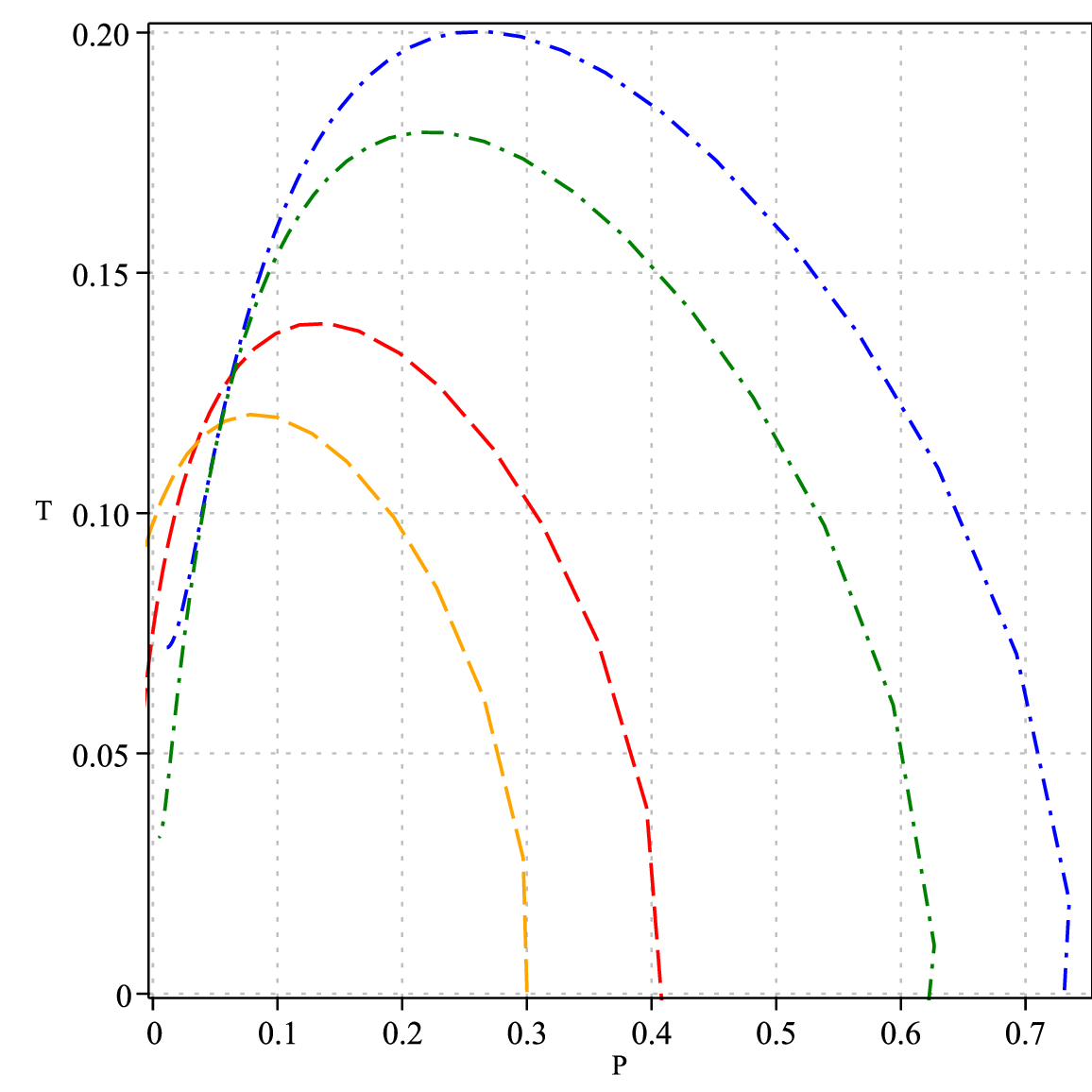}}\hfill
\caption{ $M=2$, $Q=1$, $\alpha=0.5$ and $c=1$. Left panel: $c_2=0.1$ denoted by an orange dash line,  $c_2=1$ denoted by a red dash line,  $c_2=-1$ denoted by a green dash-dot line, and $c_2=-2$ denoted by a blue dash-dot line. Right panel: $c_1=1$ denoted by a red dash line,  $c_1=2$ denoted by an orange dash line,  $c_1=-1$ denoted by a green dash-dot line and $c_2=-2$ denoted by a blue dash-dot line. }\label{fig:14}
\end{figure}
Using equation \eqref{eq:6.3} and equation \eqref{eq:3.4} constant mass curve can be obtained. In Fig. \ref{fig:11}-Fig. \ref{fig:15} we plot the constant mass curve. In Fig. \ref{fig:11}, constant mass and an inverse curve are shown for different values of black hole mass. The left region of the inverse curve represents cooling and the right region represents heating. In Fig. \ref{fig:12} and Fig. \ref{fig:13} constant mass curve is shown for different values of Gauss--Bonnet coupling parameter and charge of the black hole. The effects of parameters $c_1$ and $c_2$ are shown in Fig. \ref{fig:14} - Fig. \ref{fig:16}. 
\begin{figure}[hbt]
\centering
\subfloat[$\alpha=0.5$]{\includegraphics[width=6.5cm,height=5.5cm]{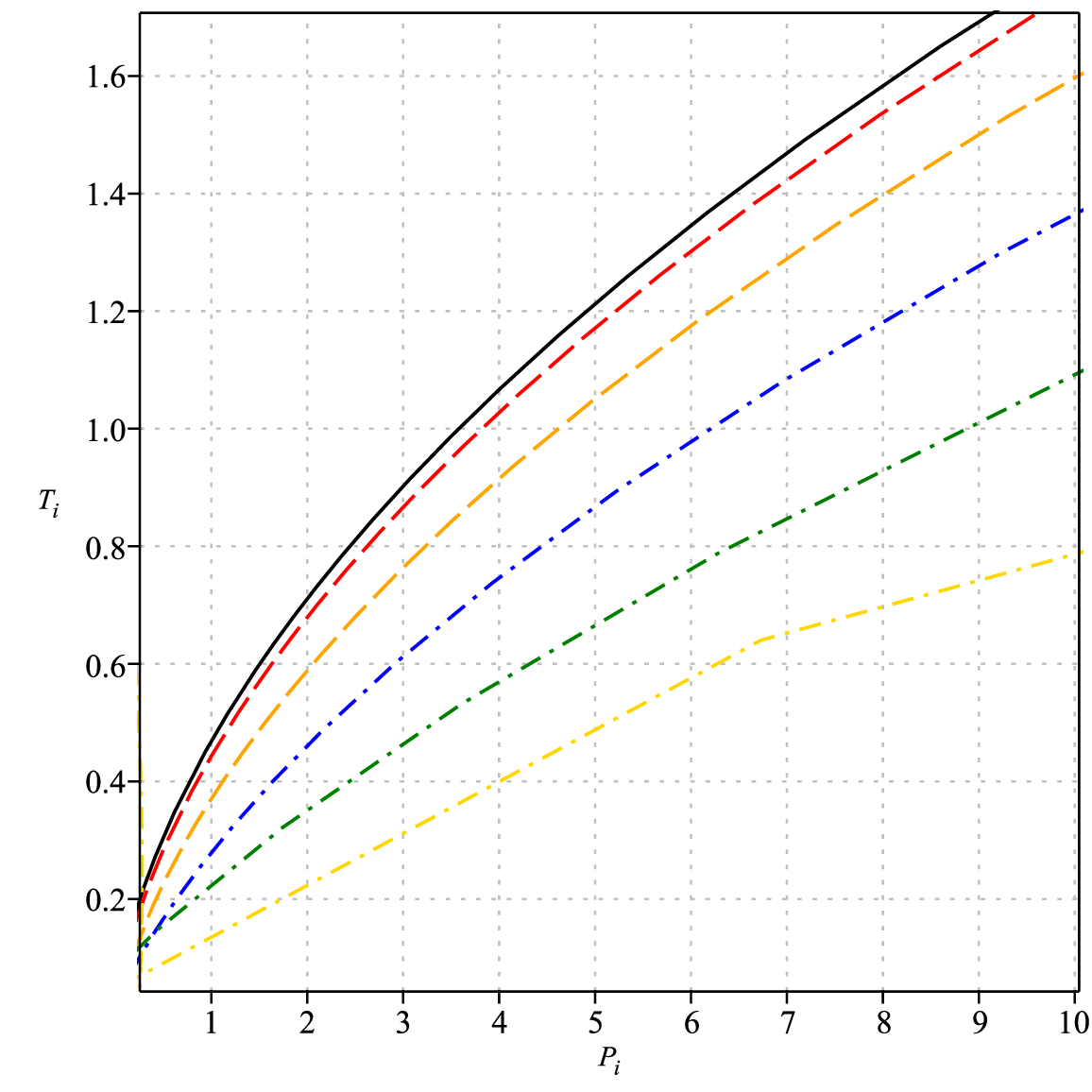}}\hfill
\subfloat[$m=1$]{\includegraphics[width=6.5cm,height=5.5cm]{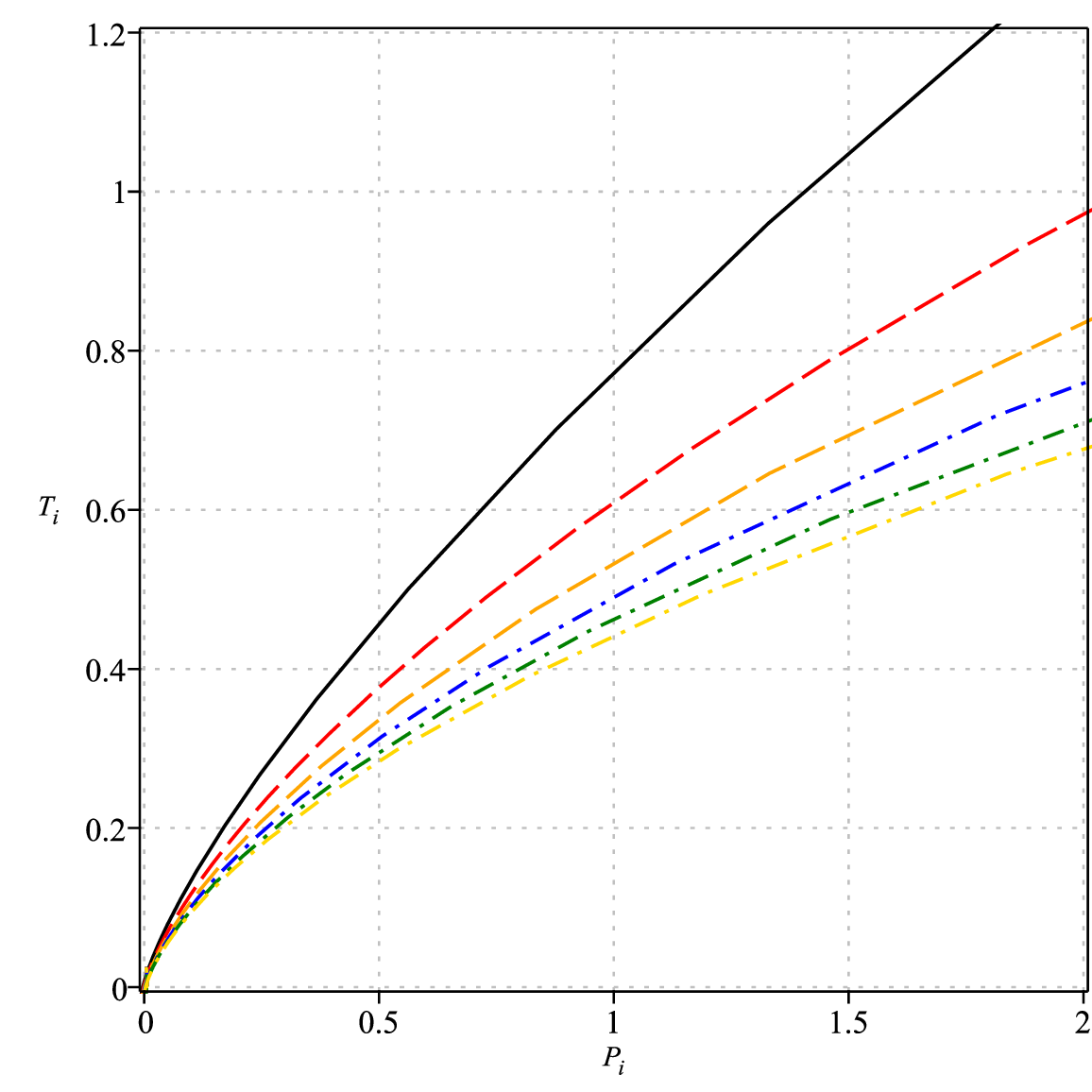}}\hfill
\caption{ $Q=1$, $c=1$, $c_1=-1$ and  $c_2=1$. Left panel: $m=0$ denoted by the solid black line,  $m=1$ denoted by red dash line,  $m=2$ denoted by orange dash line, $m=3$ denoted by blue dash-dot line, $m=4$ denoted by green dash-dot line and $m=5$ denoted by a gold dash-dot line. Right panel:  $\alpha=0$ denoted by a solid black line,  $\alpha=0.1$ denoted by red dash line,  $\alpha=0.2$ denoted by orange dash line, $\alpha=0.3$ denoted by blue dash-dot line, $\alpha=0.4$ denoted by green dash-dot line and $\alpha=0.5$ denoted by a gold dash-dot line. }\label{fig:17}
\end{figure}
 We  use equations \eqref{eq:6.2} and \eqref{eq:6.4} and  obtain the inverse pressure as
\begin{equation}\label{eq:6.5}
    P_{i}=\frac{6 Q^{2} r_{+}^{2}-4 r^{4}+8 Q^{2} \alpha +2 \alpha  r_{+}^{2}+8 \alpha^{2} -4 c^{2} c_{2} m^{2} r_{+}^{4}-3 c c_{1} m^{2} r_{+}^{5}-4 \alpha  c^{2} c_{2} m^{2} r_{+}^{2}-2 \alpha  c c_{1} m^{2} r_{+}^{3}}{16 \pi r_{+}^{6} }.
\end{equation}
Now,  using the relations \eqref{eq:6.4} and \eqref{eq:6.5}, we obtain
\begin{equation}\label{eq:6.6}
    T_{i}= \frac{-m^{2} c c_{1} r_{+}^{3}+(-2 c^{2} c_{2} m^{2}-2) r_{+}^{2}+4 Q^{2}+4 \alpha}{8 \pi  r_{+}^{3}}.
\end{equation}
\begin{figure}[hbt]
\centering
\subfloat[$c_2=0$]{\includegraphics[width=6.5cm,height=5.5cm]{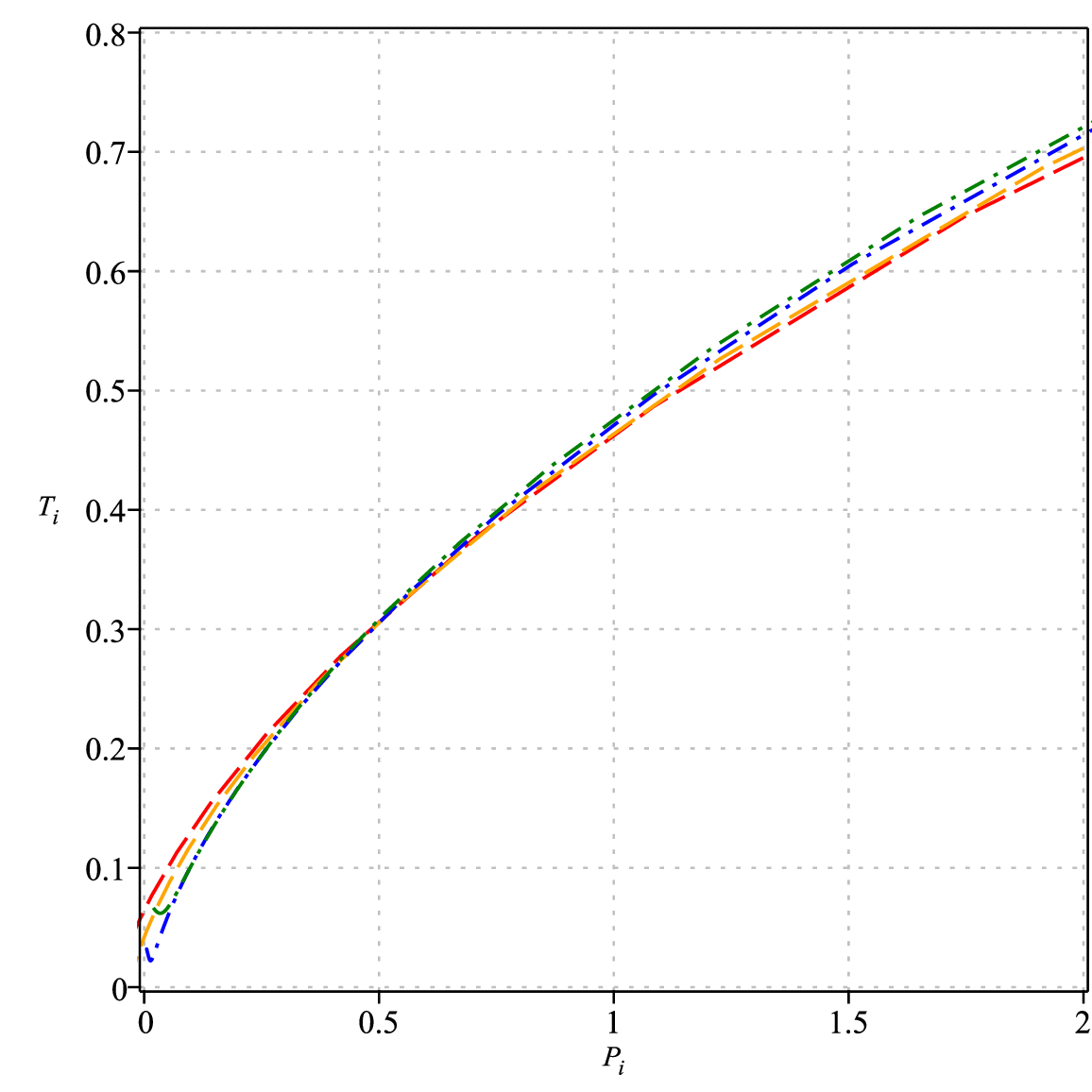}}\hfill
\subfloat[$c_1=0$]{\includegraphics[width=6.5cm,height=5.5cm]{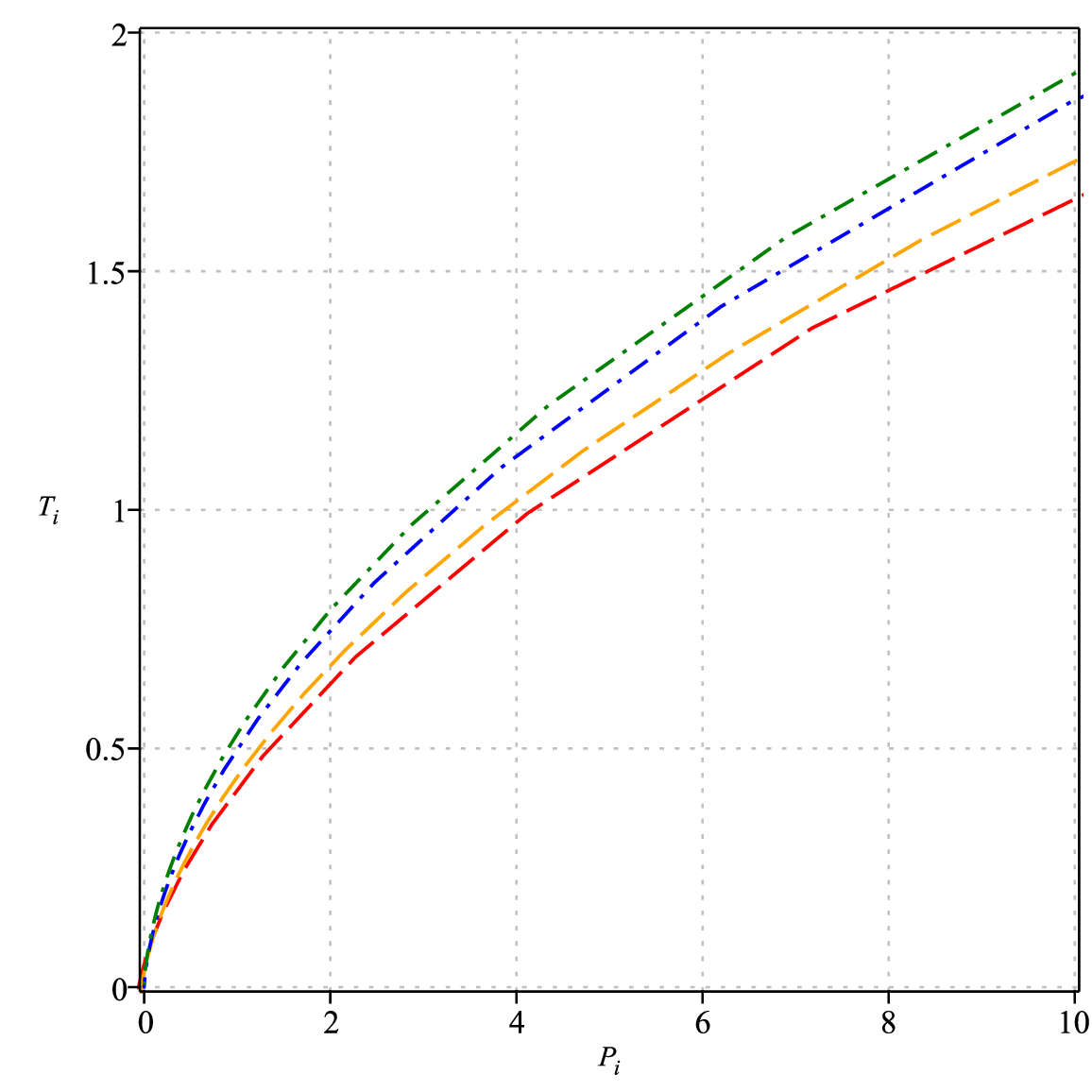}}\hfill
\caption{$Q=1$, $c=1$, $\alpha=0.5$ and $m=1$. Left panel: $c_1=2$ denoted by a red dash line,  $c_1=1$ denoted by an orange dash line,  $c_1=-1$ denoted by a blue dash-dot line, $c_1=-2$ denoted by a green dash-dot line. Right panel:  $c_2=2$ denoted by a red dash line,  $c_2=1$ denoted by an orange dash line,  $c_2=-1$ denoted by a blue dash-dot line, $c_2=-2$ denoted by a green dash-dot line.}\label{fig:18}
\end{figure}
\begin{figure}[hbt]
\centering
\subfloat[$c_1=0$]{\includegraphics[width=6.5cm,height=5.5cm]{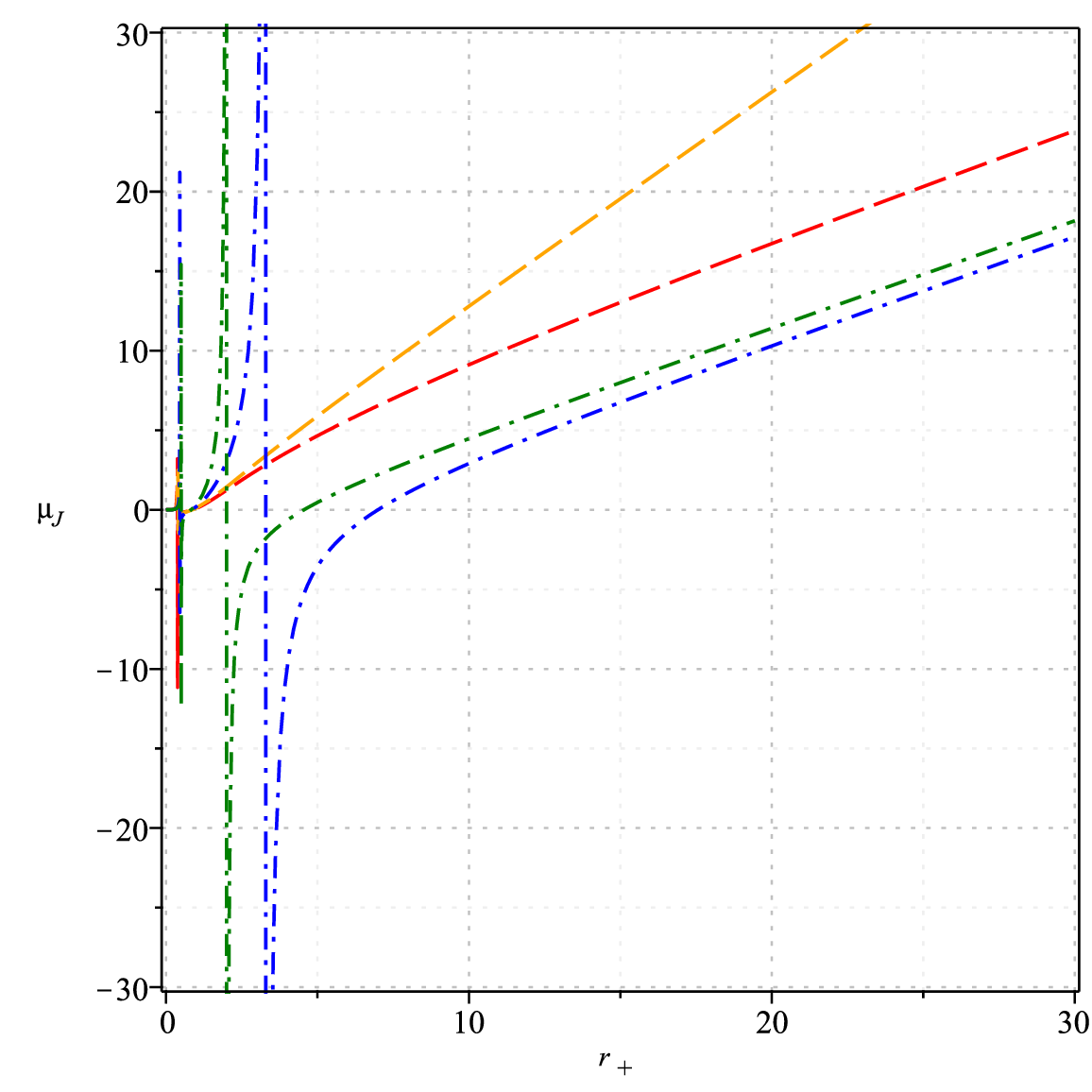}}\hfill
\subfloat[$c_2=0$]{\includegraphics[width=6.5cm,height=5.5cm]{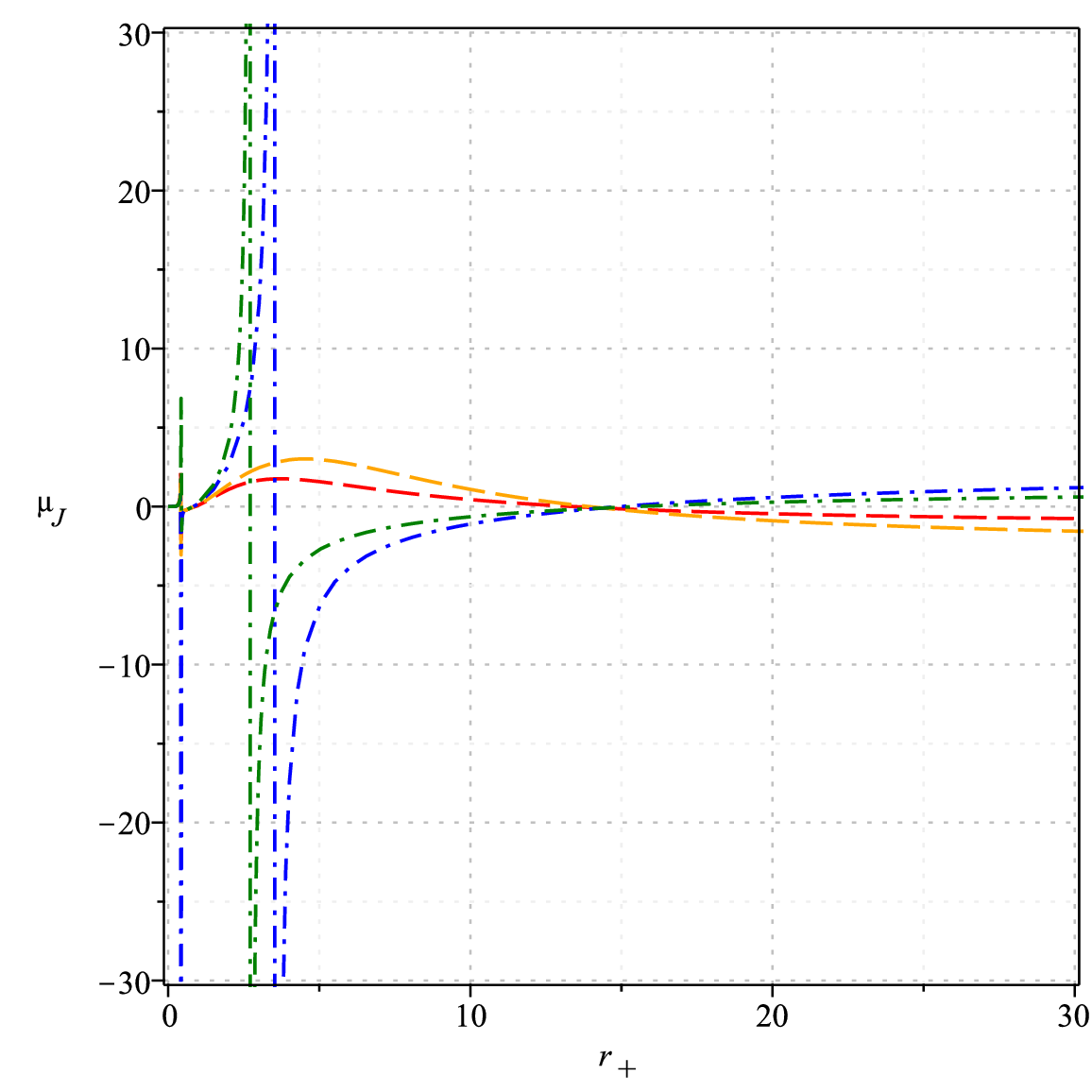}}\hfill
\caption{$M=2.5$, $Q=1$, $c=1$, $\alpha=0.5$ and $m=1$. Left panel: $c_2=-2$ denoted by a red dash line,  $c_2=-1$ denoted by an orange dash line,  $c_2=1$ denoted by a blue dash-dot line, $c_2=2$ denoted by a green dash-dot line. Right panel:  $c_1=-2$ denoted by a red dash line,  $c_1=-1$ denoted by an orange dash line,  $c_1=1$ denoted by a blue dash-dot line, $c_1=2$ denoted by a green dash-dot line.}\label{fig:19}
\end{figure}
Finally, we will drive the Joule - Thomson thermodynamic coefficient using equation \eqref{eq:6.1}, equation \eqref{eq:6.3} and  equation \eqref{eq:6.4}
\begin{equation}\label{eq:6.7}
    \mu_{J}= \frac{8r_{+}^{3} \Bigl( 6 M r_{+}^{3}-6 Q^{2} r_{+}^{2}-r_{+}^{4}-4 Q^{2} \alpha -4 \alpha  r_{+}^{2}-4 \alpha^{2}  -c^{2} c_{2} m^{2} r_{+}^{4}+2 \alpha  c^{2} c_{2} m^{2} r_{+}^{2}+\alpha  c c_{1} m^{2} r_{+}^{3}\Bigl)}{3 (r_{+}^{2}+2 \alpha )^{2} \Bigl(12 M r_{+} -8 Q^{2}-4 r_{+}^{2}-8 \alpha -4 c^{2} c_{2} m^{2} r_{+}^{2}-m^{2} c c_{1} r_{+}^{3} \Bigl)}.
\end{equation}
\begin{figure}[hbt]
\centering
\subfloat[$c_1=1$ and $c_2=1$]{\includegraphics[width=6.5cm,height=5cm]{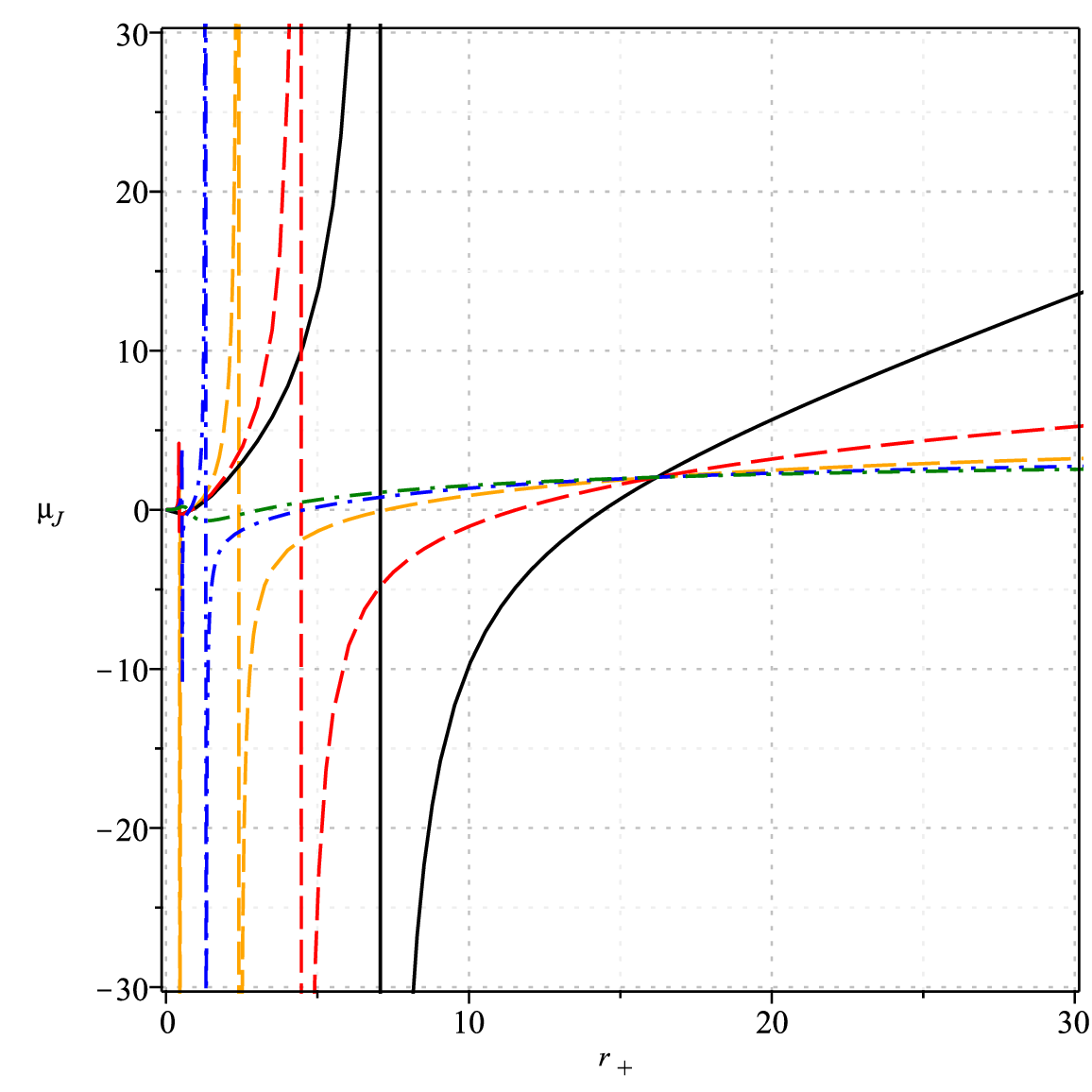}}\hfill
\subfloat[$c_1=1$ and $c_2=-1$]{\includegraphics[width=6.5cm,height=5cm]{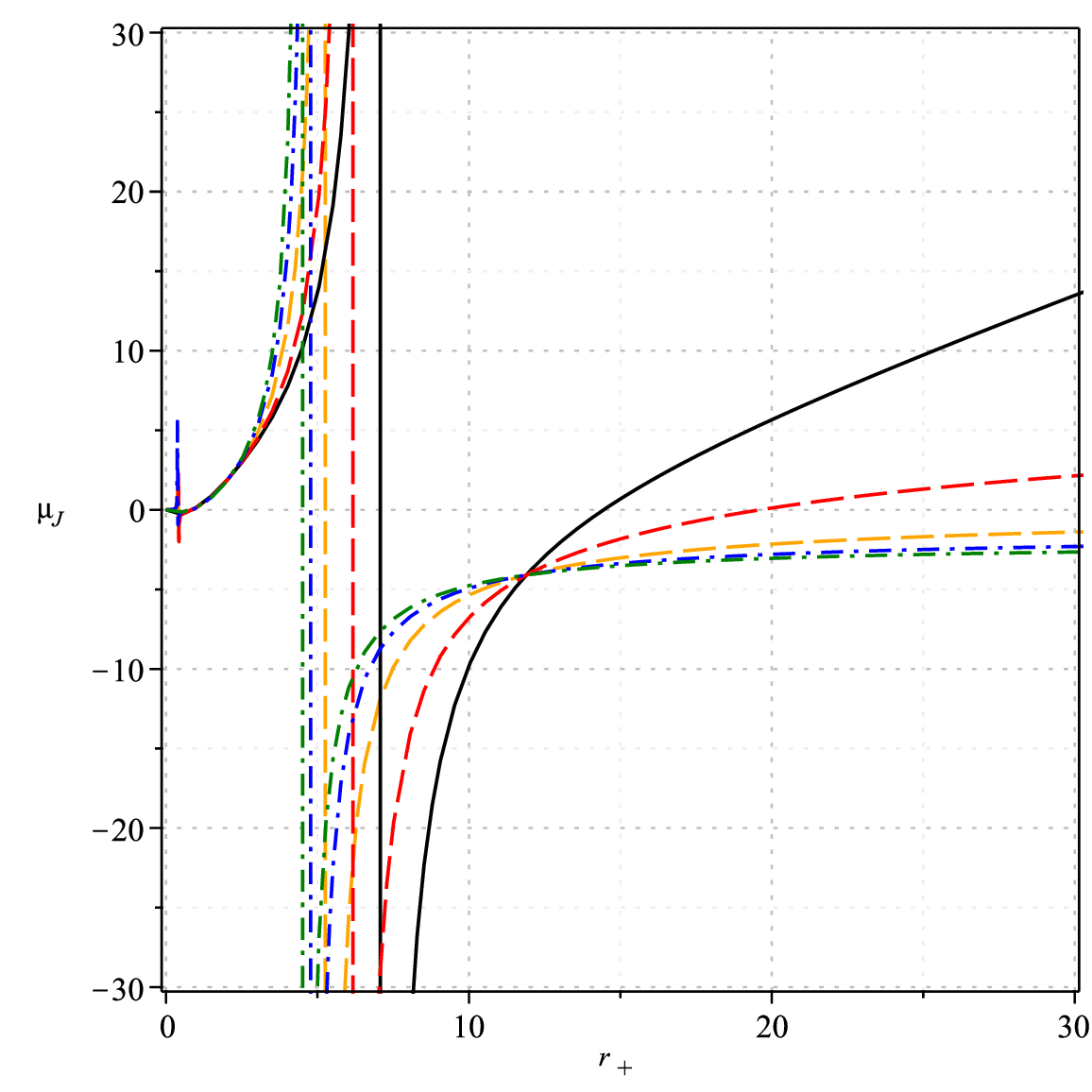}}\hfill
\subfloat[$c_1=-1$ and $c_2=1$]{\includegraphics[width=6.5cm,height=5cm]{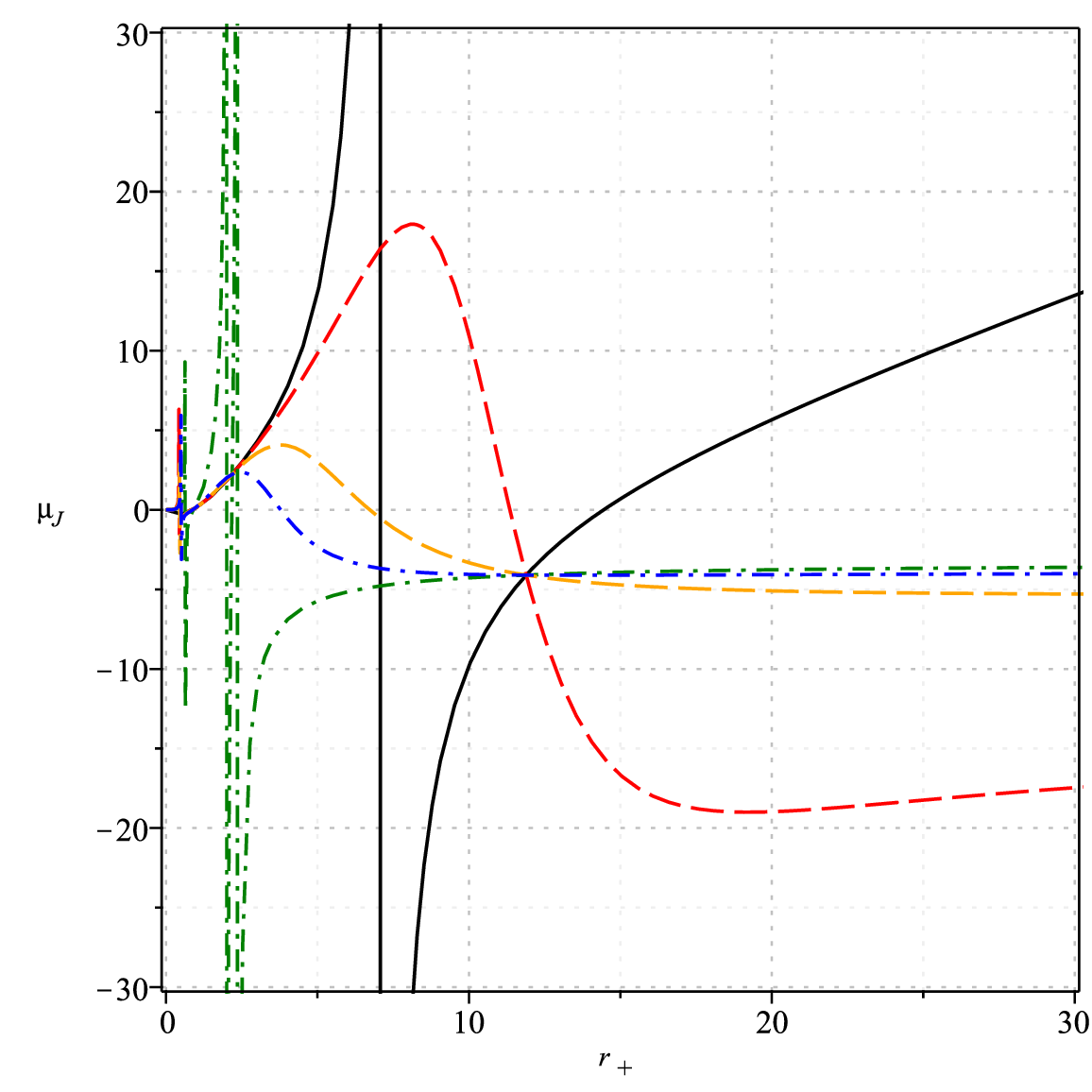}}\hfill
\subfloat[$c_1=-1$ and $c_2=-1$]{\includegraphics[width=6.5cm,height=5cm]{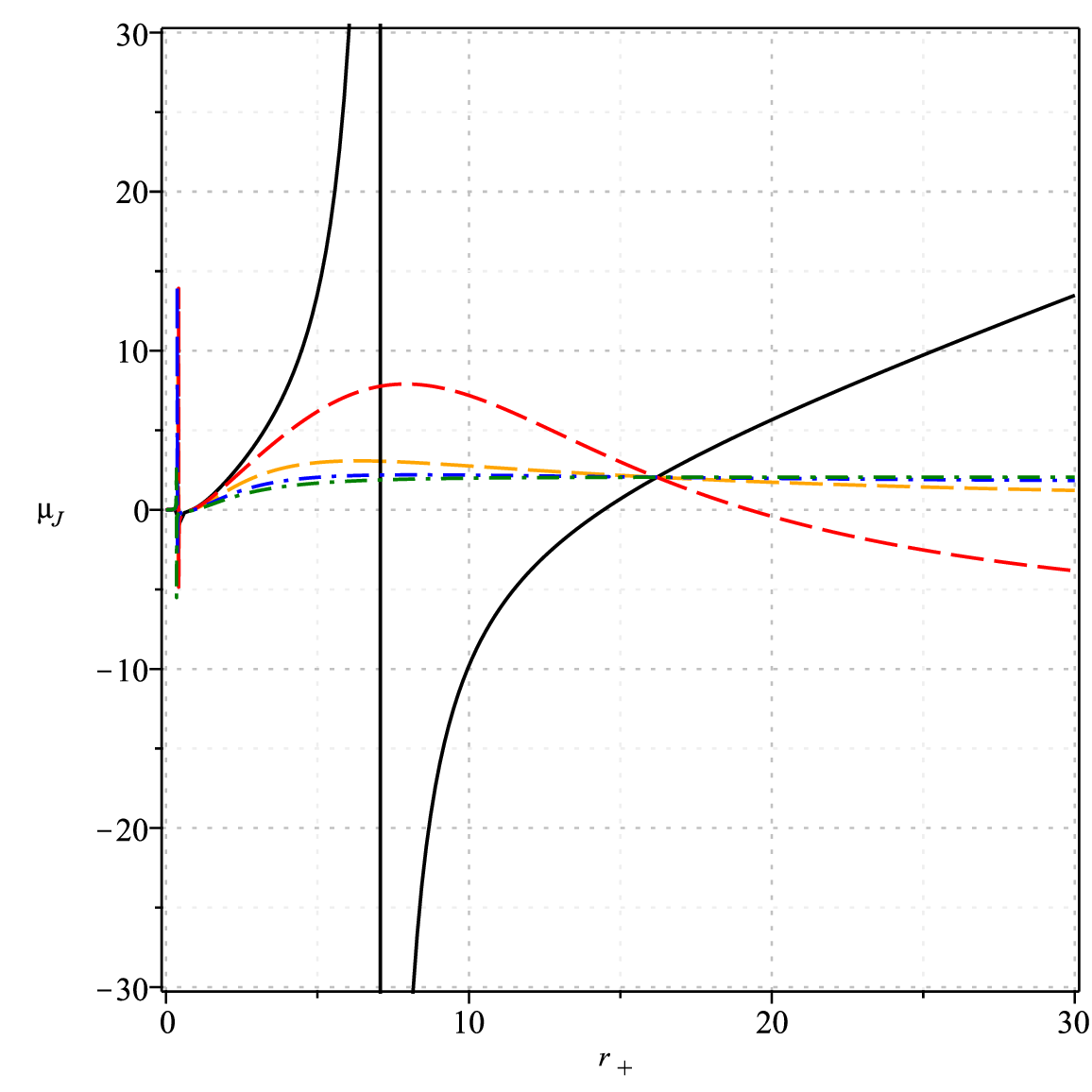}}\hfill
\caption{$m=0$ denoted by solid black line,  $m=0.5$ denoted by red dash line,  $m=1$ denoted by orange dash line, $m=1.5$ denoted by blue dash-dot line, $m=2$ denoted by green dash-dot line with $M=2.5$, $Q=1$, $c=1$, and  $\alpha=0.5$.}\label{fig:20}
\end{figure}
In Fig. \ref{fig:20} and Fig. \ref{fig:19}, Joule - Thomson thermodynamic coefficient is plotted. $\mu<0$ represent heating phase and $\mu>0$ represent cooling phase.

\section{Conclusions}\label{sec:7}
In this work, we have found an exact solution of Einstein-Gauss--Bonnet  massive gravity with charge in $4D$ AdS space and the horizon structure of the black holes is discussed. The physical mass (enthalpy) and Hawking temperature of the black holes were computed. Treating the cosmological constant as the pressure, we drive the first law of black hole thermodynamics. To check global stability, the Gibbs free energy of the black hole was computed. For the local stability of the black holes we estimated the specific heat. Here, we found that as the mass of the graviton increases one divergent point appears. For $m=0,1$, the specific heat of the black hole goes from a negative value (unstable phase) to a positive one (stable phase) and when $m>1$ a second-order phase transition occurs. Furthermore, we investigated the Van der Walls-like phase transition of the black holes. The effects of graviton mass, the charge of the black holes and the Gauss--Bonnet  coupling parameter on the critical points were also studied. As the mass of the graviton increase, critical pressure also increases and critical temperature shows the opposite behavior. When charge and Gauss--Bonnet  coupling parameters of the black hole increase, we found that both critical pressure and temperature decrease. From the $G-T$ plot, we observed that a swallow tail appears below the critical point which indicates a first-order phase transition, and at the critical point, a second-order phase transition occurs. Finally, Joule-Thomson expansion of charged AdS black hole in $4D$ Einstein--Gauss--Bonnet  massive gravity was studied. The effects of the Gauss--Bonnet  coupling parameter and massive gravity parameters on the constant mass and inverse curve are also depicted. 

\section*{Acknowledgement}
D.V.S. thanks University Grant Commission for the Start-Up Grant No. 30-600/2021(BSR)/1630.
\section*{Data Availability Statement}
No Data is associated with the manuscript.

 \appendix
\section{}
    The Ricci and Kretschmann scalar for the metric function \eqref{eq:2.11} are given by
\begin{eqnarray}
     R&=& -\frac{f_{{rr}}(r) r^{2}+4 f_{r}(r) r +2 f(r) -2}{r^{2}},
\\
    K&=&f_{{rr}}^{2}(r)+\frac{4 f_{r}^{2}(r)}{r^{2}}+\frac{4 (-1+f(r) )^{2}}{r^{4}},
\end{eqnarray}
where we use $f(r)= e^{2A(r)}$, $f_{rr}(r)$ stands for $d^2f(r)/dr^2$ and  $f_r(r)$ stands for $df(r)/dr$. 

The differentiation of $f(r)$ with respect to $r$ is given by
\begin{eqnarray}
f_{r}&=& \frac{r}{\alpha}\left[  1-\sqrt{ 1+4 \alpha \left\{ \frac{2 M}{r^{3}}-\frac{Q^{2}}{r^{4}}-\frac{1}{l^{2}}-\frac{m^{2} (2 c^{2} c_{2} +c c_{1} r )}{2 r^{2}} \right\} } \right]\nonumber\\
&-&\frac{r^{2} \Bigr[ -\frac{6 M}{r^{4}}+\frac{4 Q^{2}}{r^{5}}-\frac{m^{2} c c_{1}}{2 r^{2}}+\frac{m^{2} (2 c^{2} c_{2} +c c_{1} r )}{r^{3}}\Bigr]}{\sqrt{1+4 \alpha  \biggl\{ \frac{2 M}{r^{3}}-\frac{Q^{2}}{r^{4}}-\frac{1}{l^{2}}-\frac{m^{2} (2 c^{2} c_{2} +c c_{1} r )}{2 r^{2}} \biggl\}}},\label{eq:0.3}\\
f_{rr}&=& \frac{1}{\alpha}\Biggr[ 1-\sqrt{1+4 \alpha  \biggl\{ \frac{2 M}{r^{3}}-\frac{Q^{2}}{r^{4}}-\frac{1}{l^{2}}-\frac{m^{2} (2 c^{2} c_{2} +c c_{1} r )}{2 r^{2}} \biggl\} }\Biggr]\nonumber\\
&+&\frac{2 \alpha r^{2} \Bigr[ -\frac{6 M}{r^{4}}+\frac{4 Q^{2}}{r^{5}}-\frac{m^{2} c c_{1}}{2 r^{2}}+\frac{m^{2} (2 c^{2} c_{2} +c c_{1} r )}{r^{3}}\Bigr]^{2}}{\Bigr[ 1+4 \alpha  \Bigl\{ \frac{2 M}{r^{3}}-\frac{Q^{2}}{r^{4}}-\frac{1}{l^{2}}-\frac{m^{2} (2 c^{2} c_{2} +c c_{1} r )}{2 r^{2}}\Bigl\}\Bigr]^{\frac{3}{2}}}\nonumber\\
&+&\frac{l (-2 c^{2} c_{2} m^{2} r^{2}-m^{2} c c_{1} r^{3}+4 Q^{2})}{r^{2} \sqrt{\Bigl\{ (-4 c^{2} c_{2} m^{2} r^{2}-2 m^{2} c c_{1} r^{3}+8 M r -4 Q^{2}) l^{2}-4 r^{4}\Bigl\} \alpha +r^{4} l^{2}}}.\label{eq:0.4} 
\end{eqnarray}
Therefore, using \eqref{eq:0.3} and \eqref{eq:0.4}, Ricci and Kretschmann scalar are given by 
\begin{eqnarray}
  R&=&\frac{6}{\alpha} \Biggr[ -1+{ \sqrt{1+\frac{8 \alpha  M}{r^{3}}-\frac{4 \alpha  Q^{2}}{r^{4}}-\frac{4 \alpha}{l^{2}}-\frac{4 \alpha  m^{2} c^{2} c_{2}}{r^{2}}-\frac{2 \alpha  m^{2} c c_{1}}{r}}}\Biggr]\nonumber\\
 &- &\frac{ \bigl( -4 c^{2} c_{2} m^{2} r^{2}-m^{2} c c_{1} r^{3}+12 M r -8 Q^{2}\bigl)^{2} \alpha  l^{3}}{2r^{2} \Bigr[ \bigl\{  (-4 c^{2} c_{2} m^{2} r^{2}-2 m^{2} c c_{1} r^{3}+8 M r -4 Q^{2}) l^{2}-4 r^{4}\bigl\} \alpha +r^{4} l^{2}\Bigr]^{\frac{3}{2}} }\nonumber\\
 &- &
\frac{l \bigl( -10 c^{2} c_{2} m^{2} r^{2}-3 m^{2} c c_{1} r^{3}+24 M r -12 Q^{2} \bigl)}{r^{2} \sqrt{ \bigl\{ (-4 c^{2} c_{2} m^{2} r^{2}-2 m^{2} c c_{1} r^{3}+8 M r -4 Q^{2}) l^{2}-4 r^{4}\bigl\} \alpha +r^{4} l^{2}}},  
\end{eqnarray} 
\begin{eqnarray}
 K&=& \Biggr[ \frac{1}{\alpha} \left[ {1-\sqrt{1+4 \alpha  \left\{ \frac{2 M}{r^{3}}-\frac{Q^{2}}{r^{4}}-\frac{1}{l^{2}}-\frac{m^{2} (2 c^{2} c_{2} +c c_{1} r )}{2 r^{2}} \right\} }} \right]\nonumber\\
 &-&\frac{4 r \Bigr[ -\frac{6 M}{r^{4}}+\frac{4 Q^{2}}{r^{5}}-\frac{m^{2} c c_{1}}{2 r^{2}}+\frac{m^{2} (2 c^{2} c_{2} +c c_{1} r )}{r^{3}}\Bigr]}{\sqrt{1+4 \alpha  \biggl\{ \frac{2 M}{r^{3}}-\frac{Q^{2}}{r^{4}}-\frac{1}{l^{2}}-\frac{m^{2} (2 c^{2} c_{2} +c c_{1} r )}{2 r^{2}}\biggl\}}}  \nonumber\\
 &+&\frac{2\alpha r^{2} \left[ -\frac{6 M}{r^{4}}+\frac{4 Q^{2}}{r^{5}}-\frac{m^{2} c c_{1}}{2 r^{2}}+\frac{m^{2} (2 c^{2} c_{2} +c c_{1} r )}{r^{3}}\right]^{2} }{ \Bigr[ 1+4 \alpha  \Bigl\{ \frac{2 M}{r^{3}}-\frac{Q^{2}}{r^{4}}-\frac{1}{l^{2}}-\frac{m^{2} (2 c^{2} c_{2} +c c_{1} r )}{2 r^{2}}\Bigl\}\Bigr]^{\frac{3}{2}}}\nonumber\\
 &-&\frac{r^{2} \Bigr[ \frac{24 M}{r^{5}}-\frac{20 Q^{2}}{r^{6}}+\frac{2 m^{2} c c_{1}}{r^{3}}-\frac{3 m^{2} (2 c^{2} c_{2} +c c_{1} r )}{r^{4}}\Bigr]}{\sqrt{1+4 \alpha  \Bigl\{ \frac{2 M}{r^{3}}-\frac{Q^{2}}{r^{4}}-\frac{1}{l^{2}}-\frac{m^{2} (2 c^{2} c_{2} +c c_{1} r )}{2 r^{2}}\Bigl\}}} \Biggr]^{2} \nonumber\\
 &+&\frac{4}{{r^{2}}} \Biggr[ \frac{r}{{\alpha}} \left[ 1-\sqrt{1+4 \alpha  \left\{ \frac{2 M}{r^{3}}-\frac{Q^{2}}{r^{4}}-\frac{1}{l^{2}}-\frac{m^{2} (2 c^{2} c_{2} +c c_{1} r )}{2 r^{2}}\right\}}\right]\nonumber\\
 &-&\frac{r^{2} \Bigr[ -\frac{6 M}{r^{4}}+\frac{4 Q^{2}}{r^{5}}-\frac{m^{2} c c_{1}}{2 r^{2}}+\frac{m^{2} (2 c^{2} c_{2} +c c_{1} r )}{r^{3}}\Bigr]}{\sqrt{1+4 \alpha  \Bigl\{ \frac{2 M}{r^{3}}-\frac{Q^{2}}{r^{4}}-\frac{1}{l^{2}}-\frac{m^{2} (2 c^{2} c_{2} +c c_{1} r )}{2 r^{2}}\Bigl\}}}\Biggr]^{2} \nonumber\\
 &+&\frac{1}{{\alpha^{2}}}\Biggr[ 1-\sqrt{1+4 \alpha  \Bigl\{ \frac{2 M}{r^{3}}-\frac{Q^{2}}{r^{4}}-\frac{1}{l^{2}}-\frac{m^{2} (2 c^{2} c_{2} +c c_{1} r )}{2 r^{2}}\Bigl\}}\Biggr]^{2}
\end{eqnarray}
\begin{figure}[hbt]
\centering
\subfloat[$\alpha=0.5$]{\includegraphics[width=6.5cm,height=5.5cm]{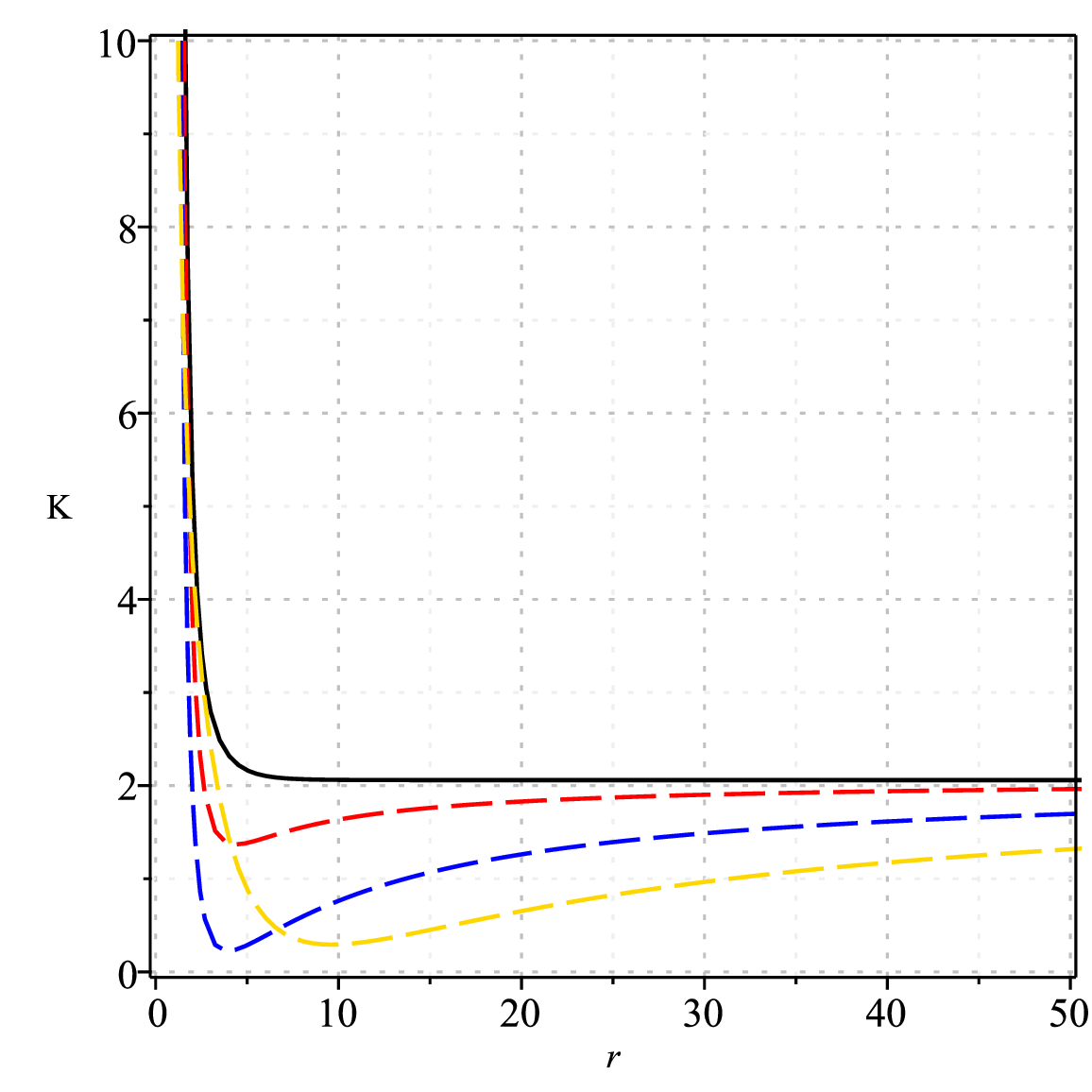}}\hfill
\subfloat[$\alpha=0.8$]{\includegraphics[width=6.5cm,height=5.5cm]{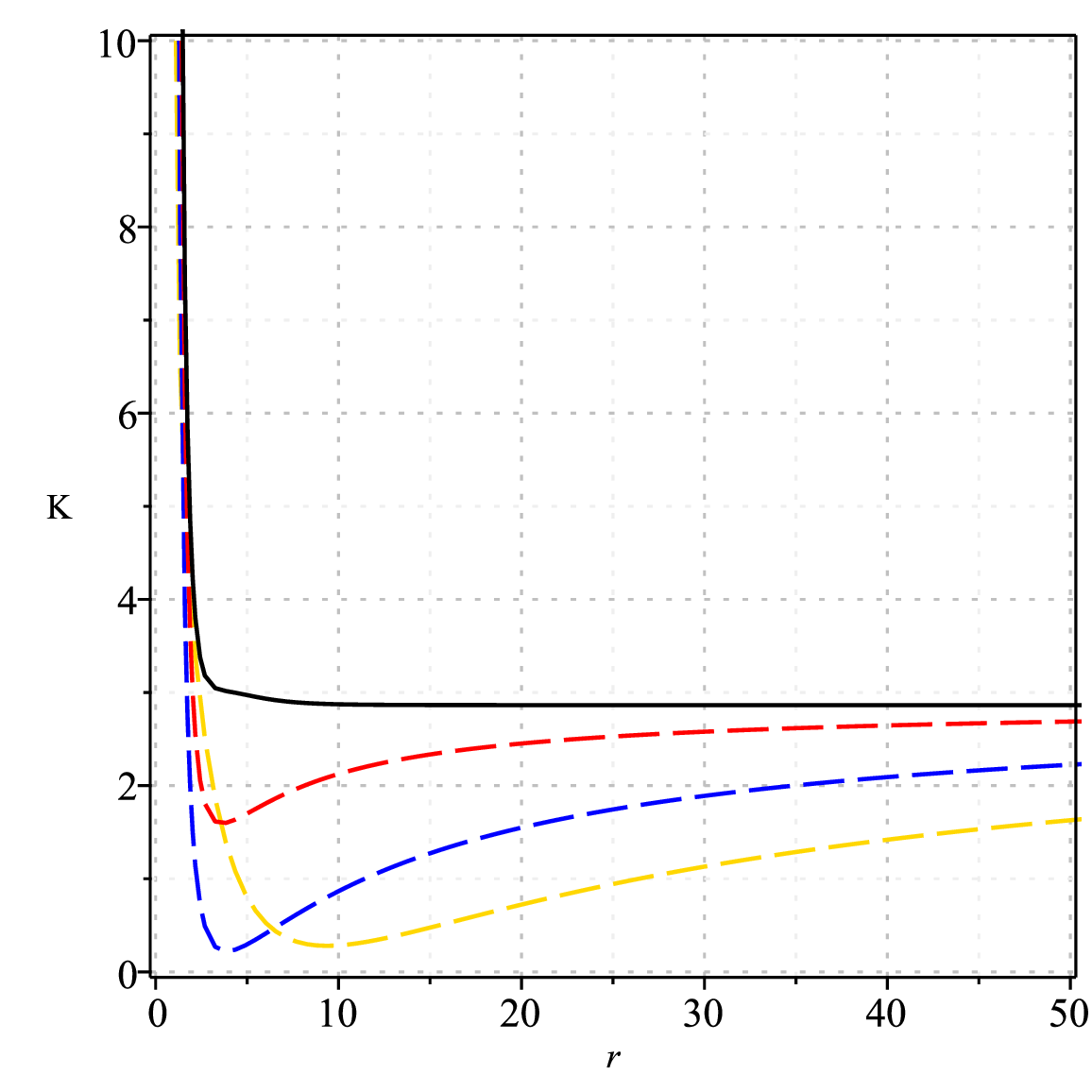}}\hfill
\caption{$M=5$, $Q=1$, $c=1$, $c_1=-1$ and $c_2=1$. $m=0$ denoted by solid black line, $m=1$ denoted by dash red line, $m=2$ denoted by blue dash line, $m=3$ denoted by gold dash line,}\label{fig:21}
\end{figure}
 
In Fig. \ref{fig:21}, Kretschmann scalar is plotted. The figure shows that $4D$ Einstein--Gauss--Bonnet  massive gravity black hole has a true singularity at $r=0$.

\end{document}